\let\orilabel\label
\documentclass[aps,pra,twocolumn,superscriptaddress,10pt]{revtex4-2}
\let\label\orilabel

\usepackage[caption=false]{subfig} 	
\usepackage{graphicx}  				
\graphicspath{{figure/}}
\usepackage{mathtools}				
\usepackage{amssymb}   				
\usepackage{amsthm}    				
\usepackage{acronym}				
\usepackage{algorithmicx}			
\usepackage[noend]{algpseudocode}	
\usepackage{tabularx}				
\usepackage{etoolbox}				
\usepackage{placeins}				
\usepackage{tcolorbox}				
\usepackage{paralist}				
\usepackage{enumitem}				
\usepackage{listings}				
\usepackage{verbatim}				
\usepackage{hyperref}
\usepackage{xcolor}
\usepackage{dsfont}

\newtheorem{lemma}{Lemma}
\newtheorem{corollary}{Corollary}

\newtheorem{theorem}{Theorem}

\makeatletter
\renewcommand{\maketag@@@}[1]{\hbox{\m@th\normalsize\normalfont#1}}%
\makeatother

\usepackage{xcolor}
\definecolor{Jblue}{HTML}{0066CC}
\definecolor{Jgrey}{HTML}{8A8888}
\definecolor{Jred}{HTML}{D70040}
\definecolor{Jgreen}{HTML}{666633}

\newif\ifcmnt
\cmnttrue

\ifcmnt
	\newenvironment{cmng}{\begingroup\color{Jgrey}}{\endgroup}
	\newenvironment{cmnb}{\begingroup\color{Jblue}}{\endgroup}
	\newenvironment{cmnr}{\begingroup\color{Jred}}{\endgroup}
\else

\fi

\usepackage{soul}
\setulcolor{red}

\allowdisplaybreaks

\begin{document}

\title{Gaussian Quantum Integrated Sensing and Communication}

\author{Tuna Erdo\u{g}an}
\affiliation{Georgia Institute of Technology}
\author{Shi-Yuan Wang}
\affiliation{Qualcomm Technologies Inc.}
\thanks{S.-Y. Wang contributed to the present work while pursuing a Ph.D. degree at the Georgia Institute of Technology.}
\thanks{This work was supported in part by the National Science Foundation (NSF) under grant 2148400 as part of the Resilient \& Intelligent NextG Systems (RINGS) program.}
\author{Shang-Jen Su}
\affiliation{Georgia Institute of Technology}
\author{Matthieu R. Bloch}
\affiliation{Georgia Institute of Technology}

\begin{abstract}
  We consider a quantum integrated sensing and communication problem over an optical link in which a low-power transmitter simultaneously communicates with a receiver and identifies the range of a defect producing a backscattered signal. We model the system as a lossy thermal-noise bosonic channel, in which the target, modeled as a beamsplitter, affects the timing of the backscattered signal through its location. Motivated by the envisioned deployment of entanglement-enabled quantum networks, we develop two protocols that allow the transmitter to leverage shared Gaussian entanglement, in the form of independent two-mode squeezed vacuum states or coupled three-mode squeezed vacuum states, to assist both sensing and communication. Since entanglement is known to enhance sensing, as demonstrated in quantum illumination, and to increase communication rates through entanglement-assisted communication, the transmitter faces a trade-off in allocating its entanglement resources between the two tasks. Our main results characterize these trade-offs under the proposed protocols in terms of achievable communication rate/sensing error-exponent regions, which outperform time-sharing schemes and demonstrate a quantum advantage.
\end{abstract}

\maketitle

\acrodef{ACDIS}[ACDIS]{Adaptive Communication Decision and Information Systems}
\acrodef{AEP}{Asymptotic Equipartition Property}
\acrodef{AoA}{Angle of Arrival}
\acrodef{AWGN}{Additive White Gaussian Noise}
\acrodef{AVC}[AVC]{Arbitrarily Varying Channel}
\acrodef{BER}{Bit-Error-Rate}
\acrodef{BEC}{Binary Erasure Channel}
\acrodef{BPSK}{Binary Phase-Shift Keying}
\acrodef{BSC}{Binary Symmetric Channel}
\acrodef{BICM}[BICM]{Bit-Interleaved Coded-Modulation}
\acrodef{C3MSV}[C3MSV]{coupled three-mode squeezed vacuum}
\acrodef{CDF}[CDF]{Cumulative Distribution Function}
\acrodef{CGF}[CGF]{Cumulant Generating Function}
\acrodef{CLT}[CLT]{Central Limit Theorem}
\acrodef{CSI}[CSI]{Channel State Information}
\acrodef{DMC}[DMC]{Discrete Memoryless Channel}
\acrodef{DMS}[DMS]{Discrete Memoryless Source}
\acrodef{ERM}[ERM]{Empirical Risk Minimization}
\acrodef{FER}[FER]{Frame Error Rate}
\acrodef{ICA}[ICA]{Independent Component Analysis}
\acrodef{iid}[i.i.d.]{independent and identically distributed}
\acrodef{IoT}[IoT]{Internet of Things}
\acrodef{KKT}[KKT]{Karush-Kuhn Tucker}
\acrodef{LASSO}[LASSO]{Least Absolute Shrinkage and Selection Operator}
\acrodef{LPD}[LPD]{Low Probability of Detection}
\acrodef{LDPC}[LDPC]{Low-Density Parity-Check}
\acrodef{LLMS}[LLMS]{Linear Least Mean Square}
\acrodef{LMS}[LMS]{Least Mean Square}
\acrodef{MAC}[MAC]{multiple-access channel}
\acrodef{MGF}[MGF]{Moment Generating Function}
\acrodef{MLC}[MLC]{Multi-Level Coding}
\acrodef{MLE}[MLE]{Maximum Likelihood Estimate}
\acrodef{MIMO}[MIMO]{Multiple-Input Multiple-Output}
\acrodef{MISO}{Multiple-Input Single-Output}
\acrodef{MSD}[MSD]{Multi-Stage Decoding}
\acrodef{MMSE}[MMSE]{Minimum Mean-Square Error}
\acrodef{NPS}[NPS]{no passive signature}
\acrodef{PAC}[PAC]{Probably Approximately Correct}
\acrodef{PCA}[PCA]{Principal Component Analysis}
\acrodef{PDF}[PDF]{Probability Density Function}
\acrodef{PMF}[PMF]{Probability Mass Function}
\acrodef{POVM}[POVM]{Positive Operator-Valued Measure}
\acrodef{PVM}[PVM]{Projection-Valued Measure}
\acrodef{PPM}[PPM]{Pulse Position Modulation}
\acrodef{PSD}{Power Spectral Density}
\acrodef{PSK}{Phase Shift Keying}
\acrodef{QI}{quantum illumination}
\acrodef{QKD}{Quantum Key Distribution}
\acrodef{ROC}{Receiver Operating Characteristic}
\acrodef{c-q}{Classical-Quantum}
\acrodef{CVQKD}{Continuous-Variable \ac{QKD}}
\acrodef{QKF}[QMF]{Quantum-Memory Free}
\acrodef{QPSK}{Quadrature Phase-Shift Keying}
\acrodef{QSDC}[QSDC]{Quantum Secure Direct Communication}
\acrodef{RV}{random variable}
\acrodef{SIMO}{Single-Input Multiple-Output}
\acrodef{SNR}{Signal-to-Noise Ratio}
\acrodef{SVM}[SVM]{Support Vector Machine}
\acrodef{TMSV}[TMSV]{two-mode squeezed vacuum}
\acrodef{TPCP}{Trace-Preserving Completely-Positive}
\acrodef{CPTP}{Completely-Positive Trace-Preserving}
\acrodef{wrt}[w.r.t.]{with respect to}
\acrodef{WSS}{Wide Sense Stationary}
\acrodef{QCB}{Quantum Chernoff Bound}
\acrodef{QISAC}{quantum integrated sensing and communication}
\acrodef{ISAC}{integrated sensing and communication}
\acrodef{RF}{radio-frequency}
\newcommand{\matA}{\mathbf{A}}
\newcommand{\bfA}{\mathbf{A}}
\newcommand{\calA}{\mathcal{A}}
\newcommand{\bbA}{\mathbb{A}}
\newcommand{\bfa}{\mathbf{a}}
\newcommand{\veca}{\mathbf{a}}
\newcommand{\matB}{\mathbf{B}}
\newcommand{\bfB}{\mathbf{B}}
\newcommand{\calB}{\mathcal{B}}
\newcommand{\bbB}{\mathbb{B}}
\newcommand{\bfb}{\mathbf{b}}
\newcommand{\vecb}{\mathbf{b}}
\newcommand{\matC}{\mathbf{C}}
\newcommand{\bfC}{\mathbf{C}}
\newcommand{\calC}{\mathcal{C}}
\newcommand{\bbC}{\mathbb{C}}
\newcommand{\bfc}{\mathbf{c}}
\newcommand{\vecc}{\mathbf{c}}
\newcommand{\matD}{\mathbf{D}}
\newcommand{\bfD}{\mathbf{D}}
\newcommand{\calD}{\mathcal{D}}
\newcommand{\bbD}{\mathbb{D}}
\newcommand{\bfd}{\mathbf{d}}
\newcommand{\vecd}{\mathbf{d}}
\newcommand{\matE}{\mathbf{E}}
\newcommand{\bfE}{\mathbf{E}}
\newcommand{\calE}{\mathcal{E}}
\newcommand{\bbE}{\mathbb{E}}
\newcommand{\bfe}{\mathbf{e}}
\newcommand{\vece}{\mathbf{e}}
\newcommand{\matF}{\mathbf{F}}
\newcommand{\bfF}{\mathbf{F}}
\newcommand{\calF}{\mathcal{F}}
\newcommand{\bbF}{\mathbb{F}}
\newcommand{\bff}{\mathbf{f}}
\newcommand{\vecf}{\mathbf{f}}
\newcommand{\matG}{\mathbf{G}}
\newcommand{\bfG}{\mathbf{G}}
\newcommand{\calG}{\mathcal{G}}
\newcommand{\bbG}{\mathbb{G}}
\newcommand{\bfg}{\mathbf{g}}
\newcommand{\vecg}{\mathbf{g}}
\newcommand{\matH}{\mathbf{H}}
\newcommand{\bfH}{\mathbf{H}}
\newcommand{\calH}{\mathcal{H}}
\newcommand{\bbH}{\mathbb{H}}
\newcommand{\bfh}{\mathbf{h}}
\newcommand{\vech}{\mathbf{h}}
\newcommand{\matI}{\mathbf{I}}
\newcommand{\bfI}{\mathbf{I}}
\newcommand{\calI}{\mathcal{I}}
\newcommand{\bbI}{\mathbb{I}}
\newcommand{\bfi}{\mathbf{i}}
\newcommand{\veci}{\mathbf{i}}
\newcommand{\matJ}{\mathbf{J}}
\newcommand{\bfJ}{\mathbf{J}}
\newcommand{\calJ}{\mathcal{J}}
\newcommand{\bbJ}{\mathbb{J}}
\newcommand{\bfj}{\mathbf{j}}
\newcommand{\vecj}{\mathbf{j}}
\newcommand{\matK}{\mathbf{K}}
\newcommand{\bfK}{\mathbf{K}}
\newcommand{\calK}{\mathcal{K}}
\newcommand{\bbK}{\mathbb{K}}
\newcommand{\bfk}{\mathbf{k}}
\newcommand{\veck}{\mathbf{k}}
\newcommand{\matL}{\mathbf{L}}
\newcommand{\bfL}{\mathbf{L}}
\newcommand{\calL}{\mathcal{L}}
\newcommand{\bbL}{\mathbb{L}}
\newcommand{\bfl}{\mathbf{l}}
\newcommand{\vecl}{\mathbf{l}}
\newcommand{\matM}{\mathbf{M}}
\newcommand{\bfM}{\mathbf{M}}
\newcommand{\calM}{\mathcal{M}}
\newcommand{\bbM}{\mathbb{M}}
\newcommand{\bfm}{\mathbf{m}}
\newcommand{\vecm}{\mathbf{m}}
\newcommand{\matN}{\mathbf{N}}
\newcommand{\bfN}{\mathbf{N}}
\newcommand{\calN}{\mathcal{N}}
\newcommand{\bbN}{\mathbb{N}}
\newcommand{\bfn}{\mathbf{n}}
\newcommand{\vecn}{\mathbf{n}}
\newcommand{\matO}{\mathbf{O}}
\newcommand{\bfO}{\mathbf{O}}
\newcommand{\calO}{\mathcal{O}}
\newcommand{\bbO}{\mathbb{O}}
\newcommand{\bfo}{\mathbf{o}}
\newcommand{\veco}{\mathbf{o}}
\newcommand{\matP}{\mathbf{P}}
\newcommand{\bfP}{\mathbf{P}}
\newcommand{\calP}{\mathcal{P}}
\newcommand{\bbP}{\mathbb{P}}
\newcommand{\bfp}{\mathbf{p}}
\newcommand{\vecp}{\mathbf{p}}
\newcommand{\matQ}{\mathbf{Q}}
\newcommand{\bfQ}{\mathbf{Q}}
\newcommand{\calQ}{\mathcal{Q}}
\newcommand{\bbQ}{\mathbb{Q}}
\newcommand{\bfq}{\mathbf{q}}
\newcommand{\vecq}{\mathbf{q}}
\newcommand{\matR}{\mathbf{R}}
\newcommand{\bfR}{\mathbf{R}}
\newcommand{\calR}{\mathcal{R}}
\newcommand{\bbR}{\mathbb{R}}
\newcommand{\bfr}{\mathbf{r}}
\newcommand{\vecr}{\mathbf{r}}
\newcommand{\matS}{\mathbf{S}}
\newcommand{\bfS}{\mathbf{S}}
\newcommand{\calS}{\mathcal{S}}
\newcommand{\bbS}{\mathbb{S}}
\newcommand{\bfs}{\mathbf{s}}
\newcommand{\vecs}{\mathbf{s}}
\newcommand{\matT}{\mathbf{T}}
\newcommand{\bfT}{\mathbf{T}}
\newcommand{\calT}{\mathcal{T}}
\newcommand{\bbT}{\mathbb{T}}
\newcommand{\bft}{\mathbf{t}}
\newcommand{\vect}{\mathbf{t}}
\newcommand{\matU}{\mathbf{U}}
\newcommand{\bfU}{\mathbf{U}}
\newcommand{\calU}{\mathcal{U}}
\newcommand{\bbU}{\mathbb{U}}
\newcommand{\bfu}{\mathbf{u}}
\newcommand{\vecu}{\mathbf{u}}
\newcommand{\matV}{\mathbf{V}}
\newcommand{\bfV}{\mathbf{V}}
\newcommand{\calV}{\mathcal{V}}
\newcommand{\bbV}{\mathbb{V}}
\newcommand{\bfv}{\mathbf{v}}
\newcommand{\vecv}{\mathbf{v}}
\newcommand{\matW}{\mathbf{W}}
\newcommand{\bfW}{\mathbf{W}}
\newcommand{\calW}{\mathcal{W}}
\newcommand{\bbW}{\mathbb{W}}
\newcommand{\bfw}{\mathbf{w}}
\newcommand{\vecw}{\mathbf{w}}
\newcommand{\matX}{\mathbf{X}}
\newcommand{\bfX}{\mathbf{X}}
\newcommand{\calX}{\mathcal{X}}
\newcommand{\bbX}{\mathbb{X}}
\newcommand{\bfx}{\mathbf{x}}
\newcommand{\vecx}{\mathbf{x}}
\newcommand{\matY}{\mathbf{Y}}
\newcommand{\bfY}{\mathbf{Y}}
\newcommand{\calY}{\mathcal{Y}}
\newcommand{\bbY}{\mathbb{Y}}
\newcommand{\bfy}{\mathbf{y}}
\newcommand{\vecy}{\mathbf{y}}
\newcommand{\matZ}{\mathbf{Z}}
\newcommand{\bfZ}{\mathbf{Z}}
\newcommand{\calZ}{\mathcal{Z}}
\newcommand{\bbZ}{\mathbb{Z}}
\newcommand{\bfz}{\mathbf{z}}
\newcommand{\vecz}{\mathbf{z}}
\newcommand{\bfalpha}{\boldsymbol{\alpha}}
\newcommand{\vecalpha}{\boldsymbol{\alpha}}
\newcommand{\bfbeta}{\boldsymbol{\beta}}
\newcommand{\vecbeta}{\boldsymbol{\beta}}
\newcommand{\bfgamma}{\boldsymbol{\gamma}}
\newcommand{\vecgamma}{\boldsymbol{\gamma}}
\newcommand{\bfdelta}{\boldsymbol{\delta}}
\newcommand{\vecdelta}{\boldsymbol{\delta}}
\newcommand{\bfepsilon}{\boldsymbol{\epsilon}}
\newcommand{\vecepsilon}{\boldsymbol{\epsilon}}
\newcommand{\bfzeta}{\boldsymbol{\zeta}}
\newcommand{\veczeta}{\boldsymbol{\zeta}}
\newcommand{\bfeta}{\boldsymbol{\eta}}
\newcommand{\veceta}{\boldsymbol{\eta}}
\newcommand{\bftheta}{\boldsymbol{\theta}}
\newcommand{\vectheta}{\boldsymbol{\theta}}
\newcommand{\bfiota}{\boldsymbol{\iota}}
\newcommand{\veciota}{\boldsymbol{\iota}}
\newcommand{\bfkappa}{\boldsymbol{\kappa}}
\newcommand{\veckappa}{\boldsymbol{\kappa}}
\newcommand{\bflamda}{\boldsymbol{\lamda}}
\newcommand{\veclamda}{\boldsymbol{\lamda}}
\newcommand{\bfmu}{\boldsymbol{\mu}}
\newcommand{\vecmu}{\boldsymbol{\mu}}
\newcommand{\bfnu}{\boldsymbol{\nu}}
\newcommand{\vecnu}{\boldsymbol{\nu}}
\newcommand{\bfxi}{\boldsymbol{\xi}}
\newcommand{\vecxi}{\boldsymbol{\xi}}
\newcommand{\bfomicron}{\boldsymbol{\omicron}}
\newcommand{\vecomicron}{\boldsymbol{\omicron}}
\newcommand{\bfpi}{\boldsymbol{\pi}}
\newcommand{\vecpi}{\boldsymbol{\pi}}
\newcommand{\bfrho}{\boldsymbol{\rho}}
\newcommand{\vecrho}{\boldsymbol{\rho}}
\newcommand{\bfsigma}{\boldsymbol{\sigma}}
\newcommand{\vecsigma}{\boldsymbol{\sigma}}
\newcommand{\bftau}{\boldsymbol{\tau}}
\newcommand{\vectau}{\boldsymbol{\tau}}
\newcommand{\bfupsilon}{\boldsymbol{\upsilon}}
\newcommand{\vecupsilon}{\boldsymbol{\upsilon}}
\newcommand{\bfphi}{\boldsymbol{\phi}}
\newcommand{\vecphi}{\boldsymbol{\phi}}
\newcommand{\bfchi}{\boldsymbol{\chi}}
\newcommand{\vecchi}{\boldsymbol{\chi}}
\newcommand{\bfpsi}{\boldsymbol{\psi}}
\newcommand{\vecpsi}{\boldsymbol{\psi}}
\newcommand{\bfomega}{\boldsymbol{\omega}}
\newcommand{\vecomega}{\boldsymbol{\omega}}
\newcommand{\bfAlpha}{\boldsymbol{\Alpha}}
\newcommand{\matAlpha}{\boldsymbol{\Alpha}}
\newcommand{\bfBeta}{\boldsymbol{\Beta}}
\newcommand{\matBeta}{\boldsymbol{\Beta}}
\newcommand{\bfGamma}{\boldsymbol{\Gamma}}
\newcommand{\matGamma}{\boldsymbol{\Gamma}}
\newcommand{\bfDelta}{\boldsymbol{\Delta}}
\newcommand{\matDelta}{\boldsymbol{\Delta}}
\newcommand{\bfEpsilon}{\boldsymbol{\Epsilon}}
\newcommand{\matEpsilon}{\boldsymbol{\Epsilon}}
\newcommand{\bfZeta}{\boldsymbol{\Zeta}}
\newcommand{\matZeta}{\boldsymbol{\Zeta}}
\newcommand{\bfEta}{\boldsymbol{\Eta}}
\newcommand{\matEta}{\boldsymbol{\Eta}}
\newcommand{\bfTheta}{\boldsymbol{\Theta}}
\newcommand{\matTheta}{\boldsymbol{\Theta}}
\newcommand{\bfIota}{\boldsymbol{\Iota}}
\newcommand{\matIota}{\boldsymbol{\Iota}}
\newcommand{\bfKappa}{\boldsymbol{\Kappa}}
\newcommand{\matKappa}{\boldsymbol{\Kappa}}
\newcommand{\bfLamda}{\boldsymbol{\Lamda}}
\newcommand{\matLamda}{\boldsymbol{\Lamda}}
\newcommand{\bfMu}{\boldsymbol{\Mu}}
\newcommand{\matMu}{\boldsymbol{\Mu}}
\newcommand{\bfNu}{\boldsymbol{\Nu}}
\newcommand{\matNu}{\boldsymbol{\Nu}}
\newcommand{\bfXi}{\boldsymbol{\Xi}}
\newcommand{\matXi}{\boldsymbol{\Xi}}
\newcommand{\bfOmicron}{\boldsymbol{\Omicron}}
\newcommand{\matOmicron}{\boldsymbol{\Omicron}}
\newcommand{\bfPi}{\boldsymbol{\Pi}}
\newcommand{\matPi}{\boldsymbol{\Pi}}
\newcommand{\bfRho}{\boldsymbol{\Rho}}
\newcommand{\matRho}{\boldsymbol{\Rho}}
\newcommand{\bfSigma}{\boldsymbol{\Sigma}}
\newcommand{\matSigma}{\boldsymbol{\Sigma}}
\newcommand{\bfTau}{\boldsymbol{\Tau}}
\newcommand{\matTau}{\boldsymbol{\Tau}}
\newcommand{\bfUpsilon}{\boldsymbol{\Upsilon}}
\newcommand{\matUpsilon}{\boldsymbol{\Upsilon}}
\newcommand{\bfPhi}{\boldsymbol{\Phi}}
\newcommand{\matPhi}{\boldsymbol{\Phi}}
\newcommand{\bfChi}{\boldsymbol{\Chi}}
\newcommand{\matChi}{\boldsymbol{\Chi}}
\newcommand{\bfPsi}{\boldsymbol{\Psi}}
\newcommand{\matPsi}{\boldsymbol{\Psi}}
\newcommand{\bfOmega}{\boldsymbol{\Omega}}
\newcommand{\matOmega}{\boldsymbol{\Omega}}
\newcommand{\bfzero}{\boldsymbol{0}}

\newcommand{\card}[1]{{|{#1}|}}
\newcommand{\abs}[1]{{\left|{#1}\right|}}
\newcommand{\sgn}[1]{{\text{sgn}\left({#1}\right)}}
\newcommand{\norm}[2][]{{\left\Vert{#2}\right\Vert}_{#1}}
\newcommand{\inprod}[3][]{{\left\langle{#2},{#3}\right\rangle}_{#1}}
\newcommand{\dotp}[3][]{{\left\langle{#2},{#3}\right\rangle}_{#1}}
\newcommand{\set}[1]{{\{#1\}}}
\newcommand{\trace}[1]{\text{trace}\left(#1\right)}
\newcommand{\argmin}{\mathop{\text{argmin}}}
\newcommand{\argmax}{\mathop{\text{argmax}}}
\newcommand{\intseq}[2]{[{#1};{#2}]}
\newcommand{\eqdef}{\triangleq}
\newcommand{\E}[2][]{{\mathbb{E}_{#1}\left[#2\right]}}
\newcommand{\Var}[1]{{\text{Var}\left(#1\right)}}
\newcommand{\Cov}[1]{{\text{Cov}\left(#1\right)}}
\renewcommand{\P}[2][]{{\mathbb{P}_{#1}\left(#2\right)}}
\newcommand{\indic}[1]{{\mathbf{1}\{#1\}}}
\newcommand{\mat}[2]{\left[\begin{array}{#1}#2\end{array}\right]}
\newcommand{\bra}[1]{{\left\langle{#1}\right\vert}}
\newcommand{\ket}[1]{{\left\vert{#1}\right\rangle}}

\newcommand{\D}[3][]{{\mathbb{D}_{#1}}\!\left(#2\,\middle\Vert\,#3\right)} 
\newcommand{\V}[3][]{{\mathbb{V}_{#1}}\!\left(#2,#3\right)} 
\newcommand{\I}[2]{{{\mathbb{I}}\!\left(#1;#2\right)}} 
\renewcommand{\H}[1]{{\mathbb{H}}\!\left(#1\right)}
\newcommand{\h}[1]{{{h}}\left(#1\right)} 
\newcommand{\Hb}[1]{{h_b}\left(#1\right)} 

\newcommand{\GF}[1]{\ensuremath{\mbox{GF}(#1)}} 
\newcommand{\F}[1]{\ensuremath{\mathbb{F}_{#1}}} 
\newcommand{\dH}[1]{\ensuremath{\mbox{d}_{\mbox{\tiny H}}(#1)}} 
\newcommand{\wt}[1]{\ensuremath{\mbox{wt}(#1)}} 

\newcommand{\dd}{\text{\textnormal{d}}} 
\renewcommand{\leq}{\leqslant} 
\renewcommand{\geq}{\geqslant} 

\newcommand{\tr}[2][]{\ensuremath{\text{\textnormal{tr}}_{#1}\left(#2\right)}} 
\newcommand{\rk}[1]{\ensuremath{\text{\textnormal{rank}}\left(#1\right)}}  
\renewcommand{\det}[1]{{\left|{#1}\right|}}                              
\renewcommand{\ker}[1]{\ensuremath{\text{\textnormal{Ker}}\left(#1\right)}}  

\newcommand{\floor}[1]{\lfloor#1\rfloor}
\newcommand{\ceil}[1]{\left\lceil#1\right\rceil}

\section{Introduction}
\label{sec:intro}

The Quantum Internet is envisioned as an entanglement sharing network allowing distant nodes to harness quantum advantages for communication, sensing, and computation~\cite{Kimble2008QuantumInternet,Wehner2018QuantumInternet}. While there has been significant experimental progress demonstrating entanglement sharing at large distances and high rates~\cite{CraddockDistributionNY2024,LuoEntanglingMemories2026,knaut2024entanglement}, the limitations of current quantum memories and nascent quantum computers prevent one from truly using entanglement as a universal resource for arbitrary quantum information processing. Experimental demonstrations have thus far focused on demonstrating specific entanglement-assisted tasks, for which quantum processing and detection can be performed with experimentally accessible hardware~\cite{XuExperimentalQuantumTarget2021,WardEntanglementNP2025,HaoEASurpassCC2021}.

There has recently been interest in exploring the entanglement benefits of situations in which distinct tasks, such as communication and sensing, try to simultaneously leverage quantum entanglement~\cite{erdogan2025,YaoJCS2025}. In general, the problem is intricate because the tasks may not be aligned, in the sense that a measurement optimal for one task may not be optimal for the other, making the characterization of the optimal joint measurement difficult. Recent works have largely circumvented this challenge by considering scenarios in which the tension could be resolved.

Our objective is to develop preliminary insights into the performance trade-offs that result from a joint operation by formulating an exemplar combining two entanglement-assisted tasks: communication and sensing. The exemplar is motivated by recent experimental results reporting \ac{QISAC} in optical fiber~\cite{LiuINtegratedQuantum2024,XuIngeratedDistributed2024}, as well as the usefulness of entanglement to improve detection in \ac{QI}~\cite{Wilde2017} and increase communication rates over bosonic channels~\cite{Bennett2002}. The model proposed also extends our prior work on \ac{QISAC} that only considered finite-dimensional Hilbert spaces and did not include entanglement assistance~\cite{Wang2022Joint}, and extends recent independently obtained results on \ac{QISAC} over lossy bosonic channels without entanglement~\cite{MunarVallespir2024Joint}. 

Specifically, we consider here a lossy thermal-noise bosonic channel in which a transmitter simultaneously wishes to communicate with a receiver and range the location of the target modeled as a beamsplitter~\cite{Zhuang2021}. We analyze three protocols: \begin{inparaenum}[(i)]
\item a fully classical baseline without entanglement assistance;
\item a protocol in which the transmitter is assisted by an entangled \ac{TMSV} resource~\cite{Weedbrook2012}, consisting of a signal-idler pair; and
\item a protocol employing a \ac{C3MSV}~\cite{Zhan2023C3MSV} resource, consisting of one signal mode and two idler modes.
\end{inparaenum}In the \ac{TMSV}-assisted protocol, the transmitter encodes information on the signal mode and faces a dilemma over the idler: it can transmit the idler to the receiver, e.g., through an entanglement-sharing network, to enable entanglement-assisted communication, or retain the idler in a local quantum memory to enable \ac{QI}. In the \ac{C3MSV}-assisted protocol, the transmitter retains one idler mode for ranging while sharing the other with the receiver to assist communication; here the trade-off is governed by how the strength of the entanglement with the signal mode is allocated between the idler modes. We characterize the resulting performance trade-off as an information-theoretic achievable communication rate/sensing error-exponent region, which reveals a quantum advantage for both entanglement-assisted protocols. Our \ac{C3MSV} protocol provides an improvement over the \ac{TMSV} protocol, further suggesting that distributing entanglement resources across jointly performed tasks is preferable to time-sharing the entanglement resource among tasks.

In classical communication, \ac{ISAC} has attracted significant interest as a potential technology harvesting the anticipated convergence between radar and communication systems in 6G, driven by the growing demand to equip interconnected devices with the ability to sense and interpret their surroundings~\cite{Paul2017Survey,Fettweis2021Joint}. Consequently, there is a vast \ac{ISAC} literature in classical information theory investigating various models: capacity/distortion/cost trade-offs for \ac{iid} channel states~\cite{Ahmadipour2024}, rate/error-exponent trade-offs for binary channel states~\cite{Joudeh2022}, and rate/error-exponent tradeoffs under monostatic and bistatic detection models~\cite{Chang2022,Chang2023}. The \ac{QISAC} literature is much less extensive, but several models have been investigated: rate/distortion trade-offs~\cite{2022PeregCommEst}, rate/error-exponent trade-offs~\cite{Munar2024}, and entanglement-assisted joint communication and change-point detection~\cite{Gong2025JcommChangePD}. In the present work, we focus on the entanglement-assisted \ac{QISAC} trade-offs in the form of sensing error-exponent and communication rate.

\section{Preliminaries}
\label{sec:prelim}
We start by reviewing necessary notation, concept, and results. A lossy thermal-noise bosonic channel is characterized by its transmissivity $\eta\in(0,1)$, the noise mean photon number $N_B$, and the input-output relations
\begin{flalign*}
    &\hat{b} = \sqrt{\eta}\hat{a}+\sqrt{1-\eta}\hat{e},\\
    &\hat{d} = -\sqrt{1-\eta}\hat{a}+\sqrt{\eta}\hat{e},
\end{flalign*}
where $\hat{a}$ is the annihilation operator of the input mode, $\hat{e}$ is the annihilation operator of the thermal environment mode with mean photon number $N_B$ and density operator in the Fock basis $\hat{\rho}_\textnormal{th}(N_B)\eqdef\sum_{n=0}^\infty\frac{N_B^n}{(N_B+1)^{n+1}}\ket{n}\bra{n}$, $\hat{b}$ is the annihilation operator of the accessible output mode and $\hat{d}$ is the inaccessible environment output. The \ac{CPTP} map corresponding to a lossy thermal-noise bosonic channel with transmissivity $\eta$ and noise mean photon number $N_B$ is denoted as $\calL^{(\eta,N_B)}_{A\to BD}$ where $A$ is the channel input system, $B$ is the accessible output system and $D$ is the inaccessible environment output system. Similarly, $\calL^{(\eta,N_B)}_{A\to B}[\cdot]\eqdef \tr[D]{\calL^{(\eta,N_B)}_{A\to BD}[\cdot]}$. The Von Neumann entropy of a quantum state $\hat{\rho}$ is defined as $S(\hat{\rho})\eqdef-\tr{\rho\log\rho}$. The Holevo information of a quantum ensemble $\left\{p_i,\hat{\rho}_i\right\}_{i}$ is denoted as $\chi\left(\left\{p_i,\hat{\rho}_i\right\}_{i}\right)\eqdef S\left(\sum_ip_i\hat{\rho}_i\right)-\sum_i p_iS\left(\hat{\rho}_i\right)$.

We leverage two Gaussian quantum resources for \ac{QISAC}. We consider a bipartite \ac{TMSV} state with Fock space representation
\begin{flalign}
    \ket{\psi}^{\textnormal{TMSV}}_{SI} \eqdef \sum_{n=0}^\infty\sqrt{\frac{N_S^n}{(N_S+1)^{n+1}}}\ket{n}_S\ket{n}_I,
\end{flalign}
and covariance matrix
\begin{align}
    V^{\textnormal{TMSV}}_{SI} &= \begin{bmatrix}
        (2N_S+1)\bfI & 2\sqrt{N_S(N_S+1)}\bfZ\\
        2\sqrt{N_S(N_S+1)}\bfZ & (2N_S+1)\bfI
    \end{bmatrix},
\end{align}
where $\bfI=\begin{bmatrix}
    1 & 0\\ 0 & 1
\end{bmatrix}$ and $\bfZ =\begin{bmatrix}
    1 & 0 \\ 0 & -1
\end{bmatrix}$. The \ac{TMSV} state is the maximally entangled state for a given mean photon number $N_S$~\cite{Wolf2006GaussianExtremality}. The \ac{TMSV} state is known to enable entanglement-assisted classical communication over lossy thermal-noise bosonic channels~\cite{Bennett2002,Holevo2002EvaluatingCapacities} and to provide a 6-dB advantage over the best classical scheme in \ac{QI}~\cite{tan2008QI} and entanglement-assisted ranging~\cite{Zhuang2021}. We also consider a tripartite \ac{C3MSV} introduced in~\cite{Zhan2023C3MSV}, which exhibits one strong mode entangled with each of the remaining modes and two weak modes that are classically maximally correlated with each other. A \ac{C3MSV} state in its most general form, up to phase shifts, is given by the covariance matrix, for $\delta\in[0,1]$,
\begin{align}
    V_{S_1S_2S_3}^{\textnormal{C3MSV}}(\delta) = \begin{bmatrix}
        S_1(\delta)\bfI & C_{12}(\delta)\bfZ & C_{13}(\delta)\bfI\\
        C_{12}(\delta)\bfZ & S_2\bfI & C_{23}(\delta)\bfZ\\
     C_{13}(\delta)\bfI & C_{23}(\delta)\bfZ & S_3(\delta)\bfI
    \end{bmatrix},
    \label{eq:cov_c3msv}
\end{align}
where $S_1(\delta)=2\delta N_S+1$, $C_{12}(\delta)=-2\sqrt{\delta N_S(N_S+1)}$, $S_2=2N_S+1$, $C_{13}(\delta) = 2\sqrt{\delta(1-\delta)}N_S$, $C_{23}(\delta)=-2\sqrt{(1-\delta) N_S(N_S+1)}$, $S_3(\delta)=2(1-\delta)N_S+1$ and the parameter $\delta$ controls the imbalance between the idlers. The corresponding density operator is denoted $\hat{\rho}_{S_1S_2S_3}(\delta)$. Notice that choosing $\delta\in\{0,1\}$ results in the states $\ket{0}\otimes\ket{\psi}^{\textnormal{TMSV}}$ and $\ket{\psi}^{\textnormal{TMSV}}\otimes\ket{0}$ up to phase shifts. A \ac{C3MSV} state can also be generated by mixing one mode of a \ac{TMSV} state with vacuum through a $\delta$ (or $1-\delta$) transmissivity beam-splitter or by consecutively two-mode squeezing $(S_1,S_2)$ and two-mode squeezing $(S_2,S_3)$ modes in the given order with appropriate squeezing parameters. The correlation coefficient $C_{12}(\delta)$ between $S_1$ and $S_2$ (similarly $C_{23}(\delta)$ for $S_2$ and $S_3$) corresponds to entanglement between the modes and the correlation coefficient $C_{13}(\delta)$ between $S_1$ and $S_3$ corresponds to maximal classical correlation. We consider mode $S_2$ as the strong arm that is entangled with the other modes and consider $(S_1,S_3)$ as the weak arms because of the mere classical correlation between them and their entanglement to $S_2$.

We will use \ac{TMSV} and \ac{C3MSV} states as bipartite and tripartite resources for our protocols. We shall utilize newly established lower bounds on the achievable communication rates with quantum ring states~\cite{su2026} that we briefly review next. Let $\hat{\rho}_{SI}$ be a bipartite quantum state consisting of a signal arm ($S$) and an idler arm ($I$), and let $\hat{\rho}_{BI}^\theta$ be the state obtained by applying a $\theta$ phase shift to the signal arm of $\hat{\rho}_{SI}$ and transmitting the signal arm through a lossy thermal-noise bosonic channel. The quantum ring state, $\hat{\rho}_{BI}$, is the mixed state obtained at the receiver by uniformly phase modulating the signal arm, $\theta\in[0,2\pi)$, and keeping the idler as is, i.e., $\hat{\rho}_{BI}=\frac{1}{2\pi}\int\displaylimits_0^{2\pi}\hat{\rho}_{BI}^\theta\,\mathrm{d}\theta$. Since the resulting ring states are non-Gaussian, there is no known closed-form expression for their entropy. However, the entropy of ring states admits a tight, simple closed-form lower bound~\cite{su2026}. This makes ring states analytically tractable without requiring asymptotic assumptions on the physical parameters of the channel, the transmissivity $\eta$ and the thermal-noise mean photon number $N_B$, and the signal mean photon number $N_S$. In particular, communication with \ac{TMSV} ring states, with $\hat{\rho}_{SI}$ chosen to be a \ac{TMSV} state, tightly approximates the entanglement-assisted classical capacity, given by $C_\textnormal{EA}(N_S,N_T,\eta)\eqdef g(N_S)+g(\eta N_S+N_T)-g\left(\frac{\mu_+-1}{2}\right)-g\left(\frac{\mu_--1}{2}\right)$, for $N_T=(1-\eta)N_B$, where $g(x)\eqdef(x+1)\log(x+1)-x\log x$, and $\mu_\pm\eqdef\sqrt{(1+(1+\eta)N_S+N_T)^2-4\eta N_S(N_S+1)}\pm (N_T-(1-\eta)N_S)$, while efficiently consuming a single \ac{TMSV} pair per channel use. Similarly, communication with coherent ring states, with $\hat{\rho}_{SI} = \ket{\sqrt{N_S}}\bra{\sqrt{N_S}} \otimes \ket{\alpha}\bra{\alpha}$, closely approximates the classical capacity of the lossy thermal-noise bosonic channel, $C(N_S,N_T,\eta) \eqdef g(\eta N_S + N_T) - g(N_T)$. In this coherent case, system $I$ serves only as a phase reference for system $B$ at the receiver and does not contribute to the Holevo information.

The last preliminary result we need relates to symmetric quantum hypothesis testing. Our characterization of the sensing performance is based on the error-exponent of $m$-ary quantum hypothesis testing~\cite{cheng2026MQHT}. Asymptotics of symmetric quantum hypothesis testing have been established for finite-dimensional quantum systems with an achievable error-exponent~\cite{Li2016Discriminating} and a converse exponent~\cite{Nussbaum2009Chernoff}. A very recent result extends these previous results to infinite-dimensional spaces~\cite{cheng2026MQHT}. Specifically, the asymptotic symmetric error-exponent for testing between the quantum states $\{\pi_i,\hat{\rho}_i^{\otimes n}\}_{i=1}^m$, where $\pi_i$ is the prior of the hypothesis corresponding to $\hat{\rho}_i^{\otimes n}$, is given by the $m$-ary \ac{QCB}~\cite{cheng2026MQHT},
\begin{align}
    E \eqdef \min_{(i,j):i\neq j}\sup_{s\in[0,1]}-\log Q_s\left(\hat{\rho}_i\Vert\hat{\rho}_j\right),\label{eq:MQCB}
\end{align}
in which $Q_s(\hat{\rho}\Vert\hat{\sigma})\eqdef \tr{\hat{\rho}^s\hat{\sigma}^{1-s}}$ for quantum states $\hat{\rho}$ and $\hat{\sigma}$. For Gaussian states, \cite{Pirandola2008} provides a technique to compute $Q_s\left(\hat{\rho}\Vert\hat{\sigma}\right)$. While a closed-form expression for $Q_s\left(\hat{\rho}\Vert\hat{\sigma}\right)$ (or $-\log Q_s\left(\hat{\rho}\Vert\hat{\sigma}\right)$) is difficult to obtain in general, \eqref{eq:MQCB} can be numerically computed using the mean vectors and symplectic decompositions of the covariance matrices of the hypotheses for Gaussian states~\cite{Pirandola2008}.

\section{Gaussian QISAC with Entanglement Assistance}\label{sec:jbcs}
\begin{figure}[!htbp]
    \centering
    \includegraphics[width=\linewidth]{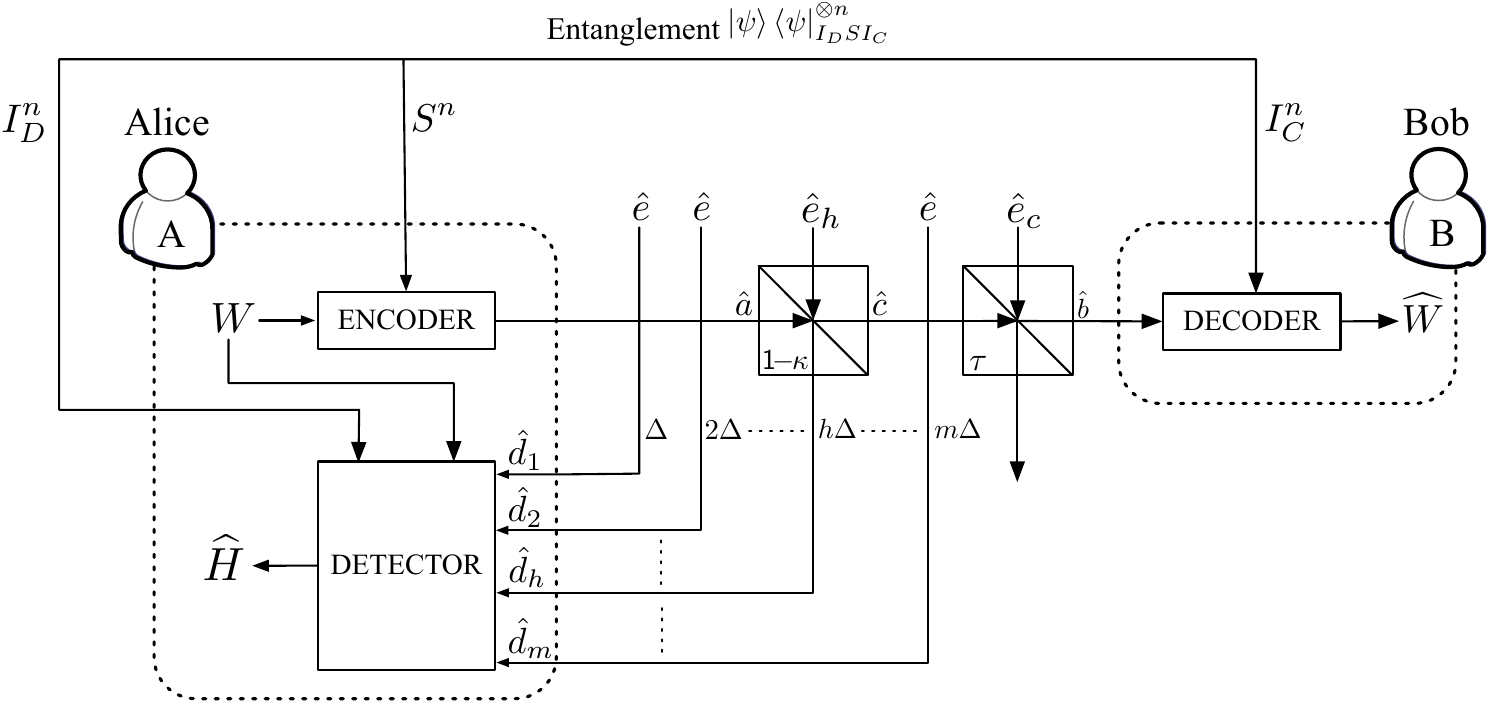}
    \caption{\ac{QISAC} model with tripartite entanglement resource.}
    \label{fig:Tripartite_model}
\end{figure}

We consider the \ac{QISAC} problem illustrated in Figure~\ref{fig:Tripartite_model}, in which the physical system consists of two concatenated single-mode lossy thermal-noise bosonic channels, $\calL^{(1-\kappa,N_\textnormal{th}/(1-\kappa))}_{A\to CD}$ and $\calL^{(\tau,N_B)}_{C\to B}$. The first channel captures the effect of a target present in the link as a beamsplitter, where $\kappa$ denotes the transmissivity of the backscattered signal and $N_\textnormal{th}/(1-\kappa)$ is the mean photon number of the environmental thermal-noise. The second channel models communication loss and noise between the transmitter and the receiver, where $\tau$ is the transmissivity and $N_B$ is the mean photon number of the thermal-noise. The transmitter (Alice) seeks to reliably communicate a message to a receiver (Bob) over multiple uses of the system while simultaneously estimating the range of the target through the backscattered signal. Alice's transmission is subject to a mean photon number constraint $N$ per channel use.

Following the ranging formulation in~\cite{Zhuang2021}, the sensing task is modeled as an $m$-ary hypothesis testing problem. The known distance between Alice and Bob is partitioned into $m+1$ intervals of equal length $\Delta$, with $m\geq2$. Under hypothesis $h\in\intseq{1}{m}$, the beamsplitter is located at position $h\Delta$. The returned mode corresponding to the $i$-th interval arrives during the time window $t\in\left[2i\frac{\Delta}{c},\,2(i+1)\frac{\Delta}{c}\right)$, for $i\in\intseq{1}{m}$, where $c$ is the speed of light.

Under hypothesis $h$, the bosonic annihilation operators satisfy
\begin{align}
    \hat{d}_h &= -\sqrt{\kappa}\hat{a}+\sqrt{1-\kappa}\hat{e}_h,\\
    \hat{d}_i &= \hat{e}_i,\quad i\neq h,\\
    \hat{c} &= \sqrt{1-\kappa}\hat{a}+\sqrt{\kappa}\hat{e}_h,\\
    \hat{b} &= \sqrt{\tau}\hat{c}+\sqrt{1-\tau}\hat{e}_c.
\end{align}

As in~\cite{Zhuang2021,tan2008QI,Nair20}, we adopt the \ac{NPS} assumption. Under hypothesis $h$, the environmental mode $\hat{e}_h$ is a thermal state with mean photon number $N_\textnormal{th}/(1-\kappa)$, while each $\hat{e}_i$ for $i\neq h$ is a thermal state with mean photon number $N_\textnormal{th}$. The \ac{NPS} assumption captures the indistinguishability of the hypotheses under a vacuum input to the channel, so that Alice must actively probe the channel in order to distinguish between the hypotheses. Without the \ac{NPS} assumption, there is no quantum advantage in detection performance in the low-$N$ regime, since both the entanglement-assisted and unassisted error-exponents converge to the constant vacuum-input performance as $N\to0$. The communication noise mode $\hat{e}_c$ is a thermal state with mean photon number $N_B$. For notational simplicity, we define the effective communication transmissivity $\eta \eqdef \tau(1-\kappa)$ and the effective communication noise photon number $N_T \eqdef \frac{\tau\kappa}{1-\kappa}N_\textnormal{th}+(1-\tau)N_B$.

Alice has access to $n$ independent pure tripartite entangled state $\hat{\rho}_{I_DSI_C}^{\otimes n}$, in which $I_D$ represents the idler mode she retains for detection to enhance sensing performance, $I_C$ is the idler mode she shares with Bob to assist communication, and $S$ is the signal mode to be encoded as the channel input. This framework includes bipartite entanglement as a special case, in which one idler mode is replaced by an independent vacuum state. Therefore, Alice's encoder is described by the encoder channels $\{ \textnormal{id}_{I_D^n}\otimes\calE^{(w)}_{S^n\to A^n}\otimes \textnormal{id}_{I_C^n}\}_{(w)}$ mapping the message $w\in\intseq{1}{M}$ to a $3n$-mode state $\hat{\rho}_{I_D^nA^nI_C^n}^{(w)}$, in which $A^n$ is entangled with the idler systems $(I^n_D,I^n_C)$. Upon transmission through the concatenated channels
$\calL^{(1-\kappa,\,N_{\textnormal{th}}/(1-\kappa))}_{A\to C}$
and
$\calL^{(\tau,\,N_B)}_{C\to B}$,
equivalently represented by the effective lossy thermal-noise channel $\calL^{(\eta,\,N_T/(1-\eta))}_{A\to B}$, the resulting state after partial tracing $I_D^n$ at the receiver is
\begin{align}
    \hat{\rho}_{B^n I_C^n}^{(w)}
    \eqdef
    \left(
    \left(\calL^{(\eta,\,N_T/(1-\eta))}_{A\to B}\right)^{\otimes n}
    \otimes \textnormal{id}_{I_C^n}
    \right)
    \left(\hat{\rho}^{(w)}_{A^n I_C^n}\right),
\end{align}
in which $\hat{\rho}^{(w)}_{A^n I_C^n}$ is obtained from $\hat{\rho}_{I_D^nA^nI_C^n}^{(w)}$ with mode $I_D^n$ traced out. Bob then applies a \ac{POVM}
$\{\Pi_w\}_{w
\in\intseq{1}{M}}$ over the modes $(B^n,I_C^n)$ to decode the message $w$, producing an estimate $\widehat{w}$. Similarly, upon receiving the returned state under hypothesis $h\in\{1,\dots,m\}$, the joint state at the detector is
\begin{flalign}
    \hat{\rho}_{\{D_j^n\}_{j=1}^m\, I_D^{n}}^{(h,w)}
    &\eqdef\;
    \left(
    \hat{\rho}_{\textnormal{th}}(N_{\textnormal{th}})^{\otimes n}
    \right)^{\otimes(h-1)}\nonumber\\
    &\otimes\left(
    \left(\calL^{(\kappa,\,N_{\textnormal{th}}/(1-\kappa))}_{A\to D}\right)^{\otimes n}
    \otimes \textnormal{id}_{I_D^n}
    \right)
    \left(\hat{\rho}^{(w)}_{A^nI_D^n}\right)
    \nonumber\\
    &\otimes
    \left(
    \hat{\rho}_{\textnormal{th}}(N_{\textnormal{th}})^{\otimes n}
    \right)^{\otimes(m-h)},
\end{flalign}
in which $\hat{\rho}^{(w)}_{A^nI_D^n}$ corresponds to $\hat{\rho}_{I_D^nA^nI_C^n}^{(w)}$ with mode $I_C^n$ traced out and modes reordered.
Alice estimates the true hypothesis by applying a \ac{POVM}
$\{\Lambda_{h}^{(w)}\}_{h\in\intseq{1}{m}}$ over the modes $(\{D_j^n\}_{j=1}^m, I_D^{n})$,
producing an estimate $\widehat{h}$. Note that message transmission is independent of the detection hypothesis; hence, there is no interaction between the tasks through communication.

We define the communication rate as $R \eqdef \lim_{n\to\infty}\frac{\log M}{n}$, and measure communication reliability using the maximum probability of error
\begin{align}
    P_{e,c}^n
    \eqdef
    \max_{w\in\intseq{1}{M}}
    \tr{
    \left(
    \textnormal{id}_{B^nI_C^n}
    -
    \Pi_{w}
    \right)
    \hat{\rho}_{B^nI_C^n}^{(w)}
    }.
\end{align}
We measure sensing performance using the maximum probability of error for hypothesis testing,
\begin{flalign}
    P_{e,d}^n &\eqdef
    \max_{h\in\intseq{1}{m}}
    \max_{w\in\intseq{1}{M}}
    \nonumber\\
    &\qquad\tr{
    \left(
    \textnormal{id}_{\{D_j^n\}_{j=1}^mI_D^{n}}
    -
    \Lambda_{h}^{(w)}
    \right)
    \hat{\rho}_{\{D_j^n\}_{j=1}^mI_D^n}^{(h,w)}
    }.
\end{flalign}
The sensing error-exponent is defined as $E\eqdef \lim_{n\to\infty}-\frac{1}{n}\log P_{e,d}^n$.

A communication rate/sensing error-exponent pair $(R,E)$ is achievable if, for every $\epsilon>0$, there exist a sufficiently large $n$, a set of encoder channels $\{\calE^{(w)}_{S^n\to A^n}\}_{w\in\intseq{1}{M}}$ and \acp{POVM} $\{\Pi_w\}_{w
\in\intseq{1}{M}}$ and $\{\Lambda_{h}^{(w)}\}_{h\in\intseq{1}{m}}$ such that
\begin{align*}
    P_{e,c}^n &\leq \epsilon,\\
    -\frac{\log P_{e,d}^n}{n} &\geq E-\epsilon,\\
    \frac{\log M}{n} &\geq R-\epsilon.
\end{align*}
The smallest region containing all the achievable rate/error-exponent pairs is called the \ac{QISAC} capacity region.

\subsection{QISAC with Classical Resource}\label{sec:classical}
We first present a classical characterization of the achievable communication rate/sensing error-exponent region for a \ac{QISAC} protocol based on coherent \ac{PSK}~\cite{su2026} and coherent detection~\cite{tan2008QI}. The resulting performance given in the following theorem, serves as a baseline for subsequent results.
\begin{theorem}
\label{thm:JBCS_class}
    The following communication rate/sensing error-exponent region is achievable without entanglement assistance under a mean photon number constraint $N$ if $N_T>\frac{\eta N}{2}-1$:
    \begin{align*}
        \calC^{\textnormal{Classical}}_{\textnormal{QISAC}}= \left\{\begin{array}{l}
            (R,E)\in\bbR^2_+:\\
            R\leq R^{\textnormal{Classical}}(N)\\
            E\leq E^{\textnormal{Classical}}(N)
        \end{array}\right\},
    \end{align*}
    where
    \begin{flalign*}
        &R^{\textnormal{Classical}}(N) = \eta N\left(\frac{1}{N_T+1}+\log\left(1+\frac{1}{N_T}\right)\right)\\
        &\phantom{=========}-(\eta N+N_T)\log\left(1+\frac{\eta N}{N_T(N_T+1)}\right),\\
        &E^{\textnormal{Classical}}(N) = \frac{2\kappa N}{1+2N_\textnormal{th}+2\sqrt{N_\textnormal{th}(N_\textnormal{th}+1)}}.
    \end{flalign*}
\end{theorem}
\begin{proof}
    See Appendix~\ref{app:class_proof}.
\end{proof}

Note that the condition $N_T > \frac{\eta N}{2} - 1$ is required for the $m$-\ac{PSK} modulated coherent-state channel output to converge to the ring state as $m \to \infty$, and for the corresponding Holevo information to converge as well. The set of achievable communication rate/sensing error-exponent pairs in Theorem~\ref{thm:JBCS_class} forms a rectangular region implying that there is no trade-off between the communication rate and the sensing error-exponent. This is consistent with classical information-theoretic results on \ac{ISAC} over additive Gaussian channels~\cite{Joudeh2022} because in both~\cite{Joudeh2022} and the present model, communication is independent of the detection hypothesis. $R^\textnormal{Classical}$ corresponds to the lower bound on the Holevo information of the coherent ring state, which tightly approximates classical capacity for a broad range of signal mean photon number and physical parameters of the bosonic channel~\cite{su2026}. $E^\textnormal{Classical}$ corresponds to the coherent state exponent, which is the optimal classical ranging error-exponent~\cite{Zhuang2021}. The capacity region of the \ac{QISAC} problem without entanglement assistance is contained within the rectangular region with its corner at $(C, E^{\mathrm{Classical}})$. Because $\calC^\textnormal{Classical}_\textnormal{QISAC}$ simultaneously achieves the best classical sensing error exponent and a very tight approximation of the classical capacity, it closely approximates the upper bound on the \ac{QISAC} capacity region.

\subsection{TMSV-QISAC Protocol}\label{sec:TMSV_model_result}
\begin{figure}[!htbp]
    \centering
    \includegraphics[width=0.5\textwidth]{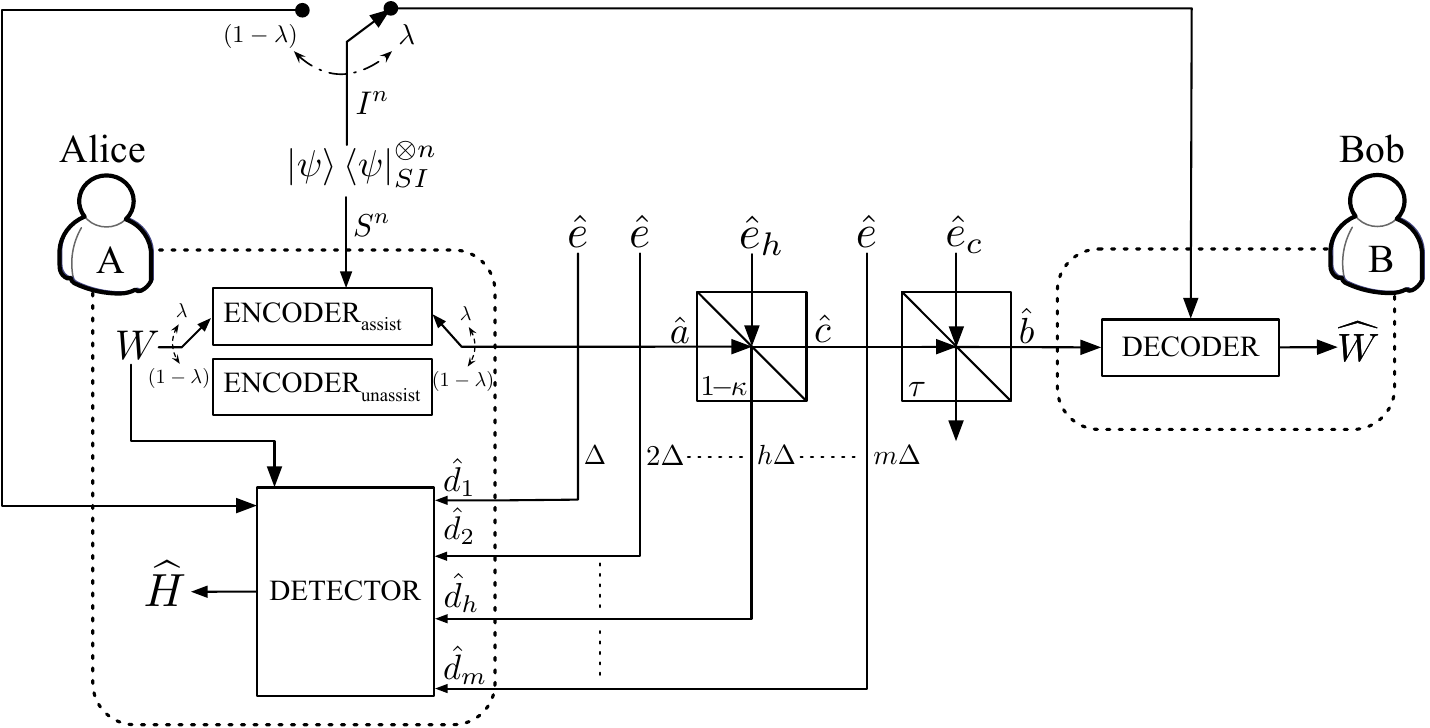}
    \caption{\ac{QISAC} model with bipartite entanglement resources over a lossy thermal-noise bosonic channel.}
    \label{fig:JBCS_model_TMSV}
\end{figure}

We now develop a \ac{QISAC} protocol motivated by ring-state communication schemes for coherent states and \ac{TMSV} state~\cite{su2026}, and \ac{TMSV} assisted ranging~\cite{Zhuang2021}. As illustrated in Figure~\ref{fig:JBCS_model_TMSV}, Alice has access to $n$ independent \ac{TMSV} states $\hat{\rho}_{SI}^{\otimes n}(N_S)$, where $\hat{\rho}_{SI}(N_S)=\ket{\psi}\bra{\psi}_{SI}^\textnormal{TMSV}$ with mean photon number $N_S$, for which she may either share the idler with Bob to assist communication or retain the idler to enhance sensing performance. The parameter $\lambda\in[0,1]$ corresponds to the portion of the blocklength $n$ in which the idler modes are shared with Bob's decoder. Without loss of generality, we can assume the modes are consecutive. In terms of the general tripartite \ac{QISAC} model in Figure~\ref{fig:Tripartite_model}, this scheme is obtained by defining the idler modes as $I_C^n\eqdef I^{1:\lambda n}$ and $I_D^n\eqdef I^{\lambda n+1:n}$.

Specifically, Alice's encoder consists of two parts:
\begin{enumerate}
    \item For the first $\lambda n$ modes, Alice encodes her classical message via \ac{PSK} by applying the single-mode phase-shift unitary $\hat{U}_{\theta_i} \eqdef \exp\left(-2\imath\theta_i\hat{a}^\dagger\hat{a}\right)$ to each signal mode, for $i\in\intseq{1}{\lambda n}$.

    \item For the remaining $(1-\lambda)n$ modes, Alice encodes the message via displaced \ac{PSK} by applying the displacement unitary $\hat{D}(\abs{\alpha}e^{\imath\theta_j})$
    to each signal mode, for $j\in\intseq{\lambda n+1}{n}$.
\end{enumerate}
Here, $\theta \in [0,2\pi)$ denotes the phase of the \ac{PSK} symbol in both schemes. We assume that the constellation is dense enough that its Holevo information is well approximated by that of the ensemble with $\theta$ uniformly distributed on $[0,2\pi)$. Explicitly, $\calE^{(w,\lambda)}_{S^n\to A^n}\left(\cdot\right)
    =
    \left(\bigotimes_{j=1}^{\lambda n}\hat{U}_{\theta_j(w)}\right)
    \otimes
    \left(\bigotimes_{k=\lambda n+1}^{n}
    \hat{D}(\abs{\alpha}e^{\imath\theta_k(w)})\right)\left(\cdot\right)\left(\bigotimes_{j=1}^{\lambda n}\hat{U}_{\theta_j(w)}^\dagger\right)
    \otimes
    \left(\bigotimes_{k=\lambda n+1}^{n}
    \hat{D}(\abs{\alpha}e^{\imath\theta_k(w)})^\dagger\right)$. Note that we are implicitly assuming the existence of a classical local oscillator shared with Bob to act as a phase reference for the displaced \ac{PSK} scheme. The displacement modulation introduces the parameter $N_{\textnormal{m}}\eqdef \abs{\alpha}^2$, which represents the additional energy injected into the channel by the displacement operation. The total mean photon number constraint becomes $N_S+N_{\textnormal{m}}\leq N$.

During the first $\lambda$ fraction of the blocklength, Alice knows the applied phase and therefore can eliminate it from the received state. The modes corresponding to all incorrect hypotheses are thermal states diagonal in the Fock basis and are therefore unaffected. During the remaning $(1-\lambda)$ fraction of the blocklength, Alice does not know which of the $\{\hat{d}_i\}_{i=1}^m$ is displaced so she cannot remove the displacement. However, she can still remove the phase of the displacement. After this phase cancellation, Alice's hypotheses become tensor products of two \ac{iid} backscattered states corresponding to the entanglement-unassisted and entanglement-assisted settings. By construction, this protocol consumes $\lambda g(N)+(1-\lambda )g(N_S)$ nats of entanglement per channel use. In this protocol, we define $\hat{\rho}_{DI}(N_S,N_\textnormal{m})\eqdef \hat{D}\left(\sqrt{2\eta  N_\textnormal{m}}\right)\otimes\textnormal{id}_I\left(\calL^{(\kappa,N_\textnormal{th}/(1-\kappa))}_{S\to D}\otimes \textnormal{id}_I\right)\left(\hat{\rho}_{SI}(N_S)\right)\hat{D}\left(\sqrt{2\eta N_\textnormal{m}}\right)^\dagger\otimes\textnormal{id}_I$, $\hat{\rho}_{BI}\eqdef \left(\calL^{(\eta,N_T/(1-\eta))}_{S\to B}\otimes \textnormal{id}_I\right)\left(\hat{\rho}_{SI}(N)\right)$, $\hat{\rho}_{BI}^\theta=\hat{U}_\theta\hat{\rho}_{BI}\hat{U}_\theta^\dagger$, $\hat{\rho}_{B}(N_S,N_\textnormal{m})= \hat{D}\left(\sqrt{2\eta N_\textnormal{m}}\right)\hat{\rho}_\textnormal{th}(\eta N_S+N_\textnormal{th})\hat{D}\left(\sqrt{2\eta N_\textnormal{m}}\right)^\dagger$, and $\hat{\rho}_{B}^\theta(N_S,N_\textnormal{m})=\hat{U}_\theta\hat{\rho}_{B}(N_S,N_\textnormal{m})\hat{U}_\theta^\dagger$. Note that displaced-\ac{PSK} and \ac{TMSV}-\ac{PSK} schemes generate ring states, for which there exist tight lower bounds of the achievable communication rates that approximate the ultimate communication rates $C$ and $C_\textnormal{EA}$, respectively.

Our main result is then the characterization of the achievable communication rate/sensing error-exponent region with bipartite entanglement assistance, given in the following theorem.
\begin{theorem}
\label{thm:JBCS_TMSV}
    Using the \ac{TMSV}-\ac{QISAC} protocol, if $N_T>\max\{\eta,\eta N-1\}$, the set of achievable communication rate/sensing error-exponent pairs includes the set
    \begin{align}
        &\calC^{\textnormal{TMSV}}_{\textnormal{QISAC}}\nonumber\\
        &=\bigcup_{\lambda,\delta\in[0,1]}\left\{\begin{array}{l}
            (R,E)\in\bbR^2_+:\\
            R\leq \lambda R^\textnormal{TMSV}(N)+(1-\lambda)R^\textnormal{Coherent}(N,\delta)\\
            E\leq \lambda E^{\textnormal{Thermal}}_{\textnormal{Ranging}}(N)+(1-\lambda)E^{\textnormal{TMSV}}_{\textnormal{Ranging}}(N,\delta)
        \end{array}\right\},
    \end{align}
    where
    \begin{align}
        &R^\textnormal{TMSV}(N) = C_\textnormal{EA}(N,N_T,\eta)+g\left(N_T\right)-g\left(\eta N+N_T\right)\nonumber\\
	&\phantom{=======}+\left(2 \eta N+N_T\right)N \log \left(\frac{\left(N_T+1\right) \left(N_T-\eta \right)}{N_T \left(N_T-\eta+1\right)}\right)\nonumber \\
&\phantom{=======}
	+\eta  N \log \left(\frac{\left(N_T+1\right)^2 \left(N_T-\eta \right)}{N_T^2 \left(N_T-\eta+1\right)}\right)\nonumber\\
    &\phantom{=======} -N \log \left(\frac{N_T-\eta+1}{N_T+1}\right),\\
    &R^\textnormal{Coherent}(N,\delta) = \frac{\eta \delta N}{N_\textnormal{eff}+1}+\eta\delta N\log\left(1+\frac{1}{N_\textnormal{eff}}\right)\nonumber\\
    &\phantom{=======}-(\eta N+N_{T})\log\left(1+\frac{\eta\delta N}{N_\textnormal{eff}(N_\textnormal{eff}+1)}\right),\\
        &E^{\textnormal{Thermal}}_{\textnormal{Ranging}}(N) = 2\log\left(\sqrt{\frac{(1+\kappa N+N_\textnormal{th})(1+N_{\textnormal{th}})}{\left(\kappa N+ N_\textnormal{th}\right)N_{\textnormal{th}}}}-1\right)\nonumber\\
        &\phantom{=========}+\log\left(\left(\kappa N+ N_\textnormal{th}\right)N_{\textnormal{th}}\right),\\
        &E^\textnormal{TMSV}_{\textnormal{Ranging}}(N,\delta)= \nonumber\\
        & -\log Q_{1/2}\left(\hat{\rho}_{TDI}((1-\delta)N,\delta N)\Vert\hat{\rho}_{DTI}((1-\delta)N,\delta N)\right)\label{eq:QCB_TM_exp},
    \end{align}
    $N_\textnormal{eff}\eqdef \eta(1-\delta) N+N_T$, $\hat{\rho}_{TDI}\eqdef\hat{\rho}_{\textnormal{th}}\left(N_\textnormal{th}\right)\otimes\hat{\rho}_{DI}((1-\delta)N,\delta N)$ and $\hat{\rho}_{DTI}(\delta)$ such that $\hat{\rho}_{DI_D}(\delta)=\tr[T]{\hat{\rho}_{DTI}(\delta)}$ and $\hat{\rho}_\textnormal{th}(N_\textnormal{th})=\tr[D,I]{\hat{\rho}_{DTI}(\delta)}$.
\end{theorem}
\begin{proof}
    See Appendix~\ref{app:proof_EA}.
\end{proof}

The condition $N_T > \max\{\eta, \eta N - 1\}$, similar to that of Theorem~\ref{thm:JBCS_class}, is required for the $m$-\ac{PSK} modulated \ac{TMSV} channel output to converge to the \ac{TMSV} ring state as $m \to \infty$, and for the corresponding Holevo information to converge as well. Note that the boundary of $\calC^\textnormal{TMSV}_\textnormal{QISAC}$ is attained when the mean photon number constraint is fully utilized, i.e., when $N_S + N_\textnormal{m} = N$,
so that the constraint can be parameterized as $N_\textnormal{m} \eqdef \delta N$ and $N_S \eqdef (1 - \delta)N$ for $\delta \in [0,1]$. Moreover, when entanglement is allocated for ranging, the communication channel is equivalent to $\calL_{A\to B}^{\left(\eta, (\eta N_S+N_T)/(1-\eta)\right)}$. For $R^{\textnormal{Coherent}}(N,\delta)$, $\eta(1-\delta) N$ is acting as an effective thermal-noise mean photon number and $\delta N$ is acting as the channel input. $R^\textnormal{Coherent}(N,\delta)$ is increasing in $\delta$ with boundaries $R^\textnormal{Coherent}(N,0)=0$ and $R^\textnormal{Coherent}(N,1)=R^\textnormal{Classical}(N)$. Note also that $E^{\textnormal{TMSV}}_\textnormal{Ranging}(N,1)=E^{\textnormal{Classical}}_\textnormal{Ranging}(N)$.

Finally, we emphasize that $R^\textnormal{TMSV}(N)$ and $R^{\textnormal{Coherent}}(N,\delta)$ are lower bounds on the Holevo information of the ensembles $\left\{\frac{1}{2\pi},\hat{\rho}_{BI}^\theta\right\}_{\theta\in[0,2\pi)}$ and $\left\{\frac{1}{2\pi},\hat{\rho}_{B}^\theta((1-\delta)N,\delta N)\right\}_{\theta\in[0,2\pi)}$, respectively, obtained without invoking any Taylor series expansion in the physical parameters. Given the complexity of the \ac{QISAC} problem described in this section, the expressions in Theorem~\ref{thm:JBCS_TMSV} are remarkably compact.

Focusing on the regime $N\to0$, in which entanglement assistance provides the greatest enhancement, we first present how the assisted and unassisted communication rates and sensing error-exponents behave in the following lemma.
\begin{lemma}\label{lm:low_N_rate_exp}
    In the regime $N\to0$, achievable communication rate and sensing error-exponent terms $R^{\textnormal{Coherent}}(N,\delta)$, $R^{\textnormal{TMSV}}(N)$, $E^{\textnormal{Thermal}}_{\textnormal{Ranging}}(N)$ and $E^{\textnormal{TMSV}}_{\textnormal{Ranging}}(N,\delta)$ behave as,
    \begin{align}
        &R^{\textnormal{TMSV}}_{\textnormal{L}}(N)
= -\frac{\eta N \log N}{N_T+1}\nonumber\\
&\phantom{======}+N \left(
    (N_T+\eta)\log \frac{(N_T+1)(N_T-\eta)}{N_T(N_T-\eta+1)}\right. \nonumber\\
&\phantom{======} \left.+ \frac{\eta}{N_T+1} \left( 1+ \log \frac{(N_T+1)^2}{N_T(N_T-\eta+1)}\right)
\right),\\
        &R^{\textnormal{Coherent}}_{\textnormal{L}}(N,\delta) \nonumber\\
        &= \delta\eta N\left(
        \log\left(1+\frac{1}{N_T+1}\right)
        -\frac{1}{N_T(N_T+1)}\right),\\
        &E^\textnormal{Thermal}_{\textnormal{Ranging,L}}(N) = \frac{\kappa^2N^2}{4 N_\textnormal{th}(N_\textnormal{th}+1)},\\
        &E^\textnormal{TMSV}_{\textnormal{Ranging,L}}(N,\delta)\nonumber\\
        &=2\kappa N\left(\frac{1-\delta}{N_{\textnormal{th}}+1}+\frac{\delta}{1+2N_\textnormal{th}+2\sqrt{N_\textnormal{th}(N_\textnormal{th}+1)}}\right).
    \end{align}
\end{lemma}
\begin{proof}
    See Appendix~\ref{app:proof_EA}.
\end{proof}

Notice the difference between the scalings of entanglement-assisted and unassisted performances $R^\textnormal{TMSV}_\textnormal{L}(N)=O(-N\log N)$ and $R^\textnormal{Coherent}_\textnormal{L}(N,\delta)=O(N)$, $E^\textnormal{TMSV}_\textnormal{Ranging,L}(N,\delta)=O(N)$ and $E^\textnormal{Thermal}_\textnormal{Ranging,L}=O(N^2)$. A further simplified and more explicit achievable communication rate/sensing error-exponent region is given in Corollary~\ref{cor:JBCS_approx}.

\begin{corollary}
\label{cor:JBCS_approx}
    Using the \ac{TMSV}-\ac{QISAC} protocol, in the regime $N\to0$, if $N_T>\eta$, the set of achievable rate/error-exponent pairs includes the set
    \begin{align}
        &\calC^{\textnormal{TMSV}}_{\textnormal{QISAC}} \nonumber\\
        &= \bigcup_{\lambda\in[0,1]}\left\{\begin{array}{l}
            (R,E)\in\bbR^2_+:\\
            R\leq \lambda R^\textnormal{TMSV}_\textnormal{L}(N)\\
            E\leq (1-\lambda)E^{\textnormal{TMSV}}_\textnormal{Ranging,L}(N,0)
        \end{array}\right\},
    \end{align}
\end{corollary}
\begin{proof}
    See Appendix~\ref{app:proof_EA}.
\end{proof}

Note that $\calC^\textnormal{TMSV}_\textnormal{QISAC}$ reduces to time-sharing between the rate/error-exponent pairs $(R^\textnormal{TMSV}_\textnormal{L},E^\textnormal{Thermal}_{\textnormal{Ranging,L}})$ and $(0,E^{\textnormal{TMSV}}_{\textnormal{L}})$, which enforces $\delta=0$ in Theorem~\ref{thm:JBCS_TMSV}. In the low mean photon number regime, entanglement assistance provides such a substantial performance improvement that devoting resources to classical operation only degrades performance. This follows from the imbalance between the scalings of $R^\textnormal{TMSV}_\textnormal{L}(N)=O\left(-N\log N\right)$ and $R^{\textnormal{Coherent}}_\textnormal{L}(N,\delta)=O\left(N\right)$ as established in Lemma~\ref{lm:low_N_rate_exp}. As a result, the time-sharing curve has slope on the order of $O\left(-1/\log N\right)$, whereas the curve $\{(R^\textnormal{Coherent}_\textnormal{L}(N,\delta),E^\textnormal{TMSV}_\textnormal{ranging,L}(N,\delta))\}_{\delta\in[0,1]}$ has slope on the order of $O(1)$. Since both slopes are negative, the $(N_S,N_\textnormal{m})$-split curve always lies below the time-sharing curve when the approximations in Lemma~\ref{lm:low_N_rate_exp} are valid. Also, $E^{\textnormal{TMSV}}_\textnormal{Ranging,L}(N)$ is consistent with the ranging exponent in~\cite{Zhuang2021} as $N_\textnormal{th}\to\infty$.

\subsection{C3MSV-QISAC Protocol}\label{sec:C3MSV_model_result}

\begin{figure}[!htbp]
    \centering
    \includegraphics[width=\linewidth]{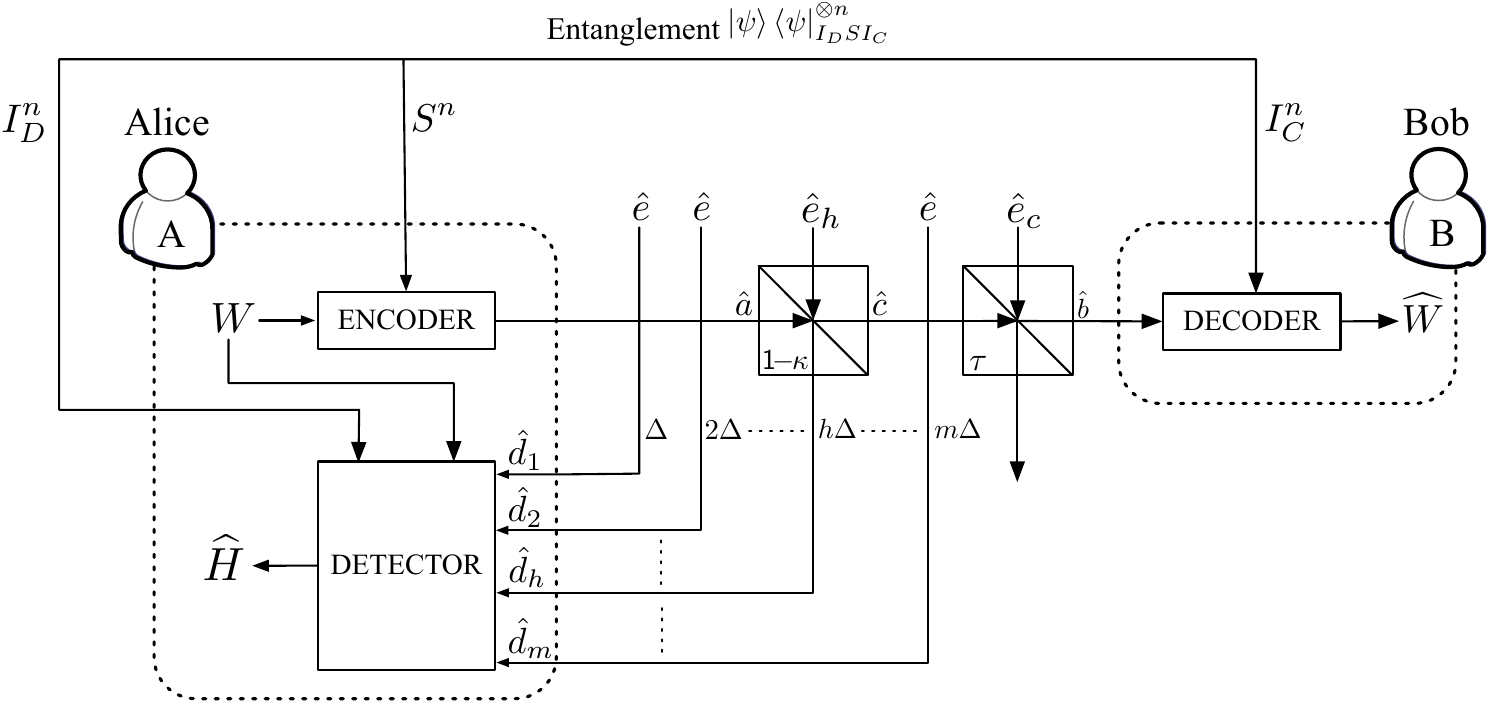}
    \caption{\ac{QISAC} with tripartite entanglement resource}
    \label{fig:C3MSV_model}
\end{figure}

We next develop a protocol using $n$ copies of a \ac{C3MSV} state as the tripartite entanglement resource, with the mean photon numbers $\left(I_C,S,I_D\right)=(\delta N,N,(1-\delta)N)$, for $\delta\in[0,1]$, of the three modes with the covariance matrix given in~\eqref{eq:cov_c3msv}. $I_C$ is the idler mode for communication, $I_D$ is the idler mode for sensing and $S$ is the signal mode to be encoded as the channel input. This assignment ensures that both the communication and sensing tasks can benefit from entanglement, since the middle strong mode of \ac{C3MSV} is entangled with weak idler modes, but the idler modes are only classically correlated with each other. The corresponding covariance matrix is given in~\eqref{eq:cov_c3msv}. Note that both the \ac{TMSV}-\ac{QISAC} protocol in Section~\ref{sec:TMSV_model_result} and the \ac{C3MSV}-\ac{QISAC} protocol use a total mean photon number of $2N$ for single channel use, while injecting $N$ photons on average into the channel, which ensures a fair comparison. Alice encodes her classical message via \ac{PSK} by applying the single-mode phase-shift unitary $\hat{U}_\theta=\exp\left(-2\imath\theta\hat{a}^\dagger\hat{a}\right)$ to each signal mode, in which $\theta\in[0,2\pi)$ is uniformly distributed. Explicitly,
$\calE^{(w)}_{S^n\to A^n}(\cdot) = \left(\bigotimes_{i=1}^n\hat{U}_{\theta_i(w)}\right)\left(\cdot\right)\left(\bigotimes_{i=1}^n\hat{U}_{\theta_i(w)}^\dagger\right)$. Again, Alice knows the applied phase and therefore can eliminate it from the received state. Hence, her observations become \ac{iid} states. By construction, this protocol consumes $g(N)$ nats of entanglement per channel use. In this protocol, we define $\hat{\rho}_{DI_D}(\delta)\eqdef \left(\calL^{(\kappa,N_\textnormal{th}/(1-\kappa))}_{S\to D}\otimes \textnormal{id}_{I_D}\right)\left(\hat{\rho}_{SI_D}(\delta)\right)$, $\hat{\rho}_{BI_C}(\delta)\eqdef \left(\calL^{(\eta,N_T/(1-\eta))}_{S\to B}\otimes \textnormal{id}_{I_C}\right)\left(\hat{\rho}_{SI_C}(\delta)\right)$, and $\hat{\rho}_{BI_C}^\theta=\hat{U}_\theta\hat{\rho}_{BI_C}\hat{U}_\theta^\dagger$.

Our main result on \ac{QISAC} with tripartite entanglement resource is the characterization of achievable communication rate/sensing error-exponent region with imbalanced \ac{C3MSV} assistance, given in the following theorem.

\begin{theorem}
\label{thm:JBCS_C3MSV}
    Using the \ac{C3MSV}-\ac{QISAC} protocol, if $N_T>\max\{\eta,\eta N-1\}$, the set of achievable communication rate/sensing error-exponent pairs includes the set
    \begin{align}
        \calC^{\textnormal{C3MSV}}_{\textnormal{QISAC}}=\bigcup_{\delta\in[0,1]}\left\{\begin{array}{l}
            (R,E)\in\bbR^2_+:\\
            R\leq R^\textnormal{C3MSV}(N,\delta)\\
            E\leq E^{\textnormal{C3MSV}}_\textnormal{Ranging}(N,\delta)
        \end{array}\right\},
    \end{align}
    where
    \begin{flalign}
        &R^\textnormal{C3MSV}(N,\delta) = \delta  N \log \left(\frac{(1-\delta) \eta  N+N_T \left(\delta  N+1\right)}{\delta  N \left(N_T-\eta\right)}\right)\nonumber\\
        &\qquad+\log \left(\left(N_T-\eta+1\right) \left(N \left((1-\delta)\eta +\delta  N_T\right)+N_T\right)\right)\nonumber\\
        &\qquad-\left(N \left(\delta  \eta +\delta +\eta +2 \delta  \eta  N+\delta  N_T\right)+N_T+1\right) \nonumber\\
        &\times\log \left(1-\frac{\eta  \left(N+1\right)}{\left(\eta -N_T-1\right) \left((\delta -1) \eta  N-N_T \left(\delta  N+1\right)\right)}\right)\nonumber\\
        &\qquad+\left(\eta  N+N_T\right)\log \left(1+\frac{1}{N_T-\eta}\right)-\log \left(N_T-\eta \right)\nonumber\\
        &\phantom{=======}-g\left(\frac{\mu_+(\delta)-1}{2}\right)-g\left(\frac{\mu_-(\delta)-1}{2}\right)\\
        &E^\textnormal{C3MSV}_{\textnormal{ranging}}(N,\delta) = -\log Q_{1/2}\left(\hat{\rho}_{TDI_D}(\delta)\Vert\hat{\rho}_{DTI_D}(\delta)\right)\label{eq:QCB_C3_exp},
    \end{flalign}
    where $\mu_\pm(\delta)\eqdef \pm(N_T-(\delta-\eta)N)+\sqrt{(1+(\delta+\eta)N+N_T)^2-4\eta\delta N(N+1)}$, $\hat{\rho}_{TDI_D}(\delta)=\hat{\rho}_{\textnormal{th}}\left(N_\textnormal{th}\right)\otimes\hat{\rho}_{DI_D}(\delta)$ and $\hat{\rho}_{DTI_D}(\delta)$ such that $\hat{\rho}_{DI_D}(\delta)=\tr[T]{\hat{\rho}_{DTI_D}(\delta)}$ and $\hat{\rho}_T=\tr[D,I_D]{\hat{\rho}_{DTI_D}(\delta)}$.
\end{theorem}
\begin{proof}
    See Appendix~\ref{app:proof_JBCS_C3}.
\end{proof}

Note that the region $\calC^\textnormal{C3MSV}_\textnormal{QISAC}$ is a union over rectangular regions with corners at $(R_\textnormal{\ac{C3MSV}}(N,\delta),E_\textnormal{\ac{C3MSV}}(N,\delta))$. Here, $R^\textnormal{C3MSV}$ is a closed form lower bound on the Holevo information of the ensemble $\left\{\frac{1}{2\pi},\hat{\rho}^\theta_{BI_C}(\delta)\right\}_{\theta\in[0,2\pi)}$. End points, $\delta\in\{0,1\}$, correspond to $(0,E^\textnormal{TMSV}(N,0))$ and $(R^\textnormal{TMSV}(N),E^\textnormal{Thermal}_\textnormal{Ranging}(N))$, respectively.

Focusing again on the region $N\to0$, in which entanglement assistance provides the greatest enhancement, we first present how the communication rate and sensing error-exponent behave in the following lemma.
\begin{lemma}\label{lm:C3MSV_low_N_rate_exp}
    In the regime $N\to0$, achievable rate and error-exponent terms $R^{\textnormal{C3MSV}}(N,\delta)$ and $E^{\textnormal{C3MSV}}_{\textnormal{Ranging}}(N,\delta)$ behave as,
    \begin{flalign*}
        &R^{\textnormal{C3MSV}}_{\textnormal{L}}(N,\delta) = \delta R^{\textnormal{TMSV}}_{\textnormal{L}}(N)+\frac{\eta N}{N_T+1}(-\delta\log\delta),\\       
        &E^\textnormal{C3MSV}_\textnormal{Ranging,L}(N,\delta)=(1-\delta)E^\textnormal{TMSV}_\textnormal{Ranging,L}(N,0).
    \end{flalign*}
\end{lemma}
\begin{proof}
    See Appendix~\ref{app:proof_EA}.
\end{proof}

Lemma~\ref{lm:C3MSV_low_N_rate_exp} shows that $E^\textnormal{C3MSV}_\textnormal{Ranging}(N,\delta)$ is behaving similar to $E^\textnormal{TMSV}_\textnormal{Ranging,L}(N)$ under time-sharing. However, $R^\textnormal{C3MSV}_\textnormal{L}(N,\delta)$ has an additional $(-\delta\log\delta)$ term. This means that for $N\to0$ when $O(N)$ terms are not fully dominated by $O(-N\log N)$ term in the rate expression, the \ac{C3MSV}-\ac{QISAC} protocol beats the \ac{TMSV}-\ac{QISAC} protocol, and hence also the time-sharing protocol. If $N\to0$ is so small that the contribution of $O(N)$ terms are fully dominated by $O(-N\log N)$, i.e., when $N\to0$ such that $O(-N\log N)$ term alone is sufficient to approximate the behavior of $R^\textnormal{C3MSV}(N,\delta)$, the \ac{C3MSV} protocol also reduces to time-sharing. In this regime, however, there is practically no tradeoff since the exponent is effectively $0$ ($O(N)$ in comparison to communication rate $O(-N\log N)$). One example for which this regime matters is covertness~\cite{su2026,Candia2021TwoWayCovert}, in which the transmitter conceals the presence of communication by approximating the output received by the eavesdropper corresponding to a vacuum input. This forces the mean photon number to scale as at most $O(1/\sqrt{n})$ per channel use, so the number of message nats transmitted over $n$ channel uses scales as $O(\sqrt{n}\log n)$. With entanglement assistance, the covert rate is normalized by $\sqrt{n}\log n$, which causes the $-\delta\log\delta$ improvement over the $O(N)$ term to vanish. In classical covert \ac{ISAC}, both the covert communication throughput $\log M$ and the sensing error exponent $-\log P_{e,d}^n$ scale as $O(\sqrt{n})$ with blocklength $n$. Hence, with covert normalization $\sqrt{n}$, a positive covert communication rate and a positive sensing error exponent are simultaneously achievable~\cite{Wang2024JCS}. In covert \ac{QISAC}, by contrast, the covert communication rate is normalized by $\sqrt{n}\log n$ while the error exponent is still normalized by $\sqrt{n}$, so any classical rate vanishes under this normalization. Moreover, the $O(N\log N)$ term depends linearly on $\delta$. Consequently, both \ac{TMSV}-\ac{QISAC} and \ac{C3MSV}-\ac{QISAC} reduce to time-sharing in the covert setting.

A further simplified and more explicit achievable communication rate/sensing error-exponent region is given in Corollary~\ref{cor:C3MSV_JBCS_approx}.

\begin{corollary}
\label{cor:C3MSV_JBCS_approx}
    Using \ac{C3MSV} \ac{QISAC} protocol, in the regime $N\to0$, if $N_T>\eta$, the set of achievable communication rate/sensing error-exponent pairs includes the set
    \begin{align}
        &\calC^{\textnormal{C3MSV}}_{\textnormal{QISAC}} = \bigcup_{\delta\in[0,1]}\left\{\begin{array}{l}
            (R,E)\in\bbR^2_+:\\
            R\leq \delta R^\textnormal{TMSV}_\textnormal{L}(N)+\frac{\eta N}{N_T+1}(-\delta\log\delta)\\
            E\leq (1-\delta) E^{\textnormal{TMSV}}_\textnormal{Ranging,L}(N,0)
        \end{array}\right\}.
    \end{align}
\end{corollary}
\begin{proof}
    See Appendix~\ref{app:proof_EA}.
\end{proof}

\subsection{Numerical Results}

\begin{figure}[!htbp]
    \centering
    \includegraphics[width=\linewidth]{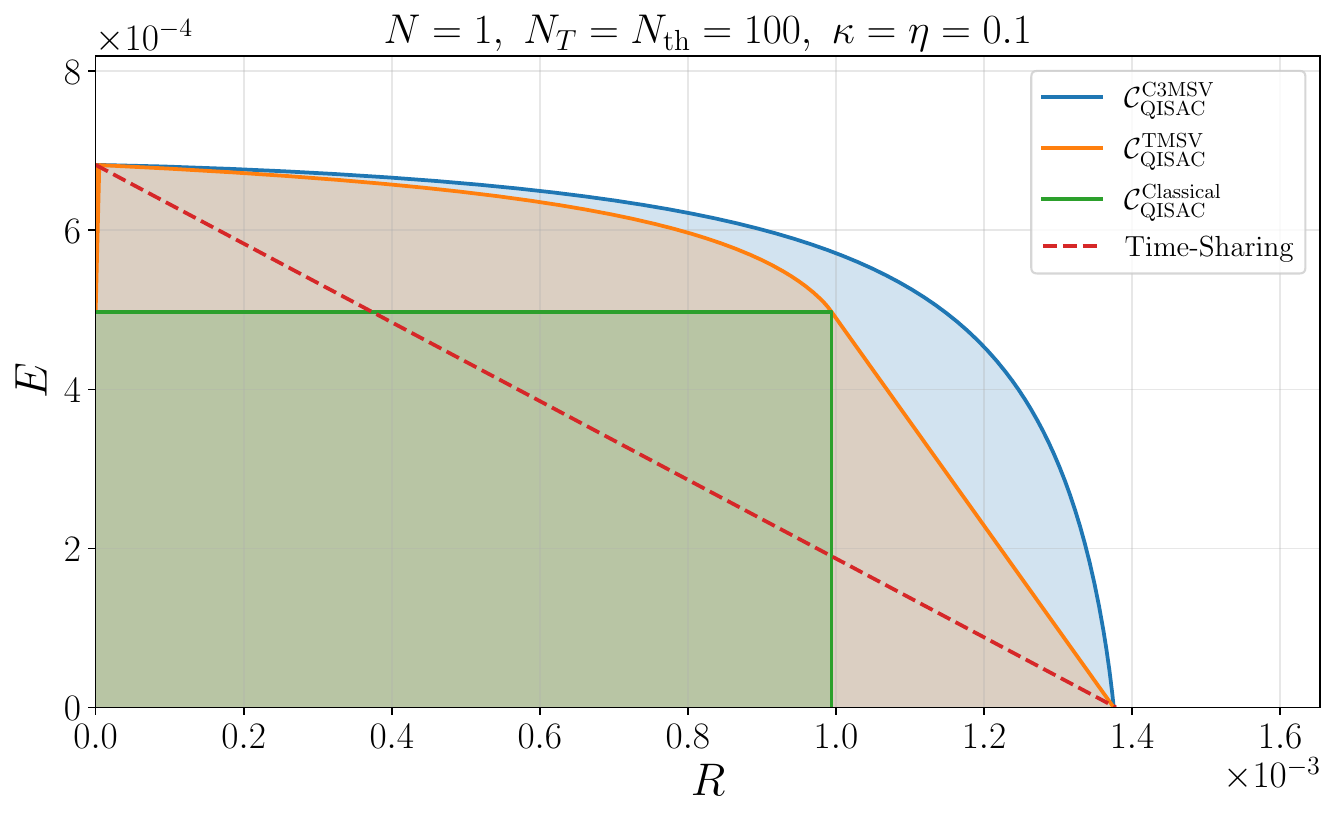}
    \caption{Comparison of $\calC^{\textnormal{TMSV}}_{\textnormal{QISAC}}$ in Theorem~\ref{thm:JBCS_TMSV} and $\calC^{\textnormal{C3MSV}}_{\textnormal{QISAC}}$ in Theorem~\ref{thm:JBCS_C3MSV}}
    \label{fig:JBCS_trade_off}
\end{figure}

We illustrate Theorems~\ref{thm:JBCS_class}, \ref{thm:JBCS_TMSV} and~\ref{thm:JBCS_C3MSV} in Figure~\ref{fig:JBCS_trade_off}. All regions $\calC^{\textnormal{Classical}}_{\textnormal{QISAC}}$, $\calC^\textnormal{TMSV}_\textnormal{QISAC}$ and $\calC^{\textnormal{C3MSV}}_{\textnormal{QISAC}}$ outperform time-sharing between the \ac{QI} exponent $E^{\textnormal{QI}}$~\cite{Zhuang2021}, which is equal to $E^\textnormal{TMSV}_\textnormal{Ranging}(N,0)=E^\textnormal{C3MSV}_\textnormal{Ranging}(N,0)$, without communication and the entanglement-assisted capacity $C_{\textnormal{EA}}$ and thermal ranging exponent $E^{\textnormal{Thermal}}_\textnormal{Ranging}$. Both $\calC^{\textnormal{TMSV}}_{\textnormal{QISAC}}$ and $\calC^{\textnormal{C3MSV}}_{\textnormal{QISAC}}$ encompass $\calC^{\textnormal{Classical}}_{\textnormal{QISAC}}$ demonstrating the quantum advantage of both protocols. Comparing performance of the \ac{TMSV}-\ac{QISAC} and the \ac{C3MSV}-\ac{QISAC} protocols, the tripartite protocol performs significantly better and contains the achievable region corresponding to the \ac{TMSV}-\ac{QISAC} protocol. This behavior can be considered as an evidence of quantum advantage, in which utilizing the available quantum resources to enhance all tasks outperforms interpolating between using all quantum resource for one task while performing the other task classically.

\begin{figure}[!htbp]
    \centering
    \includegraphics[width=\linewidth]{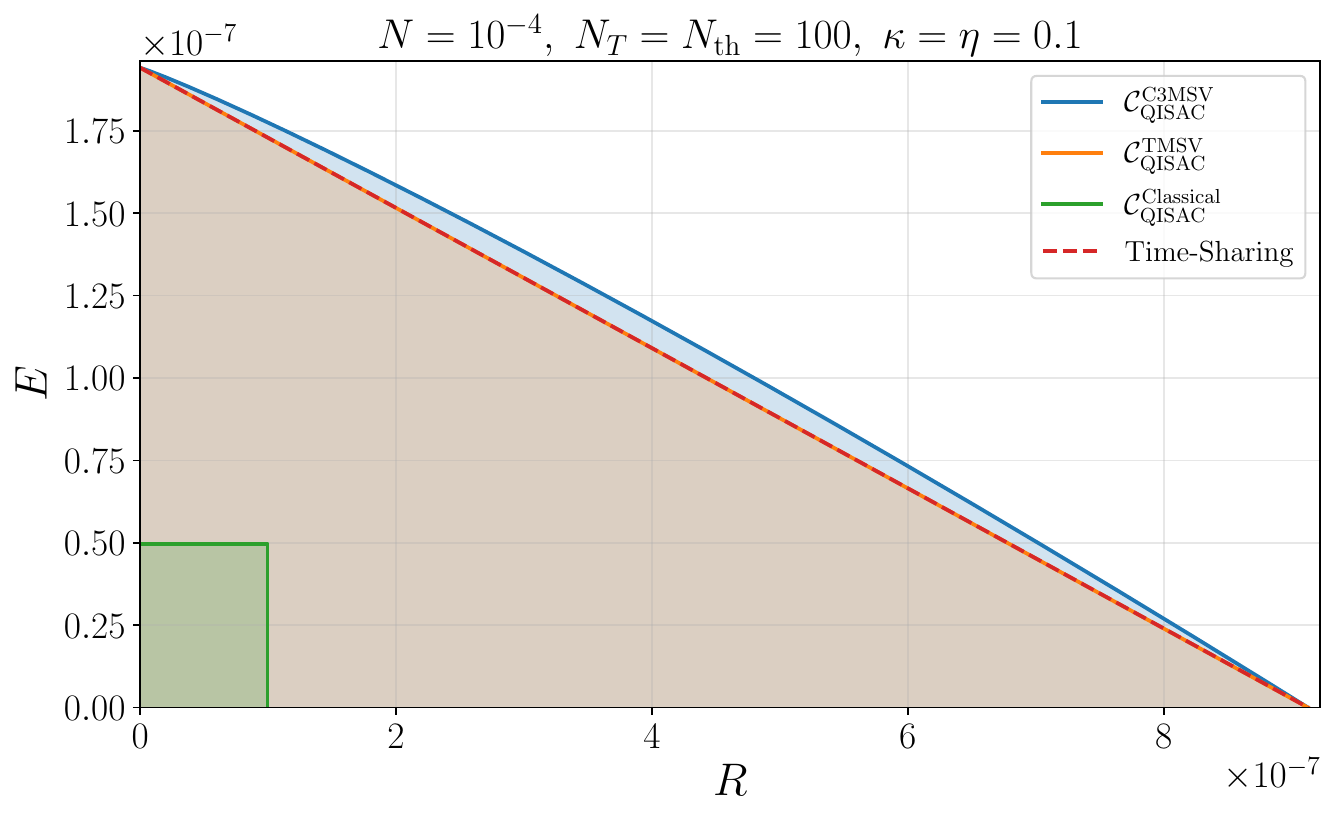}
    \caption{Comparison of $\calC^{\textnormal{TMSV}}_{\textnormal{QISAC}}$ in Corollary~\ref{cor:JBCS_approx} and $\calC^{\textnormal{C3MSV}}_{\textnormal{QISAC}}$ in Corollary~\ref{cor:C3MSV_JBCS_approx}.}
    \label{fig:JBCS_trade_off_lowN}
\end{figure}

Focusing on the low signal mean photon number regime, $N\to0$, we illustrate Corollary~\ref{cor:JBCS_approx} and~\ref{cor:C3MSV_JBCS_approx} and Theorem~\ref{thm:JBCS_class} in Figure~\ref{fig:JBCS_trade_off_lowN}. Both $\calC^{\textnormal{TMSV}}_{\textnormal{QISAC}}$ and $\calC^{\textnormal{C3MSV}}_{\textnormal{QISAC}}$ encompass again $\calC^{\textnormal{Classical}}_{\textnormal{QISAC}}$ which confirms the quantum advantage of the entanglement-assisted schemes. As shown in Corollary~\ref{cor:JBCS_approx}, $\calC^\textnormal{TMSV}_\textnormal{QISAC}$ reduces to time-sharing, while Corollary~\ref{cor:C3MSV_JBCS_approx} shows that $\calC^\textnormal{C3MSV}_\textnormal{QISAC}$ still encompasses and thus outperforms $\calC^\textnormal{TMSV}_\textnormal{QISAC}$ and consequently the time-sharing protocol. However, as seen from Corollary~\ref{cor:C3MSV_JBCS_approx}, the benefit of the \ac{C3MSV}-\ac{QISAC} protocol diminishes, and the corresponding \ac{QISAC} region reduces to time-sharing, since the rate improvement over the \ac{TMSV}-\ac{QISAC} protocol is $O(N)$, which is dominated by the leading term of the achievable rate, $O(-N\log N)$ as $N\to0$. This choice of parameters corresponds to a system operating at 300 K with a carrier frequency of approximately 56 GHz.

\begin{figure}[!htbp]
    \centering
    \includegraphics[width=\linewidth]{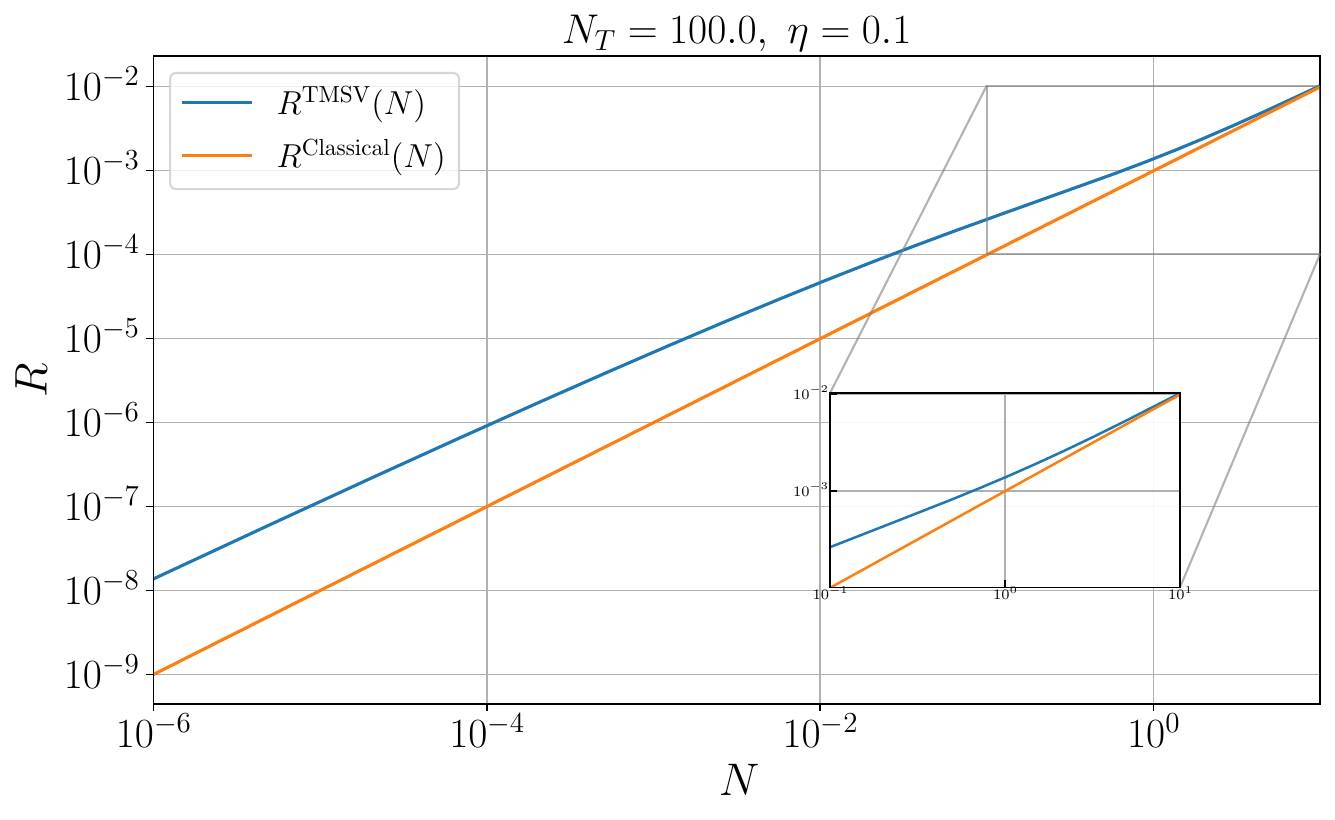}
    \caption{Comparison of achievable rates in Theorem~\ref{thm:JBCS_TMSV}.}
    \label{fig:TMSV_rates}
\end{figure}

\begin{figure}[!htbp]
    \centering
    \includegraphics[width=\linewidth]{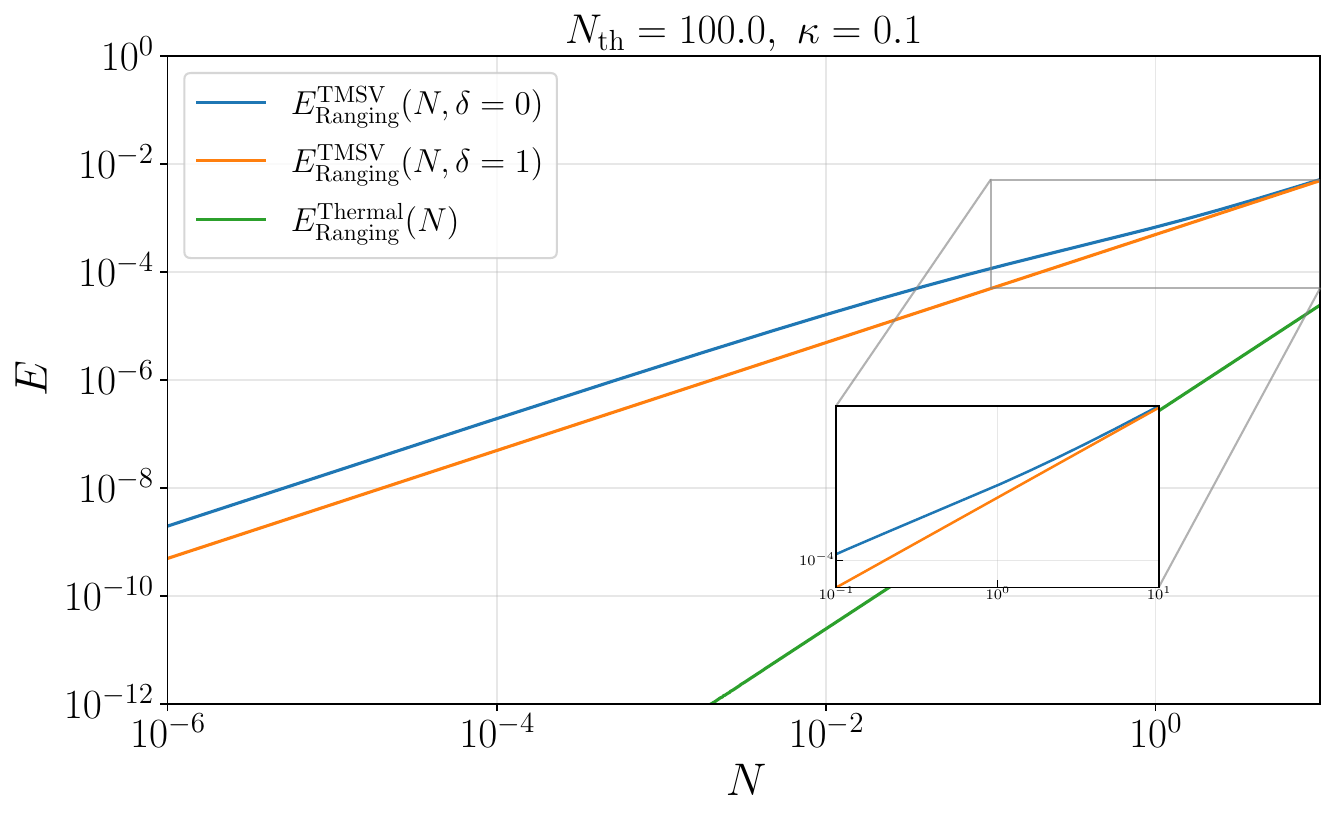}
    \caption{Comparison of achievable exponents in Theorem~\ref{thm:JBCS_TMSV}.}
    \label{fig:TMSV_exponents}
\end{figure}

Figure~\ref{fig:TMSV_rates} further highlights the behavior of the achievable rate expressions and Figure~\ref{fig:TMSV_exponents} demonstrates the behavior of achievable error-exponents in Theorem~\ref{thm:JBCS_TMSV}. Zoomed regions in both figures correspond to the signal mean photon number regime in Figure~\ref{fig:JBCS_trade_off}. In this regime, even though the effect is less prominent than in Corollary~\ref{lm:low_N_rate_exp}, a quantum advantage remains for utilizing entanglement in each task, even as the difference between entanglement-assisted and classical performance converges to zero.
\begin{figure}[!htbp]
    \centering
    \includegraphics[width=\linewidth]{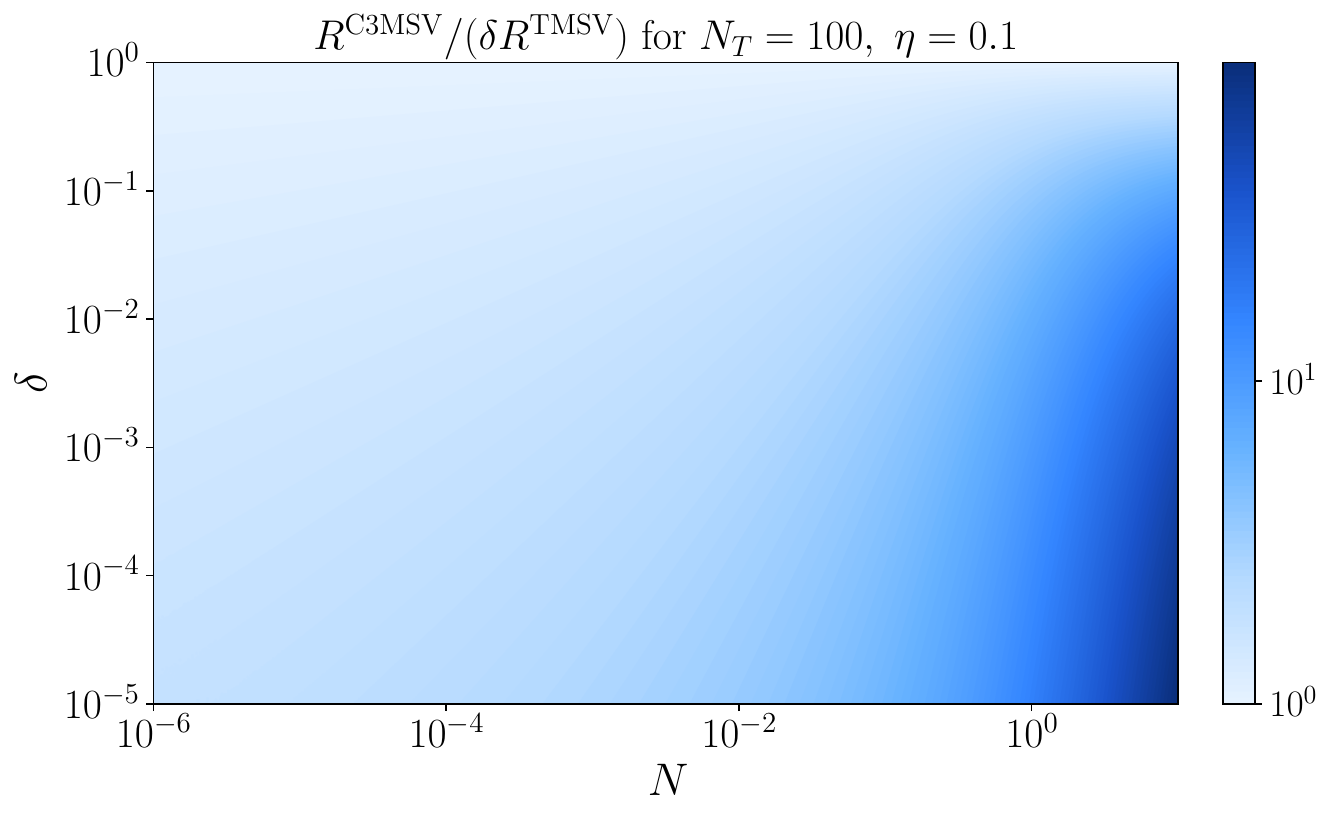}
    \caption{Comparison of the ratio of the achievable rates $R^{\textnormal{C3MSV}}$ and $\delta R^{\textnormal{TMSV}}$ over $(N,\delta)$.}
    \label{fig:rate_rat}
\end{figure}

\begin{figure}[!htbp]
    \centering
    \includegraphics[width=\linewidth]{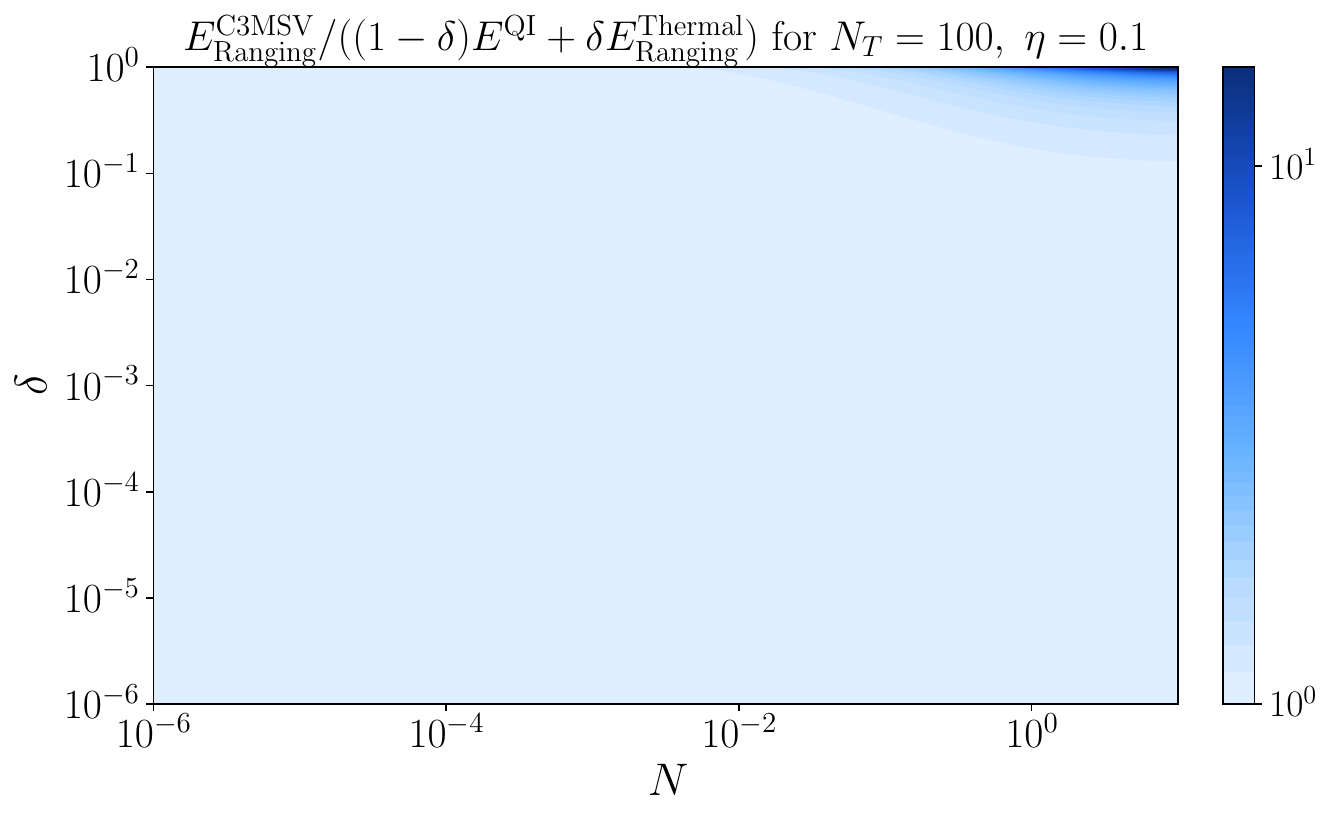}
    \caption{Ratio of the error-exponents $E^{\textnormal{C3MSV}}_\textnormal{Ranging}$ and $(1-\delta)E^{\textnormal{QI}}$ over $(N,\delta)$.}
    \label{fig:rang_exp_rat}
\end{figure}

Finally, Figures~\ref{fig:rate_rat} and~\ref{fig:rang_exp_rat} illustrate, as a function of the mean photon number $N$ and the \ac{C3MSV} idler imbalance $\delta$, the ratios of the achievable rate and error-exponent of the \ac{C3MSV} probe state to corresponding quantities along the time-sharing curve: the rate $\delta R^\textnormal{TMSV}$ and the error-exponent $(1-\delta) E^\textnormal{QI}+\delta E^\textnormal{Thermal}_\textnormal{Ranging})$, respectively. In both figures, the ratio is bounded below by one, indicating that $R^\textnormal{C3MSV}$ and $E^\textnormal{C3MSV}_\textnormal{Ranging}$ exceed the corresponding time-sharing curve rate and exponent.

\section{Conclusion}
We have analyzed two models of \ac{QISAC} over bosonic channels, a \ac{TMSV}-assisted protocol that time-shares the idler between the decoder and the detector and a \ac{C3MSV}-assisted protocol that makes the entanglement available to communication and sensing simultaneously. Our \ac{C3MSV} protocol provides an improvement over the \ac{TMSV} protocol, further suggesting that distributing entanglement resources across jointly performed tasks is preferable to time-sharing the entanglement resource among tasks. This improvement comes at no additional entanglement cost, since \ac{C3MSV} is unitarily equivalent to a \ac{TMSV} with mean photon number $N$ tensored with a vacuum mode, so that both protocols consume at most the entanglement of a single \ac{TMSV} with mean photon number $N$ per channel use. The benefit of jointly distributed entanglement is most pronounced at moderate $N$ and persists at small $N$, where the \ac{TMSV} protocol has already reduced to time-sharing. The \ac{C3MSV} protocol approaches time-sharing only as $N\to0$. These findings suggest that, in \ac{QISAC} systems, a shared entanglement source serving both functions is preferable to allocating it to one function at a time. Open questions include whether the \ac{C3MSV} protocol is optimal among entanglement-assisted strategies, and how much of the advantage survives with practical receivers and lossy entanglement distribution. Finally, in order to further asses usefulness of \ac{C3MSV} we proposed an imbalanced two-target \ac{QI} problem in Appendix~\ref{app:two_target} in which the targets are modeled as thermal-noise lossy bosonic channels with different transmissivities.

\clearpage
\appendix
\onecolumngrid

\setcounter{lemma}{0}
\renewcommand{\thelemma}{\Alph{section}\arabic{lemma}}

\section{Proofs of Theorem~\ref{thm:JBCS_class}}
\label{app:class_proof}
Throughout our \ac{QISAC} proofs, we analyze the sensing and communication protocols separately and combine the results at the end.

\noindent\textit{Classical Communication:} We assume that Alice is using the ensemble
    $\left\{\frac{1}{2\pi},\ket{\sqrt{N}e^{\imath\theta}}\bra{\sqrt{N}e^{\imath\theta}}\right\}_{\theta\in[0,2\pi)}$
    for classical communication. Let
    \[
    \hat{\rho}_B^\theta
    \eqdef
    \calL_{A\to B}^{(\eta,N_T)}
    \left(
    \ket{\sqrt{N}e^{\imath\theta}}
    \bra{\sqrt{N}e^{\imath\theta}}
    \right)
    \]
    denote the communication channel output. We use the following lemma from~\cite{su2026} as a lower bound on the achievable communication rate.

\begin{lemma}[\cite{su2026}]\label{lm:coherent_PSK}
    The Holevo information of the ensemble
    $\left\{\frac{1}{2\pi},\rho_B^\theta\right\}_{\theta\in[0,2\pi)}$
    is lower bounded by
    \begin{align}
        \chi\left(\left\{\frac{1}{2\pi},\hat{\rho}_B^\theta\right\}_{\theta\in[0,2\pi)}\right)
        &\geq
        \frac{\eta N}{N_T+1}
        +\eta N\log\left(1+\frac{1}{N_T}\right)-(N_T+\eta N)\log\left(1+\frac{\eta N}{N_T(N_T+1)}\right),
    \end{align}
    for all $N,N_{\textnormal{th}}\in\bbR_+$ and $\eta\in[0,1]$.
\end{lemma}

Note that Lemma~\ref{lm:coherent_PSK} holds without requiring approximations or assumptions such as the low-brightness or high-noise regime.

\noindent\textit{Classical Sensing:} To simplify notation, let
    $\hat{\rho}_\textnormal{th}$ denote
    $\hat{\rho}_\textnormal{th}\left(N_\textnormal{th}\right)$,
    and let $\hat{D}_\alpha$ denote $\hat{D}(\alpha)$.
    Since Alice knows the phase of the displacement, she can eliminate the phase in the returned signal at the detector. This affects only the true hypothesis, since it is the only state with non-zero mean. The remaining states are thermal states diagonal in the Fock basis.

    The classical sensing problem is a hypothesis test between the states
    \begin{align}
    \left\{\hat{\rho}_h^{\otimes n}\right\}_{h=1}^m
    \eqdef
    \left\{
    \left(\hat{\rho}_\textnormal{th}^{\otimes n}\right)^{\otimes (h-1)}
    \otimes
    \hat{\rho}_D^{\otimes n}
    \otimes
    \left(\hat{\rho}_\textnormal{th}^{\otimes n}\right)^{\otimes (m-h)}
    \right\}_{h=1}^{m},
    \end{align}
    where
    \begin{align}
    \hat{\rho}_D
    &\eqdef
    \calL^{(\kappa,N_\textnormal{th}/(1-\kappa))}
    \left(
    \ket{\sqrt{N}}\bra{\sqrt{N}}
    \right)=
    \hat{D}_{\sqrt{\kappa N}}
    \hat{\rho}_\textnormal{th}
    \hat{D}_{\sqrt{\kappa N}}^\dagger\label{eq:rho_D}.
    \end{align}

    From the $M$-ary Chernoff bound~\cite{Li2016Discriminating} and upon defining the unitary $\hat{U}_{\sqrt{\kappa N}}\eqdef \hat{D}_{\sqrt{\kappa N}}\otimes \hat{D}_{\sqrt{\kappa N}}^\dagger$, we obtain the sensing error-exponent
    \begin{align}
        E^\textnormal{Classical}_\textnormal{Ranging}
        &=
        \min_{(i,j):i\neq j}
        \sup_{s\in[0,1]}
        -\log Q_s\left(\hat{\rho}_i\Vert\hat{\rho}_j\right)\\
        &\overset{(a)}{=}
        \sup_{s\in[0,1]}
        -\log
        Q_s\left(
        \hat{\rho}_\textnormal{th}\otimes\hat{\rho}_D
        \Vert
        \hat{\rho}_D\otimes\hat{\rho}_\textnormal{th}
        \right)\\
        &\overset{(b)}{=}
        \sup_{s\in[0,1]}
        -\log
        Q_s\left(\hat{U}_{\sqrt{\kappa N}}
        \left(\hat{\rho}_D\otimes\hat{\rho}_\textnormal{th}\right)\hat{U}_{\sqrt{\kappa N}}^\dagger
        \big\Vert
        \hat{\rho}_D\otimes\hat{\rho}_\textnormal{th}\right)\\
        &\overset{(c)}{=}
        \sup_{s\in[0,1]}
        -\log
        Q_{s}\left(\hat{\rho}_D\otimes\hat{\rho}_\textnormal{th}
        \big\Vert
        \hat{U}_{\sqrt{\kappa N}}^\dagger\left(\hat{\rho}_D\otimes\hat{\rho}_\textnormal{th}
        \right)\hat{U}_{\sqrt{\kappa N}}\right)\\
        &=
        \sup_{s\in[0,1]}
        -\log
        Q_{1-s}\left(\hat{\rho}_\textnormal{th}\otimes\hat{\rho}_D
        \Vert
        \hat{\rho}_D\otimes\hat{\rho}_\textnormal{th}
        \right)\\
        &=
        -2\log
        Q_{1/2}
        \left(
        \hat{\rho}_D
        \Vert
        \hat{\rho}_\textnormal{th}
        \right).
        \label{eq:e_classic}
    \end{align}

    Here, (a) follows from the symmetry of the problem, similarly to~\cite[Eq.~(5)]{Zhuang2021}, (b) follows from $\hat{\rho}_D$ in~\eqref{eq:rho_D} and the definition of the unitary $\hat{U}_{\sqrt{\kappa N}}$, and (c) follows from the unitary invariance of $Q_s$.
    By uniqueness of the optimizing $s$~\cite{Audenaert2007}, we obtain $s^*=1/2$.
    Using techniques from~\cite{Pirandola2008}, \eqref{eq:e_classic} simplifies to
    \begin{align}
        E^\textnormal{Classical}_\textnormal{Ranging}(N)
        =
        \frac{2\kappa N}
        {1+2N_\textnormal{th}+2\sqrt{N_\textnormal{th}(N_\textnormal{th}+1)}},
    \end{align}
    which is consistent with the optimal classical ranging exponent in~\cite[Eq.~(5)]{Zhuang2021} for general $N_\textnormal{th}\in\bbR_+$.

\section{Proofs of Theorem~\ref{thm:JBCS_TMSV}, Lemma~\ref{lm:low_N_rate_exp} and Corollary~\ref{cor:JBCS_approx} }\label{app:proof_EA}

\noindent\textit{Unassisted Communication:} For unassisted communication, the encoder applies a displaced phase modulation to the signal mode of the \ac{TMSV} state. The displacement is crucial because, from Bob's perspective, the signal state is a thermal state with mean photon number $N_S$, and pure phase modulation alone does not modify the state. Formally, the displaced phase modulation is defined as
\begin{align}
    \hat{\rho}_{A}^\theta &= \hat{D}\left(\sqrt{N}_\textnormal{m}e^{\imath\theta}\right)\hat{\rho}_{S}\hat{D}\left(\sqrt{N}_\textnormal{m}e^{\imath\theta}\right)^\dagger,
\end{align}
where $\hat{\rho}_{S}=\tr[I]{\ket{\psi}\bra{\psi}_{SI}}$ and $\theta\in[0,2\pi)$. Bob's channel output is then
\begin{align}
    \hat{\rho}_B^\theta &= \hat{D}(\sqrt{\eta}\alpha_\theta)\hat{\rho}_{\textnormal{th}}(N_\eta)\hat{D}(\sqrt{\eta}\alpha_\theta)^\dagger.
\end{align}
Using Lemma~\ref{lm:coherent_PSK} with $N_T$ replaced by $N_\eta=\eta N_S+N_T$, here $\eta N_S$ acts as additional noise term hurting the communication rate, the Holevo information of the displaced \ac{PSK} scheme is lower bounded by
\begin{align}
    \chi\left(\left\{\frac{1}{2\pi},\hat{\rho}_B^\theta\right\}\right)&\geq\frac{\eta N_\textnormal{m}}{N_\eta+1}+\eta N_\textnormal{m}\log\left(1+\frac{1}{N_\eta}\right)-(N_\eta+\eta N_\textnormal{m})\log\left(1+\frac{\eta N_\textnormal{m}}{N_\eta(N_\eta+1)}\right)\label{eq:rate_dpsk},
\end{align}
$\forall N_S,N_\textnormal{m},N_{\textnormal{th}}\in\bbR_+$ and $\eta\in[0,1]$.

In the regime $N\ll 1$, substituting the parameterization of the communication rate/sensing error-exponent region boundary $N_S=(1-\delta)N,N_\textnormal{m}=\delta N$ and taking the Taylor expansion around $N=0$, we obtain the following approximation as the leading order term:
\begin{align}
    R^\textnormal{Coherent}_{\textnormal{L}}(N,\delta) = \delta\eta N\left(
        \log\left(1+\frac{1}{N_T+1}\right)-\frac{1}{N_T(N_T+1)}\right)+O\left(N^2\right)\label{eq:r_unassist_l}.
\end{align}
No assumption is made on the physical parameters of the channel, such as $\eta\ll1$ or $N_T\gg1$, for this approximation to hold.

\noindent\textit{Assisted Sensing:} As described in the time-sharing protocol between displaced-\ac{PSK} and \ac{TMSV}-\ac{PSK} protocol in Section~\ref{sec:TMSV_model_result}, Alice applies a phase modulation to the received modes in the detector to eliminate the phase. Hence the mean and covariance of the true hypothesis
\begin{align}
    \hat{\rho}_{DI}(N_S,N_\textnormal{m})\eqdef \hat{D}\left(\sqrt{2\eta  N_\textnormal{m}}\right)\otimes\textnormal{id}_I\left(\calL^{(\kappa,N_\textnormal{th}/(1-\kappa))}_{A\to D}\otimes \textnormal{id}_I\right)\left(\hat{\rho}_{SI}(N_S)\right)\hat{D}\left(\sqrt{2\eta N_\textnormal{m}}\right)^\dagger\otimes\textnormal{id}_I,
\end{align}
become
\begin{align}
    r_{DI} &= -\sqrt{2\kappa N_{\textnormal{m}}}\begin{bmatrix}
        1\\
        0\\
        0\\
        0
    \end{bmatrix}\\
    V_{DI} &= \begin{bmatrix}
        E\bfI & C\bfZ\\
        C\bfZ & S\bfI
    \end{bmatrix}
\end{align}
where $\bfZ=\begin{bmatrix}1 & 0\\0 & -1\end{bmatrix},E=2(\kappa N_S+N_\textnormal{th})+1$, $C=-2\sqrt{\kappa N_S(N_S+1)}$ and $S=2N_S+1$.

The detection problem therefore reduces to hypothesis testing between the states
\begin{align}
    \{\hat{\rho}_h^{\otimes n}\}_{h=1}^m\eqdef\left\{\left(\hat{\rho}_{\textnormal{th}}^{\otimes (h-1)}\right)^{\otimes n}\otimes\hat{\rho}_{DI}^{\otimes n}\otimes\left(\hat{\rho}_{\textnormal{th}}^{\otimes (m-h)}\right)^{\otimes n}\right\}_{h=1}^m.
\end{align}

By applying the $M$-ary Quantum Chernoff bound, the sensing error-exponent is
\begin{align}
    E^{\textnormal{TMSV}}_\textnormal{Ranging}(N_S,N_{\textnormal{m}}) &= \min_{(i,j):i\neq j}\sup_{s\in[0,1]}-\log Q_s\left(\hat{\rho}_i\Vert\hat{\rho}_j\right)\\
    &=\sup_{s\in[0,1]}-\log Q_s\left(\hat{\rho}_{\textnormal{th}}(N_{\textnormal{th}})\otimes\hat{\rho}_{DI}\Vert \hat{\rho}_{DI}\otimes \hat{\rho}_{\textnormal{th}}(N_{\textnormal{th}})\right)\\
    &=\sup_{s\in[0,1]}-\log Q_s\left(\hat{\rho}_{TDI}\Vert \hat{\rho}_{DTI}\right),
\end{align}
in which $\hat{\rho}_{TDI}(N_S,N_\textnormal{m})\eqdef\hat{\rho}_{\textnormal{th}}\left(N_\textnormal{th}\right)\otimes\hat{\rho}_{DI}(N_S,N_\textnormal{m})$ and $\hat{\rho}_{DTI}(N_S,N_\textnormal{m})$ such that $\hat{\rho}_{DI}(N_S,N_\textnormal{m})=\tr[T]{\hat{\rho}_{DTI}(N_S,N_\textnormal{m})}$ and $\hat{\rho}_\textnormal{th}(N_\textnormal{th})=\tr[D,I]{\hat{\rho}_{DTI}(N_S,N_\textnormal{m})}$. The corresponding covariance matrices and means of the states $\hat{\rho}_{TDI}$ and $\hat{\rho}_{DTI}$ are
\begin{align}
    V_{TDI} &= \begin{bmatrix}
        (2N_\textnormal{th}+1)\bfI & \bfzero & \bfzero\\
        \bfzero & E\bfI & C\bfZ\\
        \bfzero & C\bfZ & S\bfI
    \end{bmatrix},\quad r_{TDI} = -\sqrt{2\kappa N_\textnormal{m}}\begin{bmatrix}
        1\\0\\0\\0\\0\\0
    \end{bmatrix},\\
    V_{DTI} &= \begin{bmatrix}
        E\bfI & \bfzero & C\bfZ\\
        \bfzero & (2N_\textnormal{th}+1)\bfI & \bfzero\\
        C\bfZ & \bfzero & S\bfI
    \end{bmatrix},\quad r_{DTI} = -\sqrt{2\kappa N_\textnormal{m}}\begin{bmatrix}
        0\\0\\1\\0\\0\\0
    \end{bmatrix}.
\end{align}
Notice that, for $P=\begin{bmatrix}
    \bfzero &\bfI&\bfzero\\
    \bfI&\bfzero &\bfzero\\
    \bfzero&\bfzero&\bfI
\end{bmatrix}$ we have $V_{DTI} = PV_{TDI}P^T$. Since $P$ is a symplectic matrix, there exists a Gaussian unitary $\hat{U}_{RT}$ such that $\hat{\rho}_{DTI}=\hat{U}_{DT}\hat{\rho}_{TDI}\hat{U}_{RDT}^\dagger$. Similar to the classical exponent, we obtain
\begin{align}
    \sup_{s\in[0,1]}-\log Q_s\left(\hat{\rho}_{TDI}\Vert\hat{\rho}_{DTI}\right) &= \sup_{s\in[0,1]}-\log\tr{\hat{\rho}_{TDI}^s\left(\hat{U}_{DT}\hat{\rho}_{TDI}\hat{U}_{DT}^\dagger\right)^{1-s}}\\
    &= \sup_{s\in[0,1]}-\log\tr{\left(\hat{U}_{DT}^\dagger\hat{\rho}_{TDI}\hat{U}_{DT}\right)^s\hat{\rho}_{TDI}^{1-s}}\\
    &= \sup_{s\in[0,1]}-\log Q_{1-s}\left(\hat{\rho}_{TDI}\Vert\hat{\rho}_{DTI}\right).
\end{align}
Once again, by symmetry of the problem, the optimal value is $s^*=1/2$.

Following the approach of~\cite{Pirandola2008} to compute the Chernoff bound between the two Gaussian states with covariance matrices $(V_{DTI},r_{DTI})$ and $(V_{TDI},r_{TDI})$, we need the symplectic decompositions $(S_1,\{\nu_\textnormal{th},\nu_1,\nu_2\})$ and $(PS_1P^T,\{\nu_1,\nu_\textnormal{th},\nu_2\})$, respectively, in which 
\begin{align}
    \nu_i=((-1)^{i+1}(N_{\textnormal{th}}-(1-\kappa)N_S)+\sqrt{(1+(1+\kappa)N_S+N_{\textnormal{th}})^2-4\kappa N_S(N_S+1)}).
\end{align}
for $i\in\{1,2\}$~\cite{tan2008QI} and $\nu_\textnormal{th}=2N_\textnormal{th}+1$. The symplectic matrix $S_1$ that diagonalizes $V_{TRI}$ is given by
\begin{align}
    S_1 &= \begin{bmatrix}
        \bfI &\bfzero&\bfzero\\
        \bfzero& x_+\bfI& x_-\bfZ\\
        \bfzero& x_-\bfZ& x_+\bfI
    \end{bmatrix},\\
    x_\pm &=\sqrt{\frac{1+(1+\kappa)N_S+N_{\textnormal{th}}\pm\sqrt{(1+(1+\kappa)N_S+N_{\textnormal{th}})^2-4\kappa N_S(N_S+1)}}{2\sqrt{(1+(1+\kappa)N_S+N_{\textnormal{th}})^2-4\kappa N_S(N_S+1)}}}.
\end{align}
Hence, the Quantum Chernoff bound~\cite{Pirandola2008} is given by
\begin{align}
    Q_{1/2} &= \overline{Q}_{1/2}\exp\left(-d^T(V_1+V_2)^{-1}d\right),\label{eq:QCB}
\end{align}
where
\begin{align}
    V_1 &= S_1\left( \Lambda\left(\nu_\textnormal{th}\right)\bfI_2\oplus\Lambda\left(\nu_1\right)\bfI_2\oplus\Lambda\left(\nu_2\right)\bfI_2\right)S_1^T,\\
    d &= r_{TRI}-r_{RTI}= \sqrt{2\kappa N_\textnormal{m}}\begin{bmatrix}
        1 & 0 & -1 & 0 & 0 & 0
    \end{bmatrix}^T\label{eq:mean_diff},\\
    V_2 &= PV_1P^T,\\
    \overline{Q}_{1/2} &= \frac{4\Lambda\left(\nu_1\right)\Lambda\left(\nu_2\right)\Lambda\left(\nu_\textnormal{th}\right)}{E_R(E_RS_R-C_R^2)},\\
    E_R&\eqdef\Lambda\left(\nu_\textnormal{th}\right)+x_+^2\Lambda\left(\nu_1\right)+x_-^2\Lambda\left(\nu_2\right),\\
    S_R &\eqdef x_-^2\Lambda\left(\nu_1\right)+x_+^2\left(\nu_2\right),\\
    C_R &\eqdef x_+x_-\left(\Lambda\left(\nu_1\right)+\Lambda\left(\nu_2\right)\right),\\
    \det{V_1+V_2} &= \left(2E_R\left(E_RS_R-C_R^2\right)\right)^2.
\end{align}
With to the parameterization $N_\textnormal{m}=\delta N$, $N_S=(1-\delta)N$, the Taylor expansion of the error-exponent corresponding to mean independent part of quantum Chernoff bound in~\cite{Pirandola2008}, denoted $-\log \overline{Q}_{1/2}$, around $N=0$ becomes
\begin{align}
    -\log\overline{Q}_\frac{1}{2}&=\frac{2\kappa(1-\delta) N}{1+N_{\textnormal{th}}}+o\left(N\right).
\end{align}
The Taylor series of the displacement term around $N=0$ is given by
\begin{align}
    \frac{1}{2}d^T\left(V_1+V_2\right)^{-1}d &= \frac{2\delta\kappa N}{1+2N_{\textnormal{th}}+2\sqrt{N_{\textnormal{th}}(N_{\textnormal{th}}+1)}}+O\left(N^2\right).
\end{align}

Hence, the error-exponent for $N\to0$ becomes 
\begin{align}
    E^{\textnormal{TMSV}}_\textnormal{Ranging,L}(N,\delta) &= 2\kappa N\left(\frac{\delta}{{N_{\textnormal{th}}}+1}+\frac{1-\delta}{1+2N_{\textnormal{th}}+2\sqrt{N_{\textnormal{th}}(N_{\textnormal{th}}+1)}}\right)+o\left(N\right)\label{eq:exp_assist_l}.
\end{align}

\noindent\textit{Assisted Communication:} In order the lower bound the achievable rate using phase modulated TMSV we use the following lemma from~\cite{su2026}, in which $\hat{\rho}_{BI}^\theta=\hat{U}_\theta\otimes \textnormal{id}_I\left(\calL^{(\eta,N_T/(1-\eta))}\otimes \textnormal{id}_I\right)\left(\rho_{SI}\right)\hat{U}_\theta^\dagger\otimes \textnormal{id}_I$
\begin{lemma}[\cite{su2026}]
    Under uniform phase modulation of \ac{TMSV}, the Holevo information of the ensemble $\left\{\frac{1}{2\pi},\hat{\rho}_{BI}^\theta\right\}_{\theta\in[0,2\pi)}$ is lower bounded by
    \begin{align}
        \chi\left(\left\{\frac{1}{2\pi},\hat{\rho}_{BI}^\theta\right\}_{\theta\in[0,2\pi)}\right)&\geq C_\textnormal{EA}(N,N_T,\eta)+g\left(N_T\right)-g\left(\eta N+N_T\right)
	+\left(2 \eta N+N_T\right)N \log \left(\frac{\left(N_T+1\right) \left(N_T-\eta \right)}{N_T \left(N_T-\eta+1\right)}\right)\nonumber \\
&\phantom{===============}
	-N \log \left(\frac{N_T-\eta+1}{N_T+1}\right)
	+\eta  N \log \left(\frac{\left(N_T+1\right)^2 \left(N_T-\eta \right)}{N_T^2 \left(N_T-\eta+1\right)}\right),
    \end{align}
    $\forall N,N_T\in\bbR_+$ such that $N_T>\max\{\eta,\eta N-1\}$ and $\eta\in[0,1]$ where $\mu_\pm = \pm(N_T-(1-\eta)N)+\sqrt{(1+(1+\eta)N+N_T)-4\eta N(N+1)}$.
    \label{lm:HI_bound_assisted}
\end{lemma}

In the regime $N\ll 1$, we will use the following lemma from~\cite{su2026} to obtain an approximation of the lower bound in Lemma~\ref{lm:HI_bound_assisted}.
\begin{lemma}[\cite{su2026}]
    Taylor series of the lower bound of Holevo information around $N=0$ is
    \begin{align}
        -\frac{\eta N \log N}{N_T+1}+N \left(
    (N_T+\eta)\log \frac{(N_T+1)(N_T-\eta)}{N_T(N_T-\eta+1)}+ \frac{\eta}{N_T+1} \left( 1+ \log \frac{(N_T+1)^2}{N_T(N_T-\eta+1)}\right)\label{eq:r_assist_L}
\right).
    \end{align}
\end{lemma}
We denote the expansion in \eqref{eq:r_assist_L} by $R^\textnormal{TMSV}_\textnormal{L}(N)$.

\noindent\textit{Unassisted Sensing:} Since we are using phase modulation for entanglement-assisted communication, the hypothesis testing amounts to distinguishing between the following states
\begin{align}
    \left\{\hat{\rho}_h^{\otimes n}\right\}_{h=1}^m\eqdef\left\{\left(\hat{\rho}_{\textnormal{th}}(N_{\textnormal{th}})^{\otimes n}\right)^{\otimes (h-1)}\otimes\hat{\rho}_{\textnormal{th}}(\kappa N+N_\textnormal{th})^{\otimes n}\otimes\left(\hat{\rho}_{\textnormal{th}}\left(N_\textnormal{th}\right)^{\otimes n}\right)^{\otimes (m-h)}\right\}_{h=1}^m.
\end{align}

$E^{\textnormal{Thermal}}_\textnormal{Ranging}$ is given by the $M$-ary Quantum Chernoff bound
\begin{align}
    E^{\textnormal{Thermal}}_\textnormal{Ranging} &= \min_{(i,j):i\neq j}\sup_{s\in[0,1]}-\log Q_s\left(\hat{\rho}_i\Vert \hat{\rho}_j\right)\\
    &= \sup_{s\in[0,1]}-\log Q_s\left(\hat{\rho}_{\textnormal{th}}(N_{\textnormal{th}})\otimes\hat{\rho}_{\textnormal{th}}(\kappa N+N_\textnormal{th})\big\Vert\hat{\rho}_{\textnormal{th}}(\kappa N+N_\textnormal{th})\otimes\hat{\rho}_{\textnormal{th}}(N_\textnormal{th})\right)\\
    &= -2\log Q_{1/2}\left(\hat{\rho}_{\textnormal{th}}(\kappa N+N_\textnormal{th})\Vert\hat{\rho}_{\textnormal{th}}(N_\textnormal{th})\right)
\end{align}

Defining the geometric parameters of the thermal states as $r_1\eqdef\frac{\kappa N+N_\textnormal{th}}{\kappa N+N_\textnormal{th}+1}$ and $r_2\eqdef\frac{N_{\textnormal{th}}}{N_{\textnormal{th}}+1}$, we obtain
\begin{align}
    Q_{1/2}\left(\hat{\rho}_{\textnormal{th}}(\kappa N+N_{\textnormal{th}})\Vert \hat{\rho}_{\textnormal{th}}(N_{\textnormal{th}})\right) &= \sqrt{(1-r_1)(1-r_2)}\sum_{n=0}^\infty\left(\sqrt{r_1r_2}\right)^n\\
    &=\frac{\sqrt{(1-r_1)(1-r_2)}}{1-\sqrt{r_1r_2}}\\
    &= \frac{1}{\sqrt{(\kappa N+N_\textnormal{th}+1)(N_\textnormal{th}+1)}-\sqrt{(\kappa N+N_\textnormal{th})N_\textnormal{th}}}.
\end{align}

In the regime $N\to0$ we obtain the approximation
\begin{align}
    E^\textnormal{Thermal}_{L}(N)= \frac{\kappa^2 N^2}{4N_\textnormal{th}(N_\textnormal{th}+1)}+O\left(N^3\right).
\end{align}

\noindent\textit{\ac{QISAC} region:} For a fixed fraction $\lambda$ of the blocklength $n$, the idler is shared with the receiver to enhance communication, while for the remaining fraction $1-\lambda$, the idler is retained to enhance sensing performance. In each case, the detection problem reduces to hypothesis testing with \ac{iid} states, since the phase can be eliminated. To combine the error exponents of entanglement-assisted and unassisted sensing, we use the infinite-dimensional extension of the controlled sensing error exponent for quantum hypothesis testing~\cite{Wang2022Joint}, as stated in the following theorem.
    \begin{theorem}\label{thm:contr_sensing}
        Let $\calH$ be a separable Hilbert space, and let $\{\hat{\rho}_i^{x^n}\}_{i=1}^m$ be a finite set of quantum states on $\calH^{\otimes n}$ such that $\hat{\rho}_i^{x^n}\eqdef\bigotimes_{k=1}^n\hat{\rho}_i^{x_k}$ for every $i\in\intseq{1}{m}$, $x^n=(x_1,\cdots,x_n)$ is a known control sequence with $x_k\in\calX$ for $k\in\intseq{1}{n}$ and $|\calX|<\infty$. The asymptotic exponent of the probability of error of the optimal test distinguishing $\{\hat{\rho}_i^{x^n}\}_{i=1}^m$, 
        \begin{align}
            P_e^{n^*}(\{(\pi_i,\hat{\rho}_i^{x^n})\}_{i=1}^m)=\inf_{\textnormal{POVM}\{\Lambda_i^n\}_{i=1}^m}\sum_{i=1}^m\pi_i\tr{(\textnormal{id}_\calH^{\otimes n}-\Lambda_i^n)\hat{\rho}_i^{x^n}}, 
        \end{align}
            for an arbitrary non-degenerate prior $\{\pi_i\}_{i=1}^m$ and a control sequence $\{x_n\}_{n=1}^\infty$ with fixed empirical distribution $p_X$ is
        \begin{align}
            \lim_{n\to\infty}-\frac{1}{n}\log P_e^{n^*}(\{(\pi_i,\hat{\rho}_i^{x^n})\}_{i=1}^m) &= \min_{(i,j):i\neq j}\sup_{s\in[0,1]}-\sum_{x\in\calX}p_X(x)\log Q_s\left(\hat{\rho}_i^x\Vert\hat{\rho}_j^x\right).
        \end{align}
    \end{theorem}

    \begin{proof}
        \textbf{Achievability:} Using the one-shot bound in~\cite[Theorem 3.1]{cheng2026MQHT}, we obtain
        \begin{align}
            P_e^{n^*}(\{(\pi_i,\hat{\rho}_i^{x^n})\}_{i=1}^m)\leq 4(m-1)\max_{(i,j):i\neq j}\inf_{s\in[0,1]}Q_s\left(\left(\hat{\rho}_i^{x^n}\right)^s\left(\hat{\rho}_j^{x^n}\right)^{1-s}\right).
        \end{align}
        Therefore, the exponent of the optimal probability of error satisfies
        \begin{align}
            -\frac{1}{n}\log P_e^{n^*}(\{(\pi_i,\hat{\rho}_i^{x^n})\}_{i=1}^m) &\geq\frac{1}{n} \min_{(i,j):i\neq j}\sup_{s\in[0,1]}-\log Q_s\left(\left(\hat{\rho}_i^{x^n}\right)^s\left(\hat{\rho}_j^{x^n}\right)^{1-s}\right)-\frac{1}{n}\log(4(m-1))\\
            &= \min_{(i,j):i\neq j}\sup_{s\in[0,1]}-\frac{1}{n}\sum_{k=1}^n\log Q_s\left(\left(\hat{\rho}_i^{x_k}\right)^s\left(\hat{\rho}_j^{x_k}\right)^{1-s}\right)-\frac{1}{n}\log(4(m-1))\\
            &= \min_{(i,j):i\neq j}\sup_{s\in[0,1]}-\sum_{x\in\calX}p_X(x)\log Q_s\left(\left(\hat{\rho}_i^{x}\right)^s\left(\hat{\rho}_j^{x}\right)^{1-s}\right)-\frac{1}{n}\log(4(m-1))\label{eq:fix_type},
        \end{align}
        where the last equality follows from the control sequence $\{x_n\}_{n=1}^\infty$ having fixed empirical distribution. Taking the limit $n\to\infty$ concludes the proof.

        \textbf{Converse:} We use the proof technique in~\cite[Theorem 1]{Nussbaum2009Chernoff} to show that for any sequence of \ac{POVM} discriminating the $M$-ary hypotheses, probability of error can be lower bounded by the probability of error of a binary hypothesis test between any two hypotheses.

        For a sequence $x^n\in\calX^n$, consider a \ac{POVM} $\{\Lambda_i^n\}_{i=1}^m$ and a fixed pair of indices $i,j\in\intseq{1}{m}$. Let the operators $0\leq A^n,B^n\leq \textnormal{id}_\calH^{\otimes n}$ such that $A^n+B^n=\textnormal{id}_\calH^{\otimes n}-\Lambda_i^n-\Lambda_j^n$. Define the new \ac{POVM} for testing $i$ against $j$ as $\{\Tilde{\Lambda}_i^n,\Tilde{\Lambda}_j^n\}$, in which $\Tilde{\Lambda}_i^n\eqdef \Lambda_i+A^n$ and $\Tilde{\Lambda}_j^n=\Lambda_j^n+B^n$. Since $\Tilde{\Lambda}_k^n\geq \Lambda_k^n$ for $k\in\{i,j\}$, we have
        \begin{align}
            \sum_{k=1}^m\pi_k\tr{\left(\textnormal{id}_\calH^{\otimes n}-\Lambda_k^n\right)\hat{\rho}_k^{x^n}} &\geq \pi_i\tr{\left(\textnormal{id}_\calH^{\otimes n}-\Lambda_i^n\right)\hat{\rho}_i^{x^n}}+\pi_j\tr{\left(\textnormal{id}_\calH^{\otimes n}-\Lambda_j^n\right)\hat{\rho}_j^{x^n}}\\
            &\geq \min\{\pi_i,\pi_j\}\left(\tr{\left(\textnormal{id}_\calH^{\otimes n}-\Tilde{\Lambda}_i^n\right)\hat{\rho}_i^{x^n}}+\tr{\left(\textnormal{id}_\calH^{\otimes n}-\Tilde{\Lambda}_j^n\right)\hat{\rho}_j^{x^n}}\right).
        \end{align}
        Since this is true for any \ac{POVM} $\{\Lambda_k^n\}_{k=1}^m$ and any pair $i,j\in\intseq{1}{m},i\neq j$, we obtain
        \begin{align}
            P_e^{n^*}(\{(\pi_k,\hat{\rho}_k^{x^n})\}_{k=1}^m) &\geq \max_{(i,j):i\neq j}2\max\{\pi_i,\pi_j\}P_e^{n^*}\left(\left\{(1/2,\hat{\rho}_i^{x^n}),(1/2,\hat{\rho}_j^{x^n})\right\}\right).\label{eq:worst_pair}
        \end{align}
        Hence, the performance of the optimal test distinguishing $\{(\pi_k,\hat{\rho}_k^{x^n})\}_{k=1}^m$ is lower bounded by the performance of the optimal binary hypothesis test distinguishing worst pair $(i,j)$ with uniform priors. The optimal test distinguishing binary hypotheses $\hat{\rho}_i^{x^n}$ and $\hat{\rho}_j^{x^n}$ is given by the Holevo-Helstrom test $\Tilde{\Lambda}_i^{n^*}\eqdef\{\hat{\rho}_i^{x^n}\geq \hat{\rho}_j^{x^n}\}$~\cite[Chapter IV.2]{Helstrom1976}, which is a \ac{PVM}. Without loss of generality we shall develop a lower bound on the probability of error for any \ac{PVM} $\{\Pi_i^n,\textnormal{id}_\calH^{\otimes n}-\Pi_i^n\}$, which then also holds for the optimal test $\{\Tilde{\Lambda}_i^{n^*},\textnormal{id}_\calH^{\otimes n}-\Tilde{\Lambda}_i^{n^*}\}$.

        Consider the spectral decompositions
        \begin{align}
            \hat{\rho}_k^{x} &= \sum_{\ell} \nu_{k,\ell}^xE_{k,\ell}^x,\quad k\in\{i,j\},\quad x\in\calX,\\
            \hat{\rho}_k^{x^n} &= \sum_{\ell^n} \nu_{k,\ell^n}^{x^n}E_{k,\ell^n}^{x^n},\quad k\in\{i,j\},\quad x^n\in\calX^n,
        \end{align}
        in which $\nu_{k,\ell^n}^{x^n}=\prod_{p=1}^n\nu_{k,\ell_p}^{x_p}$ and $E_{k,\ell^n}^{x^n}=\bigotimes_{p=1}^nE_{k,\ell}^x$, $\left\{\left(\nu_{k,\ell}^x,E_{k,\ell}^x\right)\right\}_\ell$ are the eigenvalues and the projectors to the corresponding eigenspaces for $k\in\{i,j\}$.

        Observe that,
        \begin{align}
            \tr{(\textnormal{id}_\calH^{\otimes n}-\Pi_i^n)\hat{\rho}_i^{x^n}} &= \sum_{\ell^n}\nu_{i,\ell^n}^{x^n}\tr{(\textnormal{id}_\calH^{\otimes n}-\Pi_i^n)E_{i,\ell^n}^{x^n}}\\
            &= \sum_{\ell^n,k^n}\nu_{i,\ell^n}^{x^n}\tr{(\textnormal{id}_\calH^{\otimes n}-\Pi_i^n)E_{i,\ell^n}^{x^n}E_{j,k^n}^{x^n}},\\
            \tr{\Pi_i^n\hat{\rho}_j^{x^n}} &= \sum_{k^n}\nu_{j,k^n}^{x^n}\tr{\Pi_i^nE_{j,\ell^n}^{x^n}}\\
            &= \sum_{\ell^n,k^n}\nu_{j,k^n}^{x^n}\tr{\Pi_i^nE_{i,\ell^n}^{x^n}E_{j,k^n}^{x^n}}.
        \end{align}
        Now, define the associated Nussbaum-Szko\l a distributions~\cite{Nussbaum2009Chernoff}
        \begin{align}
            P_i^{x^n}(\ell^n,k^n) &\eqdef \nu_{i,\ell^n}^{x^n}\tr{E_{i,\ell^n}^{x^n}E_{j,k^n}^{x^n}} = \prod_{p=1}^n P_i^{x_p}(\ell_p,k_p),\quad x^n\in\calX^n\\
            P_j^{x^n}(\ell^n,k^n) &\eqdef \nu_{j,\ell^n}^{x^n}\tr{E_{i,\ell^n}^{x^n}E_{j,k^n}^{x^n}}= \prod_{p=1}^n P_j^{x_p}(\ell_p,k_p),\quad x^n\in\calX^n,
        \end{align}
        it can be verified that $\sum_{\ell^n,k^n}p_r^{x^n}(\ell^n,k^n)=1$ for $r\in\{i,j\}$. Also note that, $p_r^{x^n}$ is a product distribution for $r\in\{i,j\}$ by construction. Then,
        \begin{align}
            \tr{(\textnormal{id}_\calH^{\otimes n}-\Pi_i^n)\hat{\rho}_i^{x^n}}+\tr{\Pi_i^n\hat{\rho}_j^{x^n}} &= \sum_{\ell^n,k^n}\left(\nu_{i,\ell^n}^{x^n}\tr{(\textnormal{id}_\calH^{\otimes n}-\Pi_i^n)E_{i,\ell^n}^{x^n}E_{j,k^n}^{x^n}}+\nu_{j,k^n}^{x^n}\tr{\Pi_i^nE_{i,\ell^n}^{x^n}E_{j,k^n}^{x^n}}\right)\\
            &\geq \sum_{\ell^n,k^n}\min\{\nu_{i,k^n}^{x^n},\nu_{j,k^n}^{x^n}\}\tr{E_{i,\ell^n}^{x^n}E_{j,k^n}^{x^n}}\\
            &= \sum_{\ell^n,k^n}\min\{P_i^{x^n}(\ell^n,k^n),P_j^{x^n}(\ell^n,k^n)\}\label{eq:opt_class_test}.
        \end{align}
        Note that \eqref{eq:opt_class_test} is the optimal probability of error to the classical hypothesis testing problem $H_0:p_i^{x^n}$ against $H_1:p_j^{x^n}$. We then proceed in the same way as done in~\cite[Appendix A]{Nitinawarat2013} to find a lower bound.

        Define the \acp{PMF} for the tilted distribution
        \begin{align}
            b_{i,j}^{x,s}(\ell,k) &\eqdef\frac{P_i^x(\ell,k)^sP_j^x(\ell,k)^{1-s}}{\sum_{\ell,k}P_i^x(\ell,k)^sP_j^x(\ell,k)^{1-s}},\\
            b_{i,j}^{x^n,s}(\ell^n,k^n) &\eqdef\prod_{p=1}b_{i,j}^{x_p,s}(\ell_p,k_p).
        \end{align}
        Then the maximizing $s$ of the function
        \begin{align}
            -\sum_{p=1}^n\log\left(\sum_{\ell,k}P_i^x(\ell,k)^sP_j^x(\ell,k)^{1-s}\right),
        \end{align}
        $s^*$ occurs when
        \begin{align}
            -\sum_{p=1}^n\log\left(\sum_{\ell,k}P_i^x(\ell,k)^{s^*}P_j^x(\ell,k)^{1-s^*}\right) &= \sum_{p=1}^n\D{b_{i,j}^{x_p,s^*}}{P_i^{x_p}}= \sum_{p=1}^n\D{b_{i,j}^{x_p,s^*}}{P_j^{x_p}}.
        \end{align}
        Let $T:(\ell^n,k^n)\mapsto\{i,j\}$ be any test for discriminating between $i$ and $j$, we have either
        \begin{align}
            \P[b_{i,j}^{x^n,s^*}]{T\neq i}\geq\frac{1}{2}\quad\textnormal{or}\quad\P[b_{i,j}^{x^n,s^*}]{T\neq j}\geq\frac{1}{2}.\label{eq:tilted_any_test}
        \end{align}
        For the former of~\eqref{eq:tilted_any_test}, we define the partial sum
        \begin{align}
            S_n\eqdef\sum_{p=1}^n\left(\log\left(\frac{b_{i,j}^{x_p,s^*}(\ell_p,k_p)}{P_i^{x_p,s^*}(\ell,k)}\right)-\E[b_{i,j}^{x^n,s^*}]{\log\left(\frac{b_{i,j}^{x_p,s^*}(\ell_p,k_p)}{P_i^{x_p,s^*}(\ell,k)}\right)}\right).
        \end{align}
        By the weak law of large numbers $\frac{1}{n}S_n\xrightarrow[P]{}0$. From the definition of convergence in probability for any $\eta>0$ and $\xi>0$, there exists a number $N\in\bbN$ such that
        \begin{align}
            \P[b_{i,j}^{x^n,s^*}]{\frac{1}{n}S_n>\eta}\leq\xi,\quad\forall n\geq N.\label{eq:conv_in_prob}
        \end{align}
        Combining \eqref{eq:tilted_any_test} and \eqref{eq:conv_in_prob} with the monotonicity of probability measure, we obtain that, for $n$ sufficiently large,
        \begin{align}
            \P[b_{i,j}^{x^n,s^*}]{T\neq i,\frac{1}{n}S_n}\leq\xi.
        \end{align}
        Therefore, we can use the change of measure to obtain
        \begin{align}
            \frac{1}{2}-\xi &\overset{(a)}{\leq} \frac{1}{2}-\P[b_{i,j}^{x^n,s^*}]{T\neq i,\frac{1}{n}S_n>\eta}\\
            &\overset{(b)}{\leq}\P[b_{i,j}^{x^n,s^*}]{T\neq i,\frac{1}{n}S_n\leq\eta}\\
            &= \P[b_{i,j}^{x^n,s^*}]{T\neq i,b_{i,j}^{x^n,s^*}\leq P_i^{x^n}e^{n\left(\frac{1}{n}\sum_{k=1}\D{b_{i,j}^{x_k,s^*}}{P_i^{x_k}}+\eta\right)}}\\
            &\leq \P[P_i^{x^n}]{T\neq i,b_{i,j}^{x^n,s^*}\leq P_i^{x^n}e^{n\left(\frac{1}{n}\sum_{k=1}\D{b_{i,j}^{x_k,s^*}}{P_i^{x_k}}+\eta\right)}}e^{n\left(\frac{1}{n}\sum_{k=1}\D{b_{i,j}^{x_k,s^*}}{P_i^{x_k}}+\eta\right)}\\
            &\leq \P[P_i^{x^n}]{T\neq i}e^{n\left(\frac{1}{n}\sum_{k=1}\D{b_{i,j}^{x_k,s^*}}{P_i^{x_k}}+\eta\right)}
        \end{align}
        where $(a)$ follows from \eqref{eq:conv_in_prob}, and $(b)$ follows from \eqref{eq:tilted_any_test}. Similarly, if the latter statement of \eqref{eq:tilted_any_test} is true, we obtain
        \begin{align}
            \frac{1}{2}-\xi\leq \P[P_j^{x^n}]{T\neq j}e^{n\left(\frac{1}{n}\sum_{k=1}\D{b_{i,j}^{x_k,s^*}}{P_j^{x_k}}+\eta\right)}.
        \end{align}
        Then, for any test $T$, we obtain
        \begin{align}
            \P[P_i^{x^n}]{T\neq i}+\P[P_j^{x^n}]{T\neq j}\geq\left(\frac{1}{2}-\xi\right)e^{\left(-n\left(\frac{1}{n}\sum_{p=1}^n-\log\left(\sum_{\ell,k}P_i^{x_p}(\ell,k)^{s^*}P_j^{x_p}(\ell,k)^{1-s^*}\right)+\eta\right)\right)}\label{eq:opt_class_lb}
        \end{align}
        Since \eqref{eq:opt_class_lb} holds for any test $T$ and any $\eta,\xi>0$ for $n$ large enough, by combining it with \eqref{eq:worst_pair} and \eqref{eq:opt_class_test}, we obtain
        \begin{align}
            \lim_{n\to\infty}-\frac{1}{n}\log P_e^{n^*}\left(\{(\pi_i,\hat{\rho}_i^{x^n})\}_{i=1}^m\right) &\leq \lim_{n\to\infty}\min_{(i,j):i\neq j}\sup_{s\in[0,1]}\frac{1}{n}\sum_{p=1}^n-\log\left(\sum_{k,l}P_i^{x_p}(\ell,k)^{s}P_j^{x_p}(\ell,k)^{1-s}\right)\\
            &= \min_{(i,j):i\neq j}\sup_{s\in[0,1]}-\sum_{x\in\calX}p_X(x)\log\left(\sum_{k,l}P_i^{x}(\ell,k)^{s}P_j^{x}(\ell,k)^{1-s}\right)\\
            &= \min_{(i,j):i\neq j}\sup_{s\in[0,1]}-\sum_{x\in\calX}p_X(x)\log\left(\sum_{k,l}\left(\nu_{i,\ell}^{x}\right)^s\left(\nu_{j,k}^x\right)^{1-s}\tr{E_{i,\ell}^xE_{j,k}^x}\right)\\
            &= \min_{(i,j):i\neq j}\sup_{s\in[0,1]}-\sum_{x\in\calX}p_X(x)\log\left(\tr{\left(\hat{\rho}_i^x\right)^s\left(\hat{\rho}_j^x\right)^{1-s}}\right)\\
            &= \min_{(i,j):i\neq j}\sup_{s\in[0,1]}-\sum_{x\in\calX}p_X(x)\log Q_s\left(\hat{\rho}_i^x\Vert\hat{\rho}_j^x\right).
        \end{align}
    \end{proof}

    We can treat the sharing of entanglement as a control input sequence with fixed empirical distribution $p_X$ on the alphabet $\calX=\{C,S\}$, in which $p_X(C)=\lambda,p_X(S)=\bar{\lambda}$. Here, $C$ denotes that entanglement is shared with the receiver for assisted communication, and $S$ denotes that entanglement is retained for assisted sensing. Next, we define the states in mode $\hat d_h$ corresponding to the modulation scheme.
    \begin{align}
        \hat{\sigma}^{C}_1 &\eqdef \hat{\rho}_{\textnormal{th}}\left(N_{\textnormal{th}}\right)\otimes\hat{\rho}_D,\\
        \hat{\sigma}^{C}_2 &\eqdef \hat{\rho}_D\otimes \hat{\rho}_{\textnormal{th}}\left(N_{\textnormal{th}}\right),\\
        \hat{\sigma}^{S}_1 &\eqdef \hat{\rho}_{TDI},\\
        \hat{\sigma}^{S}_2 &\eqdef \hat{\rho}_{DTI}.
    \end{align}
    Using Theorem~\ref{thm:contr_sensing}, we obtain
\begin{align}
    E &= \sup_{s\in [0,1]}-\sum_{x\in\calX}p_X(x)\log Q_{s}\left(\hat{\sigma}_1^x\Vert\hat{\sigma}_2^x\right)\\
    &=\sup_{s\in [0,1]} \left(-\lambda\log Q_s\left(\hat{\rho}_{\textnormal{th}}(N_{\textnormal{th}})\otimes\hat{\rho}_D\Vert\hat{\rho}_D\otimes\hat{\rho}_{\textnormal{th}}(N_{\textnormal{th}})\right)-(1-\lambda)\log Q_s\left(\hat{\rho}_{TDI}\Vert\hat{\rho}_{DTI}\right)\right)\\
    &\overset{(a)}{=} \lambda E^{\textnormal{Thermal}}_\textnormal{Ranging}(N)+(1-\lambda)E^{\textnormal{TMSV}}_\textnormal{Ranging}(N_S,N_{\textnormal{m}}),
\end{align}
where $(a)$ is because $s=0.5$ is optimal for both Chernoff bounds.

Since the protocol time-shares between the two schemes, the achievable communication rate is the corresponding convex combination of the rates in individual schemes~\cite{cover1991elements},
\begin{align}
    R = \lambda R^{\textnormal{TMSV}}(N)+(1-\lambda) R^{\textnormal{Coherent}}(N_S,N_\textnormal{m}).
\end{align}

Note that, in the regime where the inequality in~\eqref{eq:rate_dpsk} is tight, the displaced \ac{PSK} rate increases with $N_\textnormal{m}$ for fixed $N_S<N$. In contrast, the \ac{QCB} term $\overline{Q}_{1/2}$ in~\eqref{eq:QCB} depends only on the symplectic decomposition $(S_1,\{\nu_1,\nu_2\})$ and is therefore independent of $N_\textnormal{m}$, while the mean-dependent part of~\eqref{eq:QCB} contributes an exponent that grows linearly in $N_\textnormal{m}$. Both the communication rate and the sensing exponent are thus nondecreasing in $N_\textnormal{m}$, so, subject to the constraint $N_S+N_\textnormal{m}\le N$, choosing $N_\textnormal{m}=N-N_S$ is optimal for a given $N_S$. Taking the union over $\lambda,\delta\in[0,1]$, with the parametrization $N_S=(1-\delta)N$ and $N_\textnormal{m}=\delta N$, completes the proof of Theorem~\ref{thm:JBCS_TMSV}.

Let us focus on the regime $N\to0$. Starting from the point $(N_S,N_\textnormal{m})=(N,0)$, shifting an amount $\Delta N=N\Delta\delta$ of mean photon number from $N_S$ to $N_\textnormal{m}$ decreases the sensing exponent by
\begin{align}
\Delta E \eqdef E^\textnormal{TMSV}_\textnormal{Ranging,L}(N,\Delta\delta)-E^\textnormal{TMSV}_\textnormal{Ranging,L}(N,0)=-\Theta(N),
\end{align}
according to~\eqref{eq:exp_assist_l}, while the communication rate increases by
\begin{align}
\Delta R = R^{\textnormal{Coherent}}_L(N,\Delta\delta) = \Theta(N),
\end{align}
according to~\eqref{eq:r_unassist_l}. We quantify the tradeoff by the slope $\Delta E/\Delta R$, i.e., the change in exponent per unit change in rate. Reallocating power therefore has slope $-\Theta(1)$. In contrast, the changes along the time-sharing direction satisfy
\begin{align}
\Delta R &= R^\textnormal{TMSV}_\textnormal{L}(N,\Delta\delta) = -\Theta(N\log N),\\
\Delta E &= E^\textnormal{TMSV}_\textnormal{Ranging,L}(N,\Delta\delta)-E^\textnormal{TMSV}_\textnormal{Ranging,L}(N,0) = -\Theta(N),
\end{align}
according to~\eqref{eq:r_assist_L}, yielding a slope of $\Theta\left(1/\log N\right)$, which vanishes as $N\to0$. Since both slopes are negative, we compare their magnitudes. Per unit of rate gained, time sharing costs far less exponent than power reallocation, so the time-sharing slope is larger (less negative) than the slope of the boundary of the $(N_S,N_\textnormal{m})$ region. This imbalance between the rate scalings of entanglement-assisted ($-N\log N$) and unassisted ($N$) communication makes $(N_S,N_\textnormal{m})=(N,0)$ the optimal power allocation, so the achievable region reduces to time sharing.

\section{Proof of Theorem~\ref{thm:JBCS_C3MSV}}\label{app:proof_JBCS_C3}
\noindent\textit{Communication:}
    \begin{lemma}
        Let $\hat{\rho}_{I_CSI_D}(\delta)$ be a $\delta$-imbalanced \ac{C3MSV}. If $N_T>\max\left\{\eta,\eta N\left(1-\frac{1-\delta}{\delta N+1}\right)-1\right\}$,
        \begin{align}
            \hat{\rho}_{BI_C}(\delta) &= \sum_{n_1=0}^\infty\sum_{n_2=0}^\infty \frac{ (N_T-\eta )}{(N_T-\eta +1) (\eta N+N_T+  \delta N( N_T-\eta)}\left(\frac{N_T-\eta}{N_T-\eta+1}\right)^{n_1}\left(\frac{\delta N (N_T-\eta )}{\eta  N+N_T+\delta N(N_T-\eta)}\right)^{n_2}\nonumber\\
            &\phantom{===============}\times F\left[n_1
            +1,n_2+1;1,\frac{\delta \eta  (N+1)}{(N_T-\eta+1)(\eta N+\delta N(N_T-\eta ))}\right]\ket{n_1n_2}\bra{n_1n_2},
        \end{align}
        where $F[n_1+1,n_2+1;1;z]=\sum_{k=0}^\infty \binom{n_1+k}{k}\binom{n_2+k}{k}z^k$ for $\abs{z}<1$.
    \end{lemma}
    \begin{proof}
        Without phase shift, Bob's covariance matrix is
        \begin{align}
            V_{BI_C}(\delta) &= \begin{bmatrix}
                (2(\eta N+N_T)+1)\bfI & -2\sqrt{\eta\delta N(N+1)}\bfZ\\
                -2\sqrt{\eta\delta N(N+1)}\bfZ & (2\delta N+1)\bfI
            \end{bmatrix}.
        \end{align}
        Following the ring state analysis in~\cite{su2026,shi2020}, by adjusting the parameters of the covariance matrix we obtain the diagonal representation given above. Note that $\delta=1$ is consistent with the \ac{TMSV} ring state results~\cite{su2026,shi2020}.
    \end{proof}

    \begin{lemma}
        Consider the ensemble $\left\{\frac{1}{2\pi},\hat{\rho}^{\theta}_{BI_C}\right\}_\theta$ generated by uniform phase shifts of $\hat{\rho}_{R_CI_C}(\delta)$. If $N_T>\max\left\{\eta,\eta N\left(1-\frac{1-\delta}{\delta N+1}\right)-1\right\}$,
        \begin{align}
            \chi\left(\frac{1}{2\pi},\rho^{\theta}_{BI_C}\right)\geq &
            \delta  N \log \left(\frac{(1-\delta) \eta  N+N_T \left(\delta  N+1\right)}{\delta  N \left(N_T-\eta\right)}\right)+\log \left(\left(N_T-\eta+1\right) \left(N \left((1-\delta)\eta +\delta  N_T\right)+N_T\right)\right)\nonumber\\
        &-\left(N \left(\delta  \eta +\delta +\eta +2 \delta  \eta  N+\delta  N_T\right)+N_T+1\right) \log \left(1-\frac{\eta  \left(N+1\right)}{\left(\eta -N_T-1\right) \left((\delta -1) \eta  N-N_T \left(\delta  N+1\right)\right)}\right)\nonumber\\
        &+\left(\eta  N+N_T\right)\log \left(1+\frac{1}{N_T-\eta}\right)-\log \left(N_T-\eta \right)\nonumber\\
        &-g\left(\frac{\nu_+(\delta)-1}{2}\right)-g\left(\frac{\nu_-(\delta)-1}{2}\right),
    \end{align}
    where $\nu_\pm(\delta)\eqdef \pm(N_T-(\delta-\eta)N)+\sqrt{(1+(\delta+\eta)N+N_T)^2-4\eta\delta N(N+1)}$.
    \end{lemma}

    \begin{proof}
        Following the ring state analysis and hypergeometric bounds in~\cite{su2026}, by adjusting the parameters of the covariance matrix we obtain the diagonal representation given above. Note that $\delta=1$ is again consistent with the \ac{TMSV} ring state results~\cite{su2026,shi2020}.
    \end{proof}

    In the regime $N\to0$, we obtain the approximation
    \begin{align}
        R^\textnormal{C3MSV}(N,\delta) &=-\frac{\delta\eta N \log N}{N_T+1}+\delta N \left(
    (N_T+\eta)\log \frac{(N_T+1)(N_T-\eta)}{N_T(N_T-\eta+1)}+ \frac{\eta}{N_T+1} \left(-\log\delta 1+ \log \frac{(N_T+1)^2}{N_T(N_T-\eta+1)}\right)\label{eq:r_c3_L}
\right)\\
        &= \delta R^\textnormal{TMSV}(N)+\frac{\eta N}{N_T+1}(-\delta\log\delta),
    \end{align}
    where the last inequality is obtained by comparing against~\ref{eq:r_assist_L}.
        
\noindent\textit{Sensing:} Alice's return idler covariance matrix under true hypothesis is
    \begin{align}
        V_{DI_D}\left(\delta\right) &= \begin{bmatrix}
            (2(\kappa N+N_\textnormal{th})+1)\bfI & \sqrt{(1-\delta)\kappa N(N+1)}\bfZ\\
            \sqrt{(1-\delta)\kappa N(N+1)}\bfZ & (2((1-\delta)N+N_\textnormal{th})+1)\bfI
        \end{bmatrix}.
    \end{align}
    Using the techniques of~\cite{Pirandola2008} and modifying the symplectic decomposition in Appendix~\ref{app:proof_EA} to account for the new covariance matrix, we compute the exponent and, as $N\to 0$, obtain
    \begin{align}
        E^{\textnormal{C3MSV}}_\textnormal{Ranging} = \frac{2(1-\delta)\kappa N}{N_\textnormal{th}+1}+o\left(N\right).
    \end{align}
\textnormal{\ac{QISAC} region}: Taking the union over $\delta\in[0,1]$ over achievable communication rate and sensing error-exponent pairs completes the proof.

\section{Two-Target QI Model}\label{app:two_target}

\begin{figure}[!htbp]
    \centering
    \includegraphics[width=0.5\textwidth]{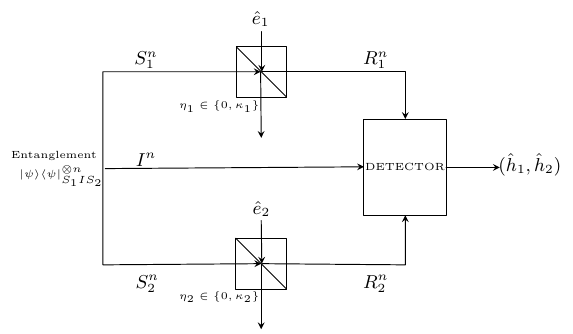}
    \caption{Two-target \ac{QI} model.}
    \label{fig:two_target_QI_model}
\end{figure}

In order to further assess the usefulness of \ac{C3MSV} states~\cite{Zhan2023C3MSV}, we also consider the imbalanced two-target illumination model shown in Figure~\ref{fig:two_target_QI_model}. In this model, $n$ copies of a probe state $\hat{\rho}_{S_1IS_2}=\ket{\psi}\bra{\psi}_{S_1IS_2}$, consisting of two signal modes $(S_1,S_2)$ and one idler mode $I$, are prepared for two-target detection. Each target is modeled as a lossy-thermal-noise bosonic channel, denoted as $\calL^{(\eta_i,N_\textnormal{th}/(1-\eta_i))}$
\begin{align}
    \hat{r}_i = \sqrt{\eta_{i}}\hat{s}_{i}+\sqrt{1-\eta_i}\hat{e}_{i},\quad i\in\{1,2\}\label{eq:two_target_channel}
\end{align}
where the transmissivity $\eta_i\in\{0,\kappa_i\}$ indicates the absence or presence of the $i^{\textnormal{th}}$ target, respectively, and $\hat{e}_{i}$ is a thermal-noise with mean photon number $N_{\textnormal{th}}/(1-\eta_{i})$. As it is standard in the \ac{QI} literature~\cite{tan2008QI,Zhuang2021} the dependence of the noise mean photon number on the channel transmissivity models the \ac{NPS} assumption, under which the transmitter must actively probe the channel to distinguish the hypotheses. The returned signal modes are denoted by $R_i$, with corresponding annihilation operators $\hat{r}_i$ for $i\in\{1,2\}$. The detection task is formulated as a hypothesis test with hypotheses $(h_1,h_2)$, in which $h_i=0$ corresponds to $\eta_i(h_i)=0$ and $h_i=1$ corresponds to $\eta_i(h_i)=\kappa_i$ for $i\in\{1,2\}$. Upon receiving the returned modes the detector applies the \ac{POVM} $\left\{\Pi_{(h_1,h_2)}\right\}_{(h_1,h_2)\in\{0,1\}^2}$ over the modes $(S_1^n,I^n,S_2^n)$. The returned state is defined as 
\begin{flalign}
    &\hat{\rho}_{R_1IR_2}^{(h_1,h_2)}\eqdef\left(\calL^{(\eta_1(h_1),N_\textnormal{th}/(1-\eta_1(h_1)))}\otimes\textnormal{id}_I\otimes \calL^{(\eta_2(h_2),N_\textnormal{th}/(1-\eta_2(h_2)))}\right) \left(\hat{\rho}_{S_1IS_2}\right).
\end{flalign}    
The average probability of sensing error is defined as $P_e\eqdef\frac{1}{4}\sum_{(h_1,h_2)\in\{0,1\}^2}\tr{\Pi_{(h_1,h_2)}\rho_{R_1IR_2}^{(h_1,h_2)}}$,.

\subsection{TMSV QI}
We first consider two independent $n$ pairs of \ac{TMSV} states for two-target \ac{QI}. In this formulation, the probe state is $\hat{\rho}_{S_1I_1I_2S_2}(\delta)\eqdef \ket{\psi}\bra{\psi}^{\ac{TMSV}}_{S_1I_1}(\delta)\otimes \ket{\psi}\bra{\psi}^{\ac{TMSV}}_{I_2S_2}(1-\delta)$, where $I$ in the original formalism is replaced by two idler modes $(I_1,I_2)$. Since the targets have different transmissivities we considered an imbalanced resource, in which modes $(S_1,S_2)$ (consequently modes $(I_1,I_2)$) have mean photon numbers $(2\delta N_S,2(1-\delta)N_S)$, respectively. Following the target model presented in \eqref{eq:two_target_channel}, the covariance matrix for the first target under both hypothesis,
\begin{align}
    V^{(0)}_{R_1I_1}(\delta)&=\begin{bmatrix}
        (2N_\textnormal{th}+1)\bfI & \bfzero\\
        \bfzero & (2\delta N_S+1)\bfI
    \end{bmatrix},\label{eq:TMSV_two_target_cm0}\\
    V^{(1)}_{R_1I_1}(\delta)&=\begin{bmatrix}
        (2\kappa \delta N_S+2N_\textnormal{th}+1)\bfI & 2\sqrt{\kappa\delta N_S(\delta N_S+1)}\bfZ\\
        2\sqrt{\kappa\delta N_S(\delta N_S+1)}\bfZ & (2N_S+1)\bfI
    \end{bmatrix}\label{eq:TMSV_two_target_cm1}
\end{align}
and the corresponding quantum state given as $\hat{\rho}^{(0)}_{R_1I_1}(\delta)$ and $\hat{\rho}^{(1)}_{R_1I_1}(\delta)$. For the second target, the only change is replacing $\delta$ with $1-\delta$. Then, the covariance matrix and the quantum state corresponding to all four modes is given as $V_{R_1I_1R_2I_2}^{(h_1,h_2)}(\delta)=V_{R_1I_1}^{(h_1)}(\delta)\bigoplus V_{R_2I_2}^{(h_2)}(\delta)$ and $\hat{\rho}^{(h_1,h_2)}_{R_1I_1R_2I_2}(\delta)=\hat{\rho}^{(h_1)}_{R_1I_1}(\delta)\otimes\hat{\rho}_{R_2I_2}^{(h_2)}(\delta)$. Under hypothesis $h_i=0$, we denote $\hat{\rho}_{R_iI_i}^{(h_i)}(\delta)\eqdef \hat{\rho}_\textnormal{th}(N_\textnormal{th})\otimes \hat{\rho}_{I_i}(\delta)$ and for $h_i=1$, we denote $\hat{\rho}_{R_iI_i}^{(h_i)}(\delta)\eqdef \left(\calL^{(\eta_i,N_\textnormal{th}/(1-\eta_i))}\otimes \textnormal{id}_I\right)\left(\hat{\rho}_{S_iI_i}(\delta)\right)$.

For each lossy thermal-noise bosonic channel, defined in~\eqref{eq:two_target_channel}, $n$ independent pairs of \ac{TMSV} states are prepared. The asymptotic error-exponent for two-target detection under two independent \ac{TMSV} sequences is given by the $m$-ary \ac{QCB}~\cite{cheng2026MQHT}.
\begin{flalign}
    E^{\textnormal{TMSV}}_\textnormal{two-target} &= \sup_{\delta\in[0,1]}\min_{(h_1,h_2)\neq(h_1',h_2')}\sup_{s\in[0,1]}-\log Q_s\left(\hat{\rho}^{(h_1,h_2)}_{R_1I_1R_2I_2}(\delta)\Vert\hat{\rho}^{(h_1',h_2')}_{R_1I_1R_2I_2}(\delta)\right)\\
      &= \sup_{\delta\in[0,1]}\min_{(h_0,h_1)\neq(h_0',h_1')}\sup_{s\in[0,1]}-\log Q_s\left(\hat{\rho}_{R_1I_1}^{(h_1)}\Vert\hat{\rho}_{R_1I_1}^{(h_1')}\right)-\log Q_s\left(\hat{\rho}_{R_2I_2}^{(h_2)}\Vert\hat{\rho}_{R_2I_2}^{(h_2')}(\delta)\right)\\
      &= \sup_{\delta\in[0,1]}\min_{i\in\{1,2\}}\sup_{s\in[0,1]}-\log Q_s\left(\hat{\rho}_{R_iI_i}^{(1)}(\delta)\Vert\hat{\rho}_{R_iI_i}^{(0)}(\delta)\right)\label{eq:TMSV_two_target_exp}\\
      &\eqdef \sup_{\delta\in[0,1]}\min_{i\in\{1,2\}} E^{\textnormal{TMSV}}_{\textnormal{two-target},i}(\delta).
\end{flalign}
Note that the inner term containing the supremum over $s$ is the same \ac{QI} exponent as in~\cite{tan2008QI}. From the covariance matrices in~\eqref{eq:TMSV_two_target_cm0} and~\eqref{eq:TMSV_two_target_cm1}, as it is not directly evident what is the optimal $\delta$ is. The expression in~\eqref{eq:TMSV_two_target_exp} is calculated using the techniques in~\cite{Pirandola2008}, however, the actual expression is too complicated to report here. Focusing on the low signal mean photon number regime, taking the Taylor series as $N_S\to0$, the leading order term is
\begin{align}
    E^{\textnormal{TMSV}}_{\textnormal{two-target},1}(\delta) &= \left(4-\left(\sqrt{1-\frac{\kappa_1}{N_\textnormal{th}+1}}+1\right)^2\right)\delta N_S,\\
    E^{\textnormal{TMSV}}_{\textnormal{two-target},2}(\delta) &= \left(4-\left(\sqrt{1-\frac{\kappa_2}{N_\textnormal{th}+1}}+1\right)^2\right)(1-\delta) N_S.\label{eq:TMSV_two_target_exp_lowNs}
\end{align}
In the low signal mean photon number regime, the optimal power allocation is given by
\begin{align}
    \delta_{\textnormal{TMSV}}^* = \frac{4-\left(\sqrt{1-\frac{\kappa_2}{N_\textnormal{th}+1}}+1\right)^2}{8-\left(\sqrt{1-\frac{\kappa_1}{N_\textnormal{th}+1}}+1\right)^2-\left(\sqrt{1-\frac{\kappa_2}{N_\textnormal{th}+1}}+1\right)^2}.
\end{align}
For the special case of targets with same reflectivity, $\kappa_1=\kappa_2=\kappa$, $\delta_{\textnormal{TMSV}}^*=1/2$ and in the limit of high $N_\textnormal{th}$ the exponent is $\frac{\kappa}{N\textnormal{th}}N_S$, consistent with~\cite{tan2008QI}.

\subsection{C3MSV Quantum Illumination}
We next consider \textnormal{an} imbalanced \ac{C3MSV} state with covariance matrix in~\eqref{eq:cov_c3msv} with $N_S$ replaced by $2N_S$ for two-target quantum illumination. Under the channel model described in~\eqref{eq:two_target_channel}, the covariance matrix of the state at the receiver for $\eta_i(h_i)\in\{0,\kappa_i\}$, for $i\in\{1,2\}$ is given by
\begin{align}
    V_{R_1IR_2}^{(h_1,h_2)}(\delta) = \begin{bmatrix}
    R_{1}^{(h_1)}(\delta)\bfI & C_{12}^{(h_1)}(\delta)\bfZ & C_{13}^{(h_1,h_2)}(\delta)\bfI\\
     C_{12}^{(h_1)}(\delta)\bfZ & S_{22}\bfI & C_{23}^{(h_2
     )}(\delta)\bfZ\\
     C_{13}^{(h_1,h_2)}(\delta)\bfI & C_{23}^{(h_2)}(\delta)\bfZ & R_2^{(h_2)}(\delta)\bfI
    \end{bmatrix},
\end{align}
where $R_1^{(h_1)}(\delta) = 2\left( \eta_1(h_1)2\delta  N_S+N_{\text{th}}\right)+1$, $R_2^{(h_2)}(\delta) = 2\left( \eta_2(h_2)2(1-\delta)  N_S+N_{\text{th}}\right)+1$, $C_{12}^{(h_1)}(\delta)=-2 \sqrt{\eta_1(h_1)2\delta N_S \left(2 N_S+1\right)}$, $C_{23}^{(h_2)}(\delta)=-2 \sqrt{\eta_2(h_2)2(1-\delta) N_S \left(2 N_S+1\right)}$ and $C_{13}^{(h_1,h_2)}(\delta)=2 \sqrt{\eta_1(h_1)\eta_2^{(h_2)} \delta(1-\delta)} N_S$.

Let $\hat{\rho}_{R_1IR_2}^{(h_0,h_1)}(\delta)$ be the density operator corresponding to Gaussian quantum state with covariance matrix $V_{R_1IR_2}^{(h_0,h_1)}(\delta)$. The asymptotic error-exponent for distinguishing $n$ copies of the states $\left\{\hat{\rho}_{R_1IR_2}^{(h_0,h_1)}(\delta)\right\}_{(h_0,h_1)\in\{0,1\}^2}$ is given by the $m$-ary \ac{QCB}
\begin{flalign}
    E^{\textnormal{C3MSV}}_\textnormal{two-target} = &\sup_{\delta\in[0,1]}\min_{(h_1,h_2)\neq(h_1',h_2')}\sup_{s\in[0,1]}-\log Q_s\left(\hat{\rho}_{R_1IR_2}^{(h_1,h_2)}(\delta)\|\hat{\rho}_{R_1IR_2}^{(h_1',h_2')}(\delta)\right).\label{eq:C3MSV_two_target_exp}
\end{flalign}
Because of the monotonicity of \ac{QCB}~\cite{Audenaert2007} under \ac{CPTP} maps, using partial trace over mode $R_2$ (similarly for $R_1$), we obtain
\begin{flalign*}
    \min_{(h_1,h_2)\neq(h_1',h_2')}\sup_{s\in[0,1]} -\log Q_s\left(\hat{\rho}_{R_1IR_2}^{(h_1,h_2)}(\delta)\|\hat{\rho}_{R_1IR_2}^{(h_1',h_2')}(\delta)\right)&\geq\min_{i\in\{1,2\}}\min_{h_i\neq h_i'}\sup_{s\in[0,1]} -\log Q_s\left(\hat{\rho}_{R_iI}^{(h_i)}(\delta)\|\hat{\rho}_{R_1I}^{(h_i')}(\delta)\right)\\
    &= \min_{i\in\{1,2\}}\sup_{s\in[0,1]} -\log Q_s\left(\hat{\rho}_{R_iI}^{(0)}(\delta)\|\hat{\rho}_{R_iI}^{(1)}(\delta)\right)\\
    &= \min_{i\in\{1,2\}}\sup_{s\in[0,1]}-\log Q_s\left(\hat{\rho}^{(0)}_{R_iI}(\delta)\otimes\hat{\rho}_\textnormal{th}\Vert\hat{\rho}^{(1)}_{R_iI}(\delta)\otimes\hat{\rho}_\textnormal{th}\right)\\
    &= \min_{i\in\{1,2\}}\sup_{s\in[0,1]}-\log Q_s\left(\hat{\rho}^{(0,0)}_{R_iIR_{3-i}}(\delta)\Vert\hat{\rho}^{(1,0)}_{R_iIR_{3-i}}(\delta)\right).
\end{flalign*}
The lower bound is achieved by the worse of the individual detection performances of each target. Hence, the minimization in~\eqref{eq:C3MSV_two_target_exp} is restricted to comparing $(h_1,h_2)=(0,0)$ against $(h_1,h_2)\in\{(0,1),(1,0)\}$, which reduces the problem to discriminating $\hat{\rho}_\textnormal{th}\left(N_\textnormal{th}\right)\otimes \hat{\rho}_I$ against $\hat{\rho}_{R_1I}(\delta)$ or $\hat{\rho}_{R_2I}(\delta)$, corresponding to covariance matrices,
\begin{align}
    V_{R_iI}^{(0)} &= \begin{bmatrix}
        (2N_\textnormal{th}+1)\bfI & \bfzero\\
        \bfzero & (4N_S+1)\bfI
    \end{bmatrix}\label{eq:C3MSV_two_target_cm0},\\
    V_{R_1I}^{(1)}(\delta) &= \begin{bmatrix}
        R_1^{(1)}(\delta)\bfI & C_{12}^{(1)}(\delta)\bfZ\\
        C_{12}^{(1)}(\delta)\bfZ & S_{22}\bfI
    \end{bmatrix},\label{eq:C3MSV_two_target_cm1}\\
    V_{R_2I}^{(1)}(\delta) &= \begin{bmatrix}
        R_2^{(1)}(\delta)\bfI & C_{23}^{(1)}(\delta)\bfZ\\
        C_{23}^{(1)}(\delta)\bfZ & S_{22}\bfI
    \end{bmatrix}\label{eq:C3MSV_two_target_cm2}.
\end{align}
Notice that, for both targets, the no-target hypothesis $h_i=0$ is independent of $\delta$. Also, $\delta\kappa_1$ and $(1-\delta)\kappa_2$ acts as the effective transmissivity. Consequently, the saddle point optimization in~\eqref{eq:C3MSV_two_target_exp} over $\delta$ and worst-pair of hypothesis is achieved by $\delta\kappa_1=(1-\delta)\kappa_2$. The optimal mean photon number allocation is given by
\begin{flalign}
    \delta^*_\textnormal{C3MSV} = \frac{\kappa_2}{\kappa_1+\kappa_2}.
\end{flalign}

Taking Taylor series as $N_S\to0$, the leading order term is
\begin{align}
    \left(4-\left(\sqrt{1-\frac{\kappa_1\kappa_2}{(N_\textnormal{th}+1)(\kappa_1+\kappa_2)}}+1\right)^2\right)N_S.\label{eq:C3MSV_two_target_exp_lowN}
\end{align}
In the limit of high $N_\textnormal{th}$ and $\kappa_1=\kappa_2=\kappa$, this term is $\frac{\kappa}{N\textnormal{th}}N_S$, identical to \ac{TMSV} result in~\eqref{eq:TMSV_two_target_exp_lowNs}.

\begin{figure}[!htbp]
    \centering
    \includegraphics[width=0.5\textwidth]{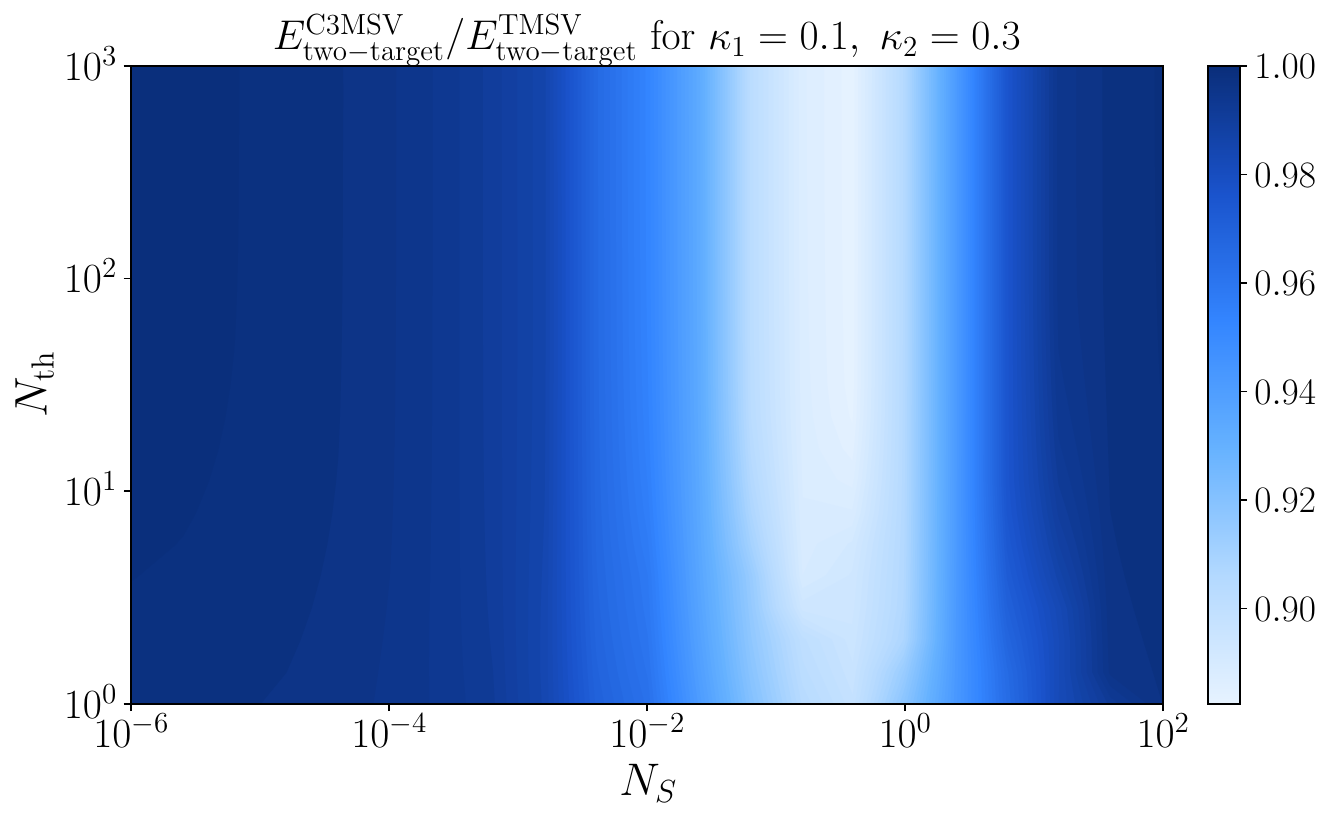}
    \caption{Ratio of the achievable exponent of \ac{C3MSV} in~\eqref{eq:C3MSV_two_target_exp} to achievable \ac{TMSV} exponent in~\eqref{eq:TMSV_two_target_exp}.}
    \label{fig:two_target_exp_comparison}
\end{figure}

In Figure~\ref{fig:two_target_exp_comparison}, we compare the exponents in \eqref{eq:TMSV_two_target_exp} and \eqref{eq:C3MSV_two_target_exp} by plotting their ratio over broad range of signal mean photon number $N_S$ and noise mean photon number $N_\textnormal{th}$. The figure shows that for a broad range of $(N_S,N_\textnormal{th})$ the exponents align and only for a small region of moderate $N_S$ regime does the \ac{TMSV} outperform \ac{C3MSV}. Even in the regime where \ac{C3MSV} underperforms \ac{TMSV} the degradation in the performance is upper bounded by $12\%$ of the \ac{TMSV} exponent. In the low signal energy, $N_S\ll1$, and high thermal-noise $N_\textnormal{th}\gg1$ the ratio converges to one, consistent with the asymptotic expressions of~\eqref{eq:TMSV_two_target_exp_lowNs} and~\eqref{eq:C3MSV_two_target_exp_lowN}. Although, the total mean photon number of the probe states is the same for \ac{TMSV} and \ac{C3MSV} ($4 N_S$), while \ac{TMSV} is using four modes, \ac{C3MSV} is using only three modes. In an experimental setting, \ac{C3MSV} would require less quantum memory to store the idler mode in comparison to \ac{TMSV}.

\bibliographystyle{apsrev4-2}
\bibliography{references.bib}

@ARTICLE{Bennett2002,
  author =       {Charles H. Bennett and Peter W. Shor and John A. Smolin and Ashish V. Thapliyal},
  journal =      {IEEE Transactions on Information Theory},
  title =        {Entanglement-assisted capacity of a quantum channel
                  and the reverse Shannon theorem},
  year =         2002,
  month =        {Oct},
  volume =       48,
  number =       10,
  pages =        {2637-2655},
  doi =          {10.1109/TIT.2002.802612}
}

@article{Wilde2017,
  title =        {Gaussian Hypothesis Testing and Quantum
                  Illumination},
  author =       {Wilde, Mark M. and Tomamichel, Marco and Lloyd, Seth
                  and Berta, Mario},
  journal =      {Phys. Rev. Lett.},
  volume =       119,
  issue =        12,
  pages =        120501,
  numpages =     6,
  year =         2017,
  month =        {Sep},
  publisher =    {American Physical Society},
  doi =          {10.1103/PhysRevLett.119.120501}
}

@article{Zhuang2021,
  title =        {Quantum Ranging with Gaussian Entanglement},
  author =       {Zhuang, Quntao},
  journal =      {Phys. Rev. Lett.},
  volume =       126,
  issue =        24,
  pages =        240501,
  numpages =     7,
  year =         2021,
  month =        {Jun},
  publisher =    {American Physical Society},
  doi =          {10.1103/PhysRevLett.126.240501}
}

@InProceedings{Wang2022Joint,
  author =       {Shi-Yuan Wang and Tuna Erdo\u{g}an and Uzi Pereg and
                  Matthieu R Bloch},
  booktitle =    {Proc. of IEEE Information Theory Workshop (ITW)},
  title =        {Joint Quantum Communication and Sensing},
  year =         2022,
  month =        aug,
  pages =        {506-511},
  doi =          {10.1109/ITW54588.2022.9965810}
}

@Article{Li2016Discriminating,
  author =       {Ke Li},
  journal =      {The Annals of Statistics},
  title =        {Discriminating quantum states: The multiple Chernoff
                  distance},
  year =         2016,
  month =        {Aug},
  number =       4,
  volume =       44,
  pages = {1661--1679},
  doi =          {10.1214/16-aos1436},
  publisher =    {Institute of Mathematical Statistics}
}

@Article{Weedbrook2012,
  author =       {Weedbrook, Christian and Pirandola, Stefano and
                  Garc\'{\i}a-Patr\'on, Ra\'ul and Cerf, Nicolas
                  J. and Ralph, Timothy C. and Shapiro, Jeffrey H. and
                  Lloyd, Seth},
  journal =      {Rev. Mod. Phys.},
  title =        {Gaussian quantum information},
  year =         2012,
  month =        {May},
  pages =        {621--669},
  volume =       84,
  doi =          {10.1103/RevModPhys.84.621},
  issue =        2,
  numpages =     0,
  publisher =    {American Physical Society},
}

@Book{Helstrom1976,
  author =	 {Helstrom, Carl W.},
  publisher =	 {Academic Press},
  title =	 {Quantum Detection and Estimation Theory},
  year =	 1976,
  address =	 {NY},
  isbn =	 9780080956329,
  issn =	 {0022-4715},
}

@article{Nair20,
author = {Ranjith Nair and Mile Gu},
journal = {Optica},
number = {7},
pages = {771--774},
publisher = {Optica Publishing Group},
title = {Fundamental limits of quantum illumination},
volume = {7},
month = {Jul},
year = {2020},
doi = {10.1364/OPTICA.391335},
}

@article{Pirandola2008,
  title = {Computable bounds for the discrimination of Gaussian states},
  author = {Pirandola, Stefano and Lloyd, Seth},
  journal = {Phys. Rev. A},
  volume = {78},
  issue = {1},
  pages = {012331},
  numpages = {8},
  year = {2008},
  month = {Jul},
  publisher = {American Physical Society}
}

@article{shi2020,
  title = {Practical Route to Entanglement-Assisted Communication Over Noisy Bosonic Channels},
  author = {Shi, Haowei and Zhang, Zheshen and Zhuang, Quntao},
  journal = {Phys. Rev. Appl.},
  volume = {13},
  issue = {3},
  pages = {034029},
  numpages = {18},
  year = {2020},
  month = {Mar},
  publisher = {American Physical Society},
  doi = {10.1103/PhysRevApplied.13.034029}
}

@InProceedings{MunarVallespir2024Joint,
  author =       {Munar-Vallespir, Pere and N\"otzel, Janis},
  booktitle =    {Proc. of IEEE World Forum on Internet of Things
                  },
  title =        {Joint communication and sensing over the lossy
                  bosonic quantum channel},
  year =         2024,
  month =        {Nov},
  pages =        {1--6},
  doi =          {10.1109/wf-iot62078.2024.10811180}
}

@article{Audenaert2007,
  title = {Discriminating States: The Quantum Chernoff Bound},
  author = {Audenaert, K. M. R. and Calsamiglia, J. and Mu\~noz-Tapia, R. and Bagan, E. and Masanes, Ll. and Acin, A. and Verstraete, F.},
  journal = {Phys. Rev. Lett.},
  volume = {98},
  issue = {16},
  pages = {160501},
  numpages = {4},
  year = {2007},
  month = {Apr},
  publisher = {American Physical Society},
  doi = {10.1103/PhysRevLett.98.160501}
}

@ARTICLE{Nitinawarat2013,
  author={Nitinawarat, Sirin and Atia, George K. and Veeravalli, Venugopal V.},
  journal={IEEE Transactions on Automatic Control}, 
  title={Controlled Sensing for Multihypothesis Testing}, 
  year={2013},
  month = {May},
  volume={58},
  number={10},
  pages={2451-2464},
  doi={10.1109/TAC.2013.2261188}}

@ARTICLE{Chang2023,
  author={Chang, Meng-Che and Wang, Shi-Yuan and Erdoğan, Tuna and Bloch, Matthieu R.},
  journal={IEEE Journal on Selected Areas in Information Theory}, 
  title={Rate and Detection-Error Exponent Tradeoff for Joint Communication and Sensing of Fixed Channel States}, 
  month={May},
  year={2023},
  volume={4},
  number={},
  pages={245-259},
  doi={10.1109/JSAIT.2023.3275877}
  }

@INPROCEEDINGS{Chang2022,
  author={Chang, Meng-Che and Erdoğan, Tuna and Wang, Shi-Yuan and Bloch, Matthieu R.},
  booktitle={Proc. of IEEE International Symposium on Joint Communications \& Sensing (JC\&S)}, 
  title={Rate and Detection Error-Exponent Tradeoffs of Joint Communication and Sensing}, 
  month={Mar},
  year={2022},
  volume={},
  number={},
  pages={1-6},
  doi={10.1109/JCS54387.2022.9743498}
  }

@ARTICLE{Joudeh2022,
  author={Joudeh, Hamdi and Willems, Frans M. J.},
  journal={IEEE Journal on Selected Areas in Information Theory}, 
  title={Joint Communication and Binary State Detection}, 
  month={Mar},
  year={2022},
  volume={3},
  number={1},
  pages={113-124},
  doi={10.1109/JSAIT.2022.3157999}
  }

@ARTICLE{Ahmadipour2024,
  author={Ahmadipour, Mehrasa and Kobayashi, Mari and Wigger, Michele and Caire, Giuseppe},
  journal={IEEE Transactions on Information Theory}, 
  title={An Information-Theoretic Approach to Joint Sensing and Communication}, 
  month = {Feb},
  year={2024},
  volume={70},
  number={2},
  pages={1124-1146},
  doi={10.1109/TIT.2022.3176139}
  }

@misc{erdogan2025,
      title={Joint Communication and Sensing with Bipartite Entanglement over Bosonic Channels}, 
      author={Tuna Erdo\u{g}an and Shi-Yuan Wang and Shang-Jen Su and Matthieu R. Bloch},
      year={2026},
      eprint={2501.15675},
      archivePrefix={arXiv},
      primaryClass={quant-ph},      
}

@article{tan2008QI,
  title = {Quantum Illumination with Gaussian States},
  author = {Tan, Si-Hui and Erkmen, Baris I. and Giovannetti, Vittorio and Guha, Saikat and Lloyd, Seth and Maccone, Lorenzo and Pirandola, Stefano and Shapiro, Jeffrey H.},
  journal = {Phys. Rev. Lett.},
  volume = {101},
  issue = {25},
  pages = {253601},
  numpages = {4},
  year = {2008},
  month = {Dec},
  publisher = {American Physical Society},
  doi = {10.1103/PhysRevLett.101.253601}
}

@article{Nussbaum2009Chernoff,
author = {Michael Nussbaum and Arleta Szko{\l}a},
title = {The Chernoff lower bound for symmetric quantum hypothesis testing},
volume = {37},
journal = {The Annals of Statistics},
number = {2},
publisher = {Institute of Mathematical Statistics},
pages = {1040 -- 1057},
month = {Apr},
year = {2009},
doi = {10.1214/08-AOS593}
}

@article{Zhan2023C3MSV,
  title = {Coupled three-mode squeezed vacuum: Gaussian steering and remote generation of Wigner negativity},
  author = {Zhan, Zi-wei and Lan, Bo and Wang, Jian and Xu, Xue-xiang},
  journal = {Phys. Rev. A},
  volume = {108},
  issue = {1},
  pages = {012436},
  numpages = {15},
  year = {2023},
  month = {Jul},
  publisher = {American Physical Society},
  doi = {10.1103/PhysRevA.108.012436},
  url = {https://link.aps.org/doi/10.1103/PhysRevA.108.012436}
}

@misc{cheng2026MQHT,
      title={Multiple Quantum Hypothesis Testing: One-Shot Pairwise Bounds and Sharp Asymptotics}, 
      author={Hao-Chung Cheng and Po-Chieh Liu},
      year={2026},
      eprint={2606.06246},
      archivePrefix={arXiv},
      primaryClass={quant-ph},
      url={https://arxiv.org/abs/2606.06246}, 
}

@article{Holevo2002EvaluatingCapacities,
  title = {Evaluating capacities of bosonic Gaussian channels},
  author = {Holevo, A. S. and Werner, R. F.},
  journal = {Phys. Rev. A},
  volume = {63},
  issue = {3},
  pages = {032312},
  numpages = {14},
  year = {2001},
  month = {Feb},
  publisher = {American Physical Society},
  doi = {10.1103/PhysRevA.63.032312},
  url = {https://link.aps.org/doi/10.1103/PhysRevA.63.032312}
}

@ARTICLE{2022PeregCommEst,
  author={Pereg, Uzi},
  journal={IEEE Transactions on Information Theory}, 
  title={Communication Over Quantum Channels With Parameter Estimation}, 
  year={2022},
  volume={68},
  number={1},
  pages={359-383},
  doi={10.1109/TIT.2021.3123221}
  }

@INPROCEEDINGS{Munar2024,
  author    = {Munar-Vallespir, Pere and N\"otzel, Janis and Seitz, Florian},
  title     = {Joint Communication and Eavesdropper Detection on the Lossy Bosonic Channel},
  booktitle = {Proc. of IEEE Global Communications Conference (GLOBECOM)},
  year      = {2024},
  pages={3473-3478},
  doi={10.1109/GLOBECOM52923.2024.10901161}
  }

@article{Gong2025JcommChangePD,
  title = {Quantum-enhanced change detection and joint communication detection},
  author = {Gong, Zihao and Guha, Saikat},
  journal = {Phys. Rev. A},
  volume = {112},
  issue = {3},
  pages = {032604},
  numpages = {8},
  year = {2025},
  month = {Sep},
  publisher = {American Physical Society},
  doi = {10.1103/fpc3-lmql},
  url = {https://link.aps.org/doi/10.1103/fpc3-lmql}
}

@ARTICLE{YaoJCS2025,
  author={Yao, Yuhang and Jafar, Syed A.},
  journal={IEEE Transactions on Communications}, 
  title={On the Utility of Quantum Entanglement for Joint Communication and Instantaneous Detection}, 
  year={2025},
  volume={73},
  number={11},
  pages={11722-11735},
  doi={10.1109/TCOMM.2025.3573393}
  }

@article{CraddockDistributionNY2024,
  title = {Automated Distribution of Polarization-Entangled Photons Using Deployed New York City Fibers},
  author = {Craddock, Alexander N. and Lazenby, Anne and Portmann, Gabriel Bello and Sekelsky, Rourke and Flament, Mael and Namazi, Mehdi},
  journal = {PRX Quantum},
  volume = {5},
  issue = {3},
  pages = {030330},
  numpages = {7},
  year = {2024},
  month = {Aug},
  publisher = {American Physical Society},
  doi = {10.1103/PRXQuantum.5.030330},
  url = {https://link.aps.org/doi/10.1103/PRXQuantum.5.030330}
}

@article{LuoEntanglingMemories2026,
  title = {Entangling Quantum Memories through a 420 km Long Fiber},
  author = {Luo, Xi-Yu and Wang, Chao-Yang and Zheng, Ming-Yang and Wang, Bin and Liu, Jian-Long and Gao, Bo-Feng and Li, Jun and Yan, Zi and Ke, Qiao-Mu and Teng, Da and Wang, Rui-Chun and Wu, Jun and Huang, Jia and Li, Hao and You, Li-Xing and Xie, Xiu-Ping and Xu, Feihu and Zhang, Qiang and Bao, Xiao-Hui and Pan, Jian-Wei},
  journal = {Phys. Rev. Lett.},
  volume = {137},
  issue = {7},
  pages = {070801},
  numpages = {7},
  year = {2026},
  month = {Aug},
  publisher = {American Physical Society},
  doi = {10.1103/ccd6-rf1s}
  }

@article{knaut2024entanglement,
  title={Entanglement of nanophotonic quantum memory nodes in a telecom network},
  author={Knaut, C. M. and Suleymanzade, A. and Wei, Y.-C. and Assumpcao, D. R. and Stas, P.-J. and Huan, Y. Q. and Machielse, B. and Knall, E. N. and Sutula, M. and Baranes, G. and Sinclair, N. and De-Eknamkul, C. and Levonian, D. S. and Bhaskar, M. K. and Park, H. and Lon{\v{c}}ar, M. and Lukin, M. D.},
  journal={Nature},
  volume={629},
  number={8012},
  pages={573--578},
  year={2024},
  publisher={Nature Publishing Group UK London}
}

@article{XuExperimentalQuantumTarget2021,
  title = {Experimental Quantum Target Detection Approaching the Fundamental Helstrom Limit},
  author = {Xu, Feixiang and Zhang, Xiao-Ming and Xu, Liang and Jiang, Tao and Yung, Man-Hong and Zhang, Lijian},
  journal = {Phys. Rev. Lett.},
  volume = {127},
  issue = {4},
  pages = {040504},
  numpages = {7},
  year = {2021},
  month = {Jul},
  publisher = {American Physical Society},
  doi = {10.1103/PhysRevLett.127.040504},
  url = {https://link.aps.org/doi/10.1103/PhysRevLett.127.040504}
}

@article{WardEntanglementNP2025,
  title = {Entanglement-enhanced Neyman-Pearson target detection},
  author = {Ward, William and Hariri, Abdulkarim and Zhang, Zheshen},
  journal = {Phys. Rev. A},
  volume = {112},
  issue = {5},
  pages = {052613},
  numpages = {13},
  year = {2025},
  month = {Nov},
  publisher = {American Physical Society},
  doi = {10.1103/cxl9-ljn8},
  url = {https://link.aps.org/doi/10.1103/cxl9-ljn8}
}

@article{HaoEASurpassCC2021,
  title = {Entanglement-Assisted Communication Surpassing the Ultimate Classical Capacity},
  author = {Hao, Shuhong and Shi, Haowei and Li, Wei and Shapiro, Jeffrey H. and Zhuang, Quntao and Zhang, Zheshen},
  journal = {Phys. Rev. Lett.},
  volume = {126},
  issue = {25},
  pages = {250501},
  numpages = {7},
  year = {2021},
  month = {Jun},
  publisher = {American Physical Society},
  doi = {10.1103/PhysRevLett.126.250501},
  url = {https://link.aps.org/doi/10.1103/PhysRevLett.126.250501}
}

@article{
XuIngeratedDistributed2024,
author = {Yuehan Xu  and Tao Wang  and Peng Huang  and Guihua Zeng },
title = {Integrated Distributed Sensing and Quantum Communication Networks},
journal = {Research},
volume = {7},
number = {},
pages = {0416},
year = {2024},
doi = {10.34133/research.0416}
}

@article{LiuINtegratedQuantum2024,
author = {Shuaishuai Liu and Yan Tian and Yu Zhang and Zhenguo Lu and Xuyang Wang and Yongmin Li},
journal = {Optica},
number = {12},
pages = {1762--1772},
publisher = {Optica Publishing Group},
title = {Integrated quantum communication network and vibration sensing in optical fibers},
volume = {11},
month = {Dec},
year = {2024},
url = {https://opg.optica.org/optica/abstract.cfm?URI=optica-11-12-1762},
doi = {10.1364/OPTICA.537655}
}

@misc{su2026,
      title={Quantum Ring States}, 
      author={Shang-Jen Su and Shi-Yuan Wang and Tuna Erdoğan and Zheshen Zhang and Matthieu R. Bloch},
      year={2026},
      eprint={2410.17181},
      archivePrefix={arXiv},
      primaryClass={quant-ph},
      url={https://arxiv.org/abs/2410.17181}, 
}

@INPROCEEDINGS{Wang2024JCS,
  author={Wang, Shi-Yuan and Chang, Meng-Che and Bloch, Matthieu R.},
  booktitle={Proc. of Annual Conference on Information Sciences and Systems (CISS)}, 
  title={Covert Joint Communication and Sensing under Variational Distance Constraint}, 
  year={2024},
  volume={},
  number={},
  pages={1-6},
  doi={10.1109/CISS59072.2024.10480161}}

@article{Wolf2006GaussianExtremality,
  title = {Extremality of Gaussian Quantum States},
  author = {Wolf, Michael M. and Giedke, Geza and Cirac, J. Ignacio},
  journal = {Phys. Rev. Lett.},
  volume = {96},
  issue = {8},
  pages = {080502},
  numpages = {4},
  year = {2006},
  month = {Mar},
  publisher = {American Physical Society},
  doi = {10.1103/PhysRevLett.96.080502},
  url = {https://link.aps.org/doi/10.1103/PhysRevLett.96.080502}
}

@Article{Paul2017Survey,
  author          = {Bryan Paul and Alex R. Chiriyath and Daniel W. Bliss},
  journal         = {{IEEE} Access},
  title           = {Survey of {RF} Communications and Sensing Convergence
                  Research},
  year            = 2017,
  pages           = {252--270},
  volume          = 5,
  doi             = {10.1109/access.2016.2639038},
  publisher       = {Institute of Electrical and Electronics Engineers ({IEEE})},
}

@TechReport{Fettweis2021Joint,
  author =	 {Gerhard Fettweis and Martin Schl\"uter and Reiner
                  Thom\"{a} and Holger Boche and Hans Schotten and
                  André Barreto and Andreas Mueller and Attila Bilgic
                  and Christoph Hildebrand and Dominic Schupke and
                  Eckhard Grass and Frank Hofmann and Gerhard Kahmen
                  and Stephan Blicker and Isabelle Korthals and Hans
                  J. Einsiedler and Ingo Rechenberger and Gerd
                  Zimmermann and Heinrich Heiss and Uli Klepser and
                  Martin Claus and Hendrik Seidel and Jan Goerlich and
                  Lars Lemke and Meik Kottkamp and Michael Meyer and
                  Padmanava Sen and Patrick Scheele and Paul Bender
                  and Alexander Deder and R\"udiger Quay and Slawomir
                  Stanczak and Snezhana Jovanoska and Stefan
                  K\"{o}psell and Thomas Dallmann and Thomas Multerer
                  and Thomas Neugebauer and Thorsten Wild and Uli
                  Barth and Wolfgang Heinrich and Milos Krstic and
                  Giovanni Del Galdo and Rahmi Salman and Stefan
                  Br\"uggenwirth and Ralf Irmer},
  institution =	 {VDE},
  title =	 {Joint Communications \& Sensing: Common Radio-Communications and sensor technology},
  year =	 2021,
  month =	 jul,
  type =	 {Position Paper}
}

@article{Kimble2008QuantumInternet,
  title={The quantum internet},
  author={Kimble, H Jeff},
  journal={Nature},
  volume={453},
  number={7198},
  pages={1023--1030},
  year={2008},
  publisher={Nature Publishing Group UK London}
}

@article{Wehner2018QuantumInternet,
  title={Quantum internet: A vision for the road ahead},
  author={Wehner, Stephanie and Elkouss, David and Hanson, Ronald},
  journal={Science},
  volume={362},
  number={6412},
  pages={eaam9288},
  year={2018},
  publisher={American Association for the Advancement of Science}
}

@article{Candia2021TwoWayCovert,
  title = {Two-Way Covert Quantum Communication in the Microwave Regime},
  author = {Di Candia, R. and Yi\ifmmode \breve{g}\else \u{g}\fi{}itler, H. and Paraoanu, G.S. and J\"antti, R.},
  journal = {PRX Quantum},
  volume = {2},
  issue = {2},
  pages = {020316},
  numpages = {24},
  year = {2021},
  month = {May},
  publisher = {American Physical Society},
  doi = {10.1103/PRXQuantum.2.020316},
  url = {https://link.aps.org/doi/10.1103/PRXQuantum.2.020316}
}

@book{cover1991elements,
  title={Elements of Information Theory},
  author={Cover, Thomas M and Thomas, Joy A},
  edition={2nd},
  year={2006},
  publisher={Wiley},
  address   = {NJ}
}
\end{document}